\documentclass[twocolumn,aps,prd,nofootinbib,floatfix,superscriptaddress]{revtex4-2}
\usepackage[caption=false]{subfig}
\usepackage{algorithm,algorithmic, refcount}
\usepackage{xcolor}
\usepackage{graphicx}	
\usepackage{amsmath}	
\usepackage{amssymb}
\usepackage{mathtools}
\usepackage{comment}
\usepackage{dsfont}
\usepackage{ulem}
\usepackage{float}
\usepackage{array}
\usepackage{multirow}
\usepackage{natbib}
\usepackage{nccmath}
\usepackage{hyperref}
\usepackage{orcidlink}
\usepackage[utf8]{inputenc}

\definecolor{burgundy}{rgb}{0.5, 0.0, 0.13}

\makeatletter
\renewcommand\@makecaption[2]{%
  \par
  \vskip\abovecaptionskip
  \begingroup
   \small\rmfamily
    \begingroup
     \samepage
     \flushing
     \let\footnote\@footnotemark@gobble
     \@make@capt@title{#1}{#2}\par
    \endgroup
  \endgroup
  \vskip\belowcaptionskip
}
\makeatother

\begin{document}


\title{Principal Components for Model-Agnostic Modified Gravity with 3$\times$2pt}

\author{C. M. A. Zanoletti \orcidlink{0009-0002-8713-6404}}
 \email{Contact author: c.m.a.zanoletti2@newcastle.ac.uk}
\author{C. D. Leonard\,\orcidlink{0000-0002-7810-6134}}%
\affiliation{School of Mathematics, Statistics and Physics, Newcastle University, Newcastle upon Tyne, United Kingdom.}

\date{\today}

\begin{abstract}
To mitigate the severe information loss arising from widely adopted linear scale cuts in constraints on modified gravity parameterisations with Weak Lensing (WL) and Large-Scale Structure (LSS) data, we introduce a novel alternative method for data reduction. This Principal Component Analysis (PCA)-based framework extracts key features in the matter power spectrum arising from nonlinear effects in a set of representative gravity theories. By performing the analysis in the space of principal components, we can replace sweeping `linear-only' scale cuts with targeted cuts on the transformed data vector, ultimately reducing parameter bias and significantly tightening constraints. We forecast constraints on a minimal parameterised extension to $\Lambda$CDM which includes modifications to the growth of structure and lensing of light ($\Lambda$CDM$+\mu_0+\Sigma_0$) using mock Stage-IV data for two simulated cosmologies: the $\Lambda$CDM model and Extended Shift Symmetric (ESS) gravity. Under the assumption of a Universe defined by $\Lambda$CDM and General Relativity, our method offers constraints on $\mu_0$ a factor of 1.65 tighter than traditional linear-only scale cuts. Crucially, our approach also provides the necessary constraining power to break key degeneracies in modified gravity without relying on $f\sigma_8$ measurements, introducing a promising new tool for the analysis of present and future WL and LSS photometric surveys.
\end{abstract}

\keywords{cosmology: cosmological parameters, nonlinear perturbations, matter power spectrum, large-scale structure of the Universe, weak gravitational lensing -- methods: numerical}

\maketitle

\section{Introduction}\label{sec:intro}

Next-generation Stage-IV late-time cosmology surveys, such as the Rubin Observatory Legacy Survey of Space and Time (LSST, \cite{Ivezi__2019}), Euclid \cite{Euclid_2011,euclidcollaboration2024euclidiovervieweuclid}, and the Roman Space Telescope \cite{Roman_2019}, promise to revolutionise our understanding of the Universe by providing unprecedented precision in cosmological measurements. These surveys will offer new opportunities to probe Dark Energy (DE), Dark Matter (DM), and modifications to gravity (MG) through a combination of enhanced statistical power and refined analysis techniques. The combination of varied and complementary datasets will enhance our ability to distinguish between competing cosmological models, potentially resolving existing tensions in the data and providing deeper insights into the fundamental physics of the universe. As these surveys begin, the landscape of cosmological research is set for a transformative decade \citep{abdalla2022cosmology}.

With the increasing volume and quality of data, cosmological parameter tensions become more pronounced \citep{Beaton_2016, freedman2017cosmologycrossroadstensionhubble, Ezquiaga_2018}. This makes the search for new physics, including signatures of alternative theories of gravity, a promising and exciting research avenue \citep{Di_Valentino_2021c}. Alternative theories of gravity have emerged as a crucial area of exploration in modern cosmology, driven by the limitations of General Relativity (GR) in explaining key phenomena such as the cosmological constant problem and the observed acceleration of the universe's expansion. A wide range of modified gravity candidates have been proposed to solve the problems and tensions that arise from the standard cosmological paradigm, but no theory has been singled out as a front-running competitor to $\Lambda$CDM  (for a review of cosmological modified gravity, see for example \cite{clifton2012modified,Koyama_2016, Nojiri_2017, Ishak_2018, Heisenberg_2019, Hohmann2021}). 

The range of available candidate gravity theories keeps growing, with the result being that testing all known viable modified gravity theories is prohibitively resource-intensive. Additionally, a purely theory-specific approach risks overlooking potential modified gravity signals in the data if the range of theories considered is too limited. We find a more attractive alternative in general parameterised frameworks. Such frameworks can in principle cover many theories at once and minimise the risk of missing potential hints of modified gravity in the data \citep{Pogosian_2016}, with the most prominent examples being the so-called `$\mu/\Sigma$' paramerisation \citep{Silvestri_2013} and the $\alpha$ parameterisation \citep{Bellini_2014}. Over the past several years, significant effort has gone into developing such frameworks and understanding requirements for their consistency, as reflected in the growing body of literature on the subject \citep{Silvestri_2013,Baker_2014,Leonard_2015,Hu_2007b, Gubitosi_2013, Gleyzes_2013, Bloomfield_2013, Hohmann2021}. 

These general parameterisation frameworks typically capture the behaviour of large classes of alternative theories of gravity, in part by restricting their {\it a priori} applicability to the linear regime. To circumvent potential model mis-specification biases arising from applying these parameterisations on nonlinear scales, the typical approach is to apply a severe cut at smaller, nonlinear scales \citep{DES_Y3_results, 2016, 2020}. However, this method, applied naively, risks discarding most if not all of our data, especially in high-precision Stage-IV surveys. Notably, over half of the signal-to-noise in some current and future surveys comes from the nonlinear regime, so cutting these scales would significantly reduce our ability to constrain cosmological parameters \citep{ishak2019modified}.

Accurate modelling of the nonlinear growth of cosmic structure is a significant challenge when testing modified gravity. In GR, we rely on fitting formulae or emulators derived from suites of N-body simulations to predict cosmological observables on small scales where nonlinear effects dominate. Creating equivalent models for each MG theory is computationally prohibitive. Ongoing attempts at mitigating this degradation of parameter constraints include improved (and faster) nonlinear modelling tools for broad classes of individual modified gravity theories \cite{Cataneo_2019,Giblin_2019,Cataneo_2019b,Bose_2020, Bose_2021, Wright_2023, Tsedrik_2024} or developing alternative general parameterisations that are able to probe further into the nonlinear regime (see e.g. \cite{Thomas_2020, Srinivasan_2021, Srinivasan_2024a, srinivasan_2024b, Thomas_2024}). The latter show very promising results but come with their own challenges, including requiring scale- as well as redshift- dependent descriptions of modifications to GR on all scales. Some of these larger MG parameter spaces are likely to introduce further issues by bringing additional degeneracies with cosmological parameters and with each other.

In this paper, we present a novel method for data reduction that aims to provide a less conservative (but still unbiased) alternative to severe linear scale cuts when constraining parameterised modified gravity with weak lensing and galaxy clustering surveys. Previous analogous studies introduce scale cuts which remove a large fraction of serviceable data (e.g. in the case of \cite{DES_Y3_results}, retaining only around 55\% of the fiducial dataset). Instead, we use a Principal Component Analysis (PCA)-based method (inspired by previous work presented in \cite{Huang_2019} and \cite{Eifler_2015} in the context of baryonic modelling) to identify the components in data space that contribute the most to deviations between the linear model and the data. To achieve this, we compute linear (approximate) and nonlinear (more accurate) model vectors for a range of alternative theories of gravity. The specific modified gravity model choices are less important, as long as they capture a diverse range of deviations from linear theory in the nonlinear regime beyond the General Relativistic prediction. These allow us to construct a reduction matrix that removes specifically these nonlinear signatures from the data. As we will describe in Section \ref{sec:results}, our PCA-based data reduction is effective at removing most of the deviations between our nonlinear vector and our linear model vector while minimising information loss.

This paper aims to introduce and analyse the feasibility of this method. We complete an LSST-like simulated parameter estimation as a proof-of-concept, and show that our method can achieve unbiased results while increasing the constraining power of current and future weak lensing and galaxy clustering surveys. We use the $\mu$-$\Sigma$ phenomenological prescription to parameterise our model (with redshift-dependence governed by the DE-like parameterisation, see for example \cite{2016, 2020, Abbott_2019, Abbott_2023}). 

The remainder of the paper is structured as follows. In Section \ref{sec:PCA_introsec}, we introduce the standard scale-cuts method, demonstrating the amount of data loss this involves, and we provide a high-level introduction to our alternative PCA-based data reduction method. In Section \ref{sec:MG}, we describe the modified gravity theories and screening mechanisms relevant to our data reduction method. In Section \ref{sec:Setup_Methodology}, we introduce the 3$\times$2pt and $f\sigma_8$ simulated data, as well as our likelihood and parameter inference methods (both for the standard method and for our novel approach). In Section \ref{sec:results}, we present the results of our proof-of-concept simulated analysis, and in Section \ref{sec:conclusion} we look at the limits and successes of this method, and the scope of future analyses. 

\section{Introducing the PCA-based Data Reduction for Modified Gravity}\label{sec:PCA_introsec}

\subsection{Conservative scale cuts}\label{sec:conservative_scalecuts}

During a standard cosmological parameter inference process, we apply scale cuts to work around limitations in current nonlinear modelling approaches (see, e.g., \citep{Abbott_2023}). This procedure is applied in $\Lambda$CDM analyses for the mitigation of primarily baryonic modelling uncertainty at small scales; in our case, we focus instead on uncertainty in nonlinear modelling for modified gravity theories and parameterisations.

The typical scale-cuts strategy for parameter inference when trying to constrain linear modified gravity parameterisations is to only keep data points in a given data vector $\mathbf{D}$ that are, by some definition,  in the linear regime. The usual criterion for the `linear regime' is determined on the basis of computing the linear (lin) and nonlinear (NL) theory prediction for our model data vectors $\mathbf{M}$ at some fiducial parameter values $\mathbf{p}^{\text{fid}}$ and calculating the quantity:

\begin{equation}
    \Delta \chi^2 = \left( \mathbf{M}_{\mathrm{NL}} - \mathbf{M}_{\mathrm{lin}} \right)^{T} 
    \mathbf{C}^{-1} 
    \left( \mathbf{M}_{\mathrm{NL}} - \mathbf{M}_{\mathrm{lin}} \right).
    \label{eq:deltachi2}
\end{equation}

We then identify the element of the data vector that contributes the most to $\Delta \chi^2$, and remove it. This process is repeated until $\Delta \chi^2 < 1$ \cite{DES_Y3_results, 2016, 2020}. 

This method for scale cuts has been commonly used in recent modified gravity analyses \cite{2016, abbott2019dark, 2020, DES_Y3_results}, but often requires discarding a significant portion of the data. This will become particularly significant in high-precision Stage-IV surveys like LSST and Euclid, where very small uncertainties on our measured data vectors require very precise modelling in order to avoid parameter inference biases. 

An issue that arises with the choice of scale cuts indicated in Equation \ref{eq:deltachi2} is the variability with which data points are selected in the case where preliminary cuts have already been applied to the data. For example, if the data vector has been pre-cut to mitigate biases due to uncertainties in baryonic modelling (see Section \ref{sec:likelihood_scalecuts}), then -- even if all points removed by these earlier, less stringent cuts are ultimately excluded by the application of Equation \ref{eq:deltachi2} -- the covariances of those points with the remaining data can affect the specifics of how the $\Delta \chi^2 < 1$ criterion is satisfied. In other words, because of the covariances in the data, the order in which data points are removed matters, which can result in differences in the choice of data points being cut.

Additionally, the assumption of a small deviation from GR is implicit in this method for scale cuts. The model data vectors ($\mathbf{M}$) used to inform these cuts are based on the $\Lambda$CDM model. These scale cuts could therefore be inadequate when testing MG theories where deviations between linear and nonlinear theory are larger (or become important at larger $k$, i.e. smaller scales, see Figure \ref{fig:Boosts} below). We will see that this is implicitly resolved in the PCA-based data reduction method we propose, where we directly and explicitly use the modified gravity power spectra difference between linear and nonlinear modelling for our data reduction.

\begin{figure}
    \centering \includegraphics[width=1.0\linewidth]{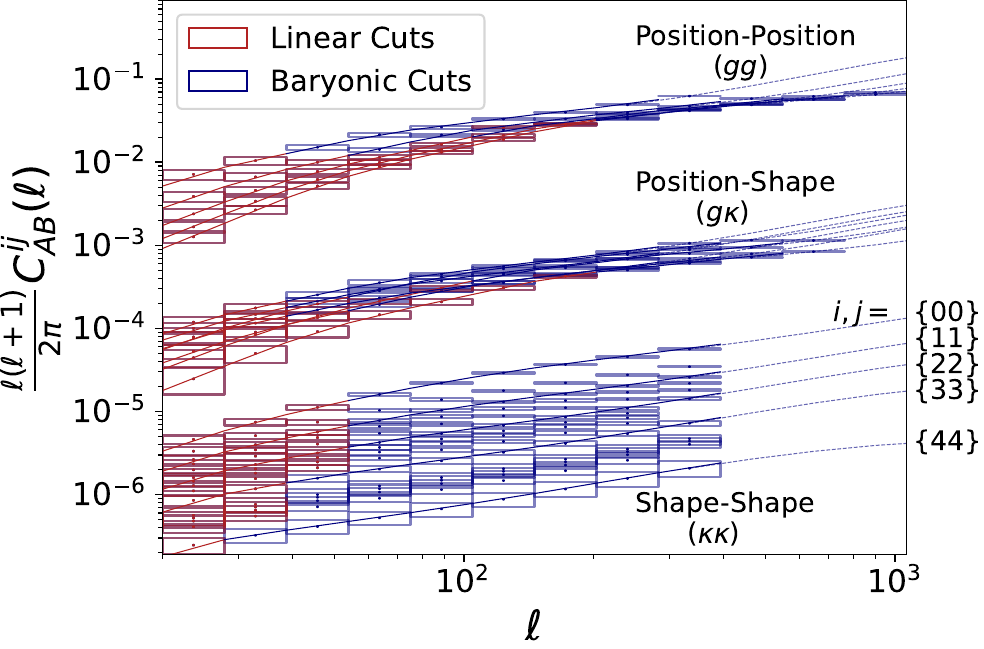}
    \caption{Illustration of scale-cuts on a simulated LSST Y1 3$\times$2pt Fourier-space data vector. With full DMO nonlinear modelling but no baryonic mitigation we could retain the navy blue data points (‘baryonic cuts’). With linear DMO modelling only, we could retain the points in red (`linear cuts').  For clarity, we only display the autocorrelations of the cosmic-shear (shape-shape) measurements.}
    \label{fig:Cuts_3x2pt}
\end{figure}

Let us look at current scale cuts for state-of-the-art constraints on alternative gravity. As an example, in \cite{DES_Y3_results}, a total of 462 data points for the fiducial case were cut down to 256 using Eq. \ref{eq:deltachi2}. Additionally, Figure \ref{fig:Cuts_3x2pt} shows LSST Year 1 (`Y1' hereafter) simulated data: in navy blue we display the datapoints that we could keep if we had adequate dark-matter-only (DMO) nonlinear modelling  for our modified gravity analysis,
and the limits in our model were driven by baryonic
effects only; in red are the datapoints we retain relying only on linear modelling and the scale cuts given by Eq. \ref{eq:deltachi2}. As a result of these intensive scale cuts, parameter constraints are significantly less precise than they would be with improved nonlinear modelling (see e.g. Figures 4 and 5 in \cite{wang2024extendingmgcambtestsgravity}). 

\subsection{PCA-based data reduction method}\label{sec:likelihood_PCA} 

To deal with the prohibitive information loss introduced by the standard linear scale cuts presented in the previous subsection, we propose a new data reduction method which takes advantage of the Principal Component Analysis (PCA) framework. We base our method closely on the one outlined in \cite{Huang_2019}, first introduced in \cite{Eifler_2015}, expanding upon that work to create a method applicable to constraints on deviations from General Relativity. By using power spectra for a representative set of models from different gravity theories of interest -- which we hereafter call our \textit{data reduction models} (for details of the specific theories used in this work, see Section \ref{sec:MG_theory}) -- we create a parameter-dependent \textit{reduction matrix}, $\mathbf{U}_{\text{ch,cut}}$ (defined below). The latter will apply a rotation and scale cut in data space, minimising both the information lost through the cut and the bias introduced by the lack of nonlinear modelling. We now outline the main steps of our method, and will expand on the method's details in Section \ref{sec:PCA_MGtheories}.

To begin, we consider a simplified analysis set-up, to illustrate how we create a reduction matrix $\mathbf{U}_{\text{ch,cut}}$ from data reduction models. Because this matrix depends on the cosmological parameters of our model vector, our scale cuts will also be cosmological parameter-dependent. This means we will find a reduction matrix and apply a scale cut to our data at each point when sampling our parameter space. We consider the process that occurs during a single step within a sampling algorithm (e.g. Markov Chain Monte Carlo (MCMC) sampling). We assume a simulated measured data vector with GR gravity and parameters $\mathbf{p}=\mathbf{p}^{\text{fid}}$ (as defined in Table \ref{table:p_fid_and_priors} below). The discussion below will refer to a step sampled at an illustrative point in parameter space: 
\begin{equation*}
\begin{split}
    \mathbf{p}^{\text{mcmc}} =& \{\Omega_c: 0.23, A_s: 2.4\times 10^{-9}, h: 0.61, \\
    &n_s: 0.962,
    \Omega_b: 0.05,b_1:1.2,b_2:1.4,\\&b_3:1.8,b_4:2.0,b_5:2.2, \mu_0:0.0,\Sigma_0:0.0)\}.
\end{split}
\end{equation*}
For definitions of these parameters, see Section \ref{sec:priors} below.

\begin{figure}
    \centering
    \includegraphics[width=1.0\linewidth]{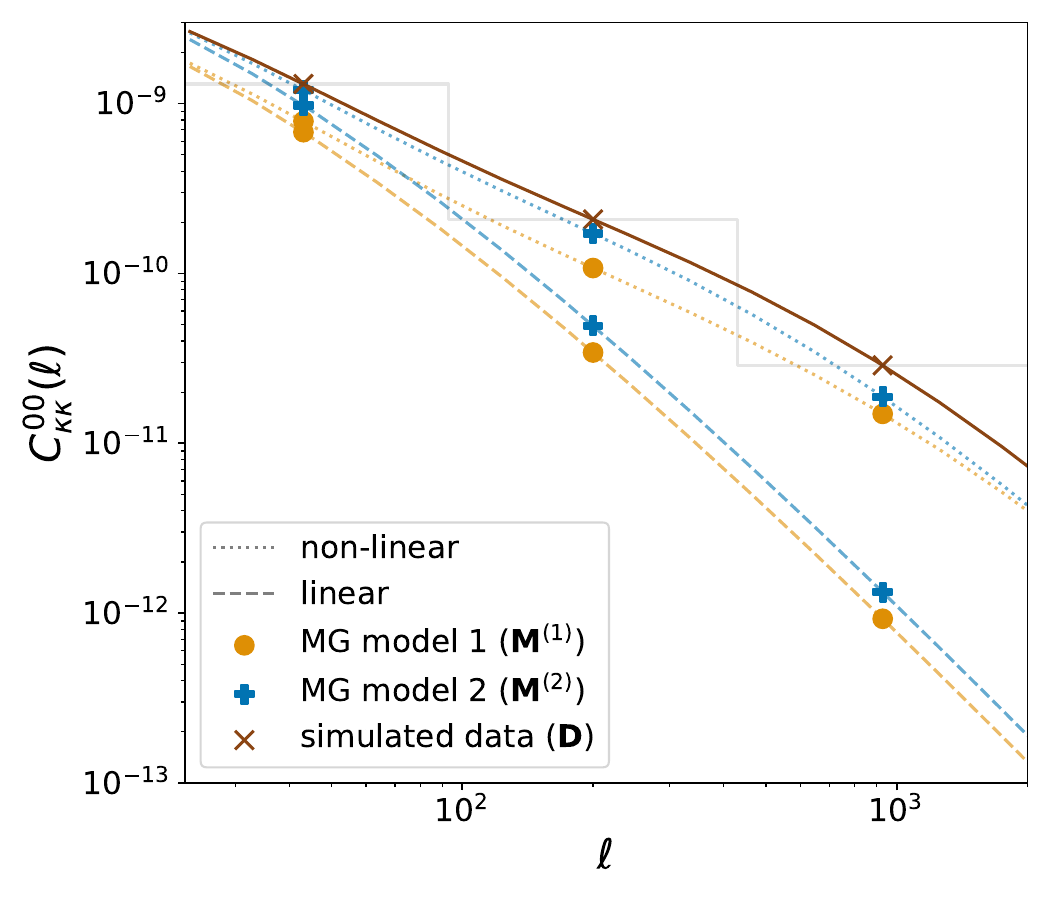}
    \caption{Simplified example analysis: the simulated data vector (brown, $(\times)$) is shown together with the model data vectors for two data reduction models (orange $(\text{o})$ and blue $(+)$) under the linear (dashed) and nonlinear (dotted) modelling.}
    \label{fig:Cell_example}
\end{figure}

As we explain further in Section \ref{sec:3x2pt_data}, in this paper we will often work with a 3$\times$2pt data vector:
\begin{equation*}
\begin{split}
     \mathbf{D} = \{C^{ij}_{A B}(\ell), i,j \in (0,...,4) \text{ and } A,B \in (\kappa, \delta_g)\},\\ \hfill\text{dim}(\mathbf{D}) = n
\end{split}
\end{equation*}
where $C$ is an angular power spectrum, $\ell$ is the angular multipole, and $i$ and $j$ label the redshift bins. $A$ and $B$ denote the type of probe being cross-correlated: either the source galaxy sample shear field ($\kappa$) or the lens galaxy sample overdensity field ($\delta_g$). However, in the illustrative scenario of this section, for the purposes of visualisation we will instead consider the (much reduced and less realistic) data vector $\mathbf{D} = C^{00}_{\kappa\kappa}([\ell_1,\ell_2,\ell_3])$. This is the angular power spectrum of the autocorrelation of the $0^{\text{th}}$ redshift bin of the shear field $\kappa$, where the data is divided in three $\ell$ bins only, spaced logarithmically between $\ell=20$ and $\ell=2000$ (centred at $\ell_1 = 43, \ell_2 = 200$ and $\ell_3 = 928$). This data vector is displayed in Figure \ref{fig:Cell_example}. We compute the covariance matrix $\mathbf{C}$\footnote{For this example, we computed a Gaussian covariance for a survey with sky fraction $f_{\text{sky}} = 0.42$, number density of sources per bin $N^i_{\text{gal, src}}= 1.78 \text{ arcmin}^{-2}$, standard deviation of the distribution of source ellipticities $\sigma_e= 0.26$, source photo-z scatter per bin $\sigma^i_z= 0.05$, photometric redshift bins from Section \ref{sec:3x2pt_data}, and cosmological and nuisance parameters $\mathbf{p}^{\text{fid}}$ as defined in Table \ref{table:p_fid_and_priors}.} for this data by using the publicly available code \texttt{TJPCov}\footnote{\hyperlink{https://github.com/LSSTDESC/tjpcov}{https://github.com/LSSTDESC/tjpcov}}.

We can then find the model vector $\mathbf{M}$ at $\mathbf{p}^{\text{mcmc}}$ under the relevant data reduction models as well as under the parameterised modified gravity model to be constrained. For the purpose of this example, assume that we use only two data reduction models to build our reduction matrix, labelled ``MG model 1'' and ``MG model 2'', respectively the orange $(\text{o})$ and blue $(+)$ markers in Figure \ref{fig:Cell_example} (concretely, these are chosen here as the null case (GR) and nDGP respectively -- more details can be found in Section \ref{sec:PCA_MGtheories} below). 

In order to find our data reduction matrix we first compute the difference between the angular power spectra in our nonlinear model ($\mathbf{M}^{(i)}_{\text{NL}} = C^{00}_{\kappa\kappa, \text{NL}}(\ell)$) and linear model ($\mathbf{M}^{(i)}_{\text{lin}} = C^{00}_{\kappa\kappa, \text{lin}}(\ell)$) \textit{for each of our data reduction models}:
$$\Delta \mathbf{M}^{(i)} = \mathbf{M}^{(i)}_{\text{NL}}(\mathbf{p}^{\text{mcmc}}) - \mathbf{M}^{(i)}_{\text{lin}}(\mathbf{p}^{\text{mcmc}})$$
with $i=1,2$ labelling the data reduction model in question. In order to incorporate available information on the relative importance of each data point on final parameter constraints, we follow \cite{Huang_2019} in defining a covariance-weighted difference of our model data vectors:
\begin{equation}
\Delta\mathbf{M}^{(i)}_{\text{ch}} = \mathbf{L}^{-1}\Delta\mathbf{M}^{(i)} \hspace{0.8em} , \hspace{0.8em} \mathbf{D}_{\text{ch}} = \mathbf{L}^{-1}\mathbf{D} \hspace{0.8em} , \hspace{0.8em} \mathbf{M}_{\text{ch}} = \mathbf{L}^{-1}\mathbf{M}
\label{eq:mod_datared_diff}
\end{equation}
where the matrix $\mathbf{L}$ is the Cholesky decomposition of $\mathbf{C}$, the data covariance matrix described above:

\begin{ceqn}
\begin{align}\label{eq:Choleski_decomposition}
    \mathbf{C} = \mathbf{L} \mathbf{L}^{T}.
\end{align}
\end{ceqn}

\begin{figure}
    \centering \includegraphics[width=1.0\linewidth]{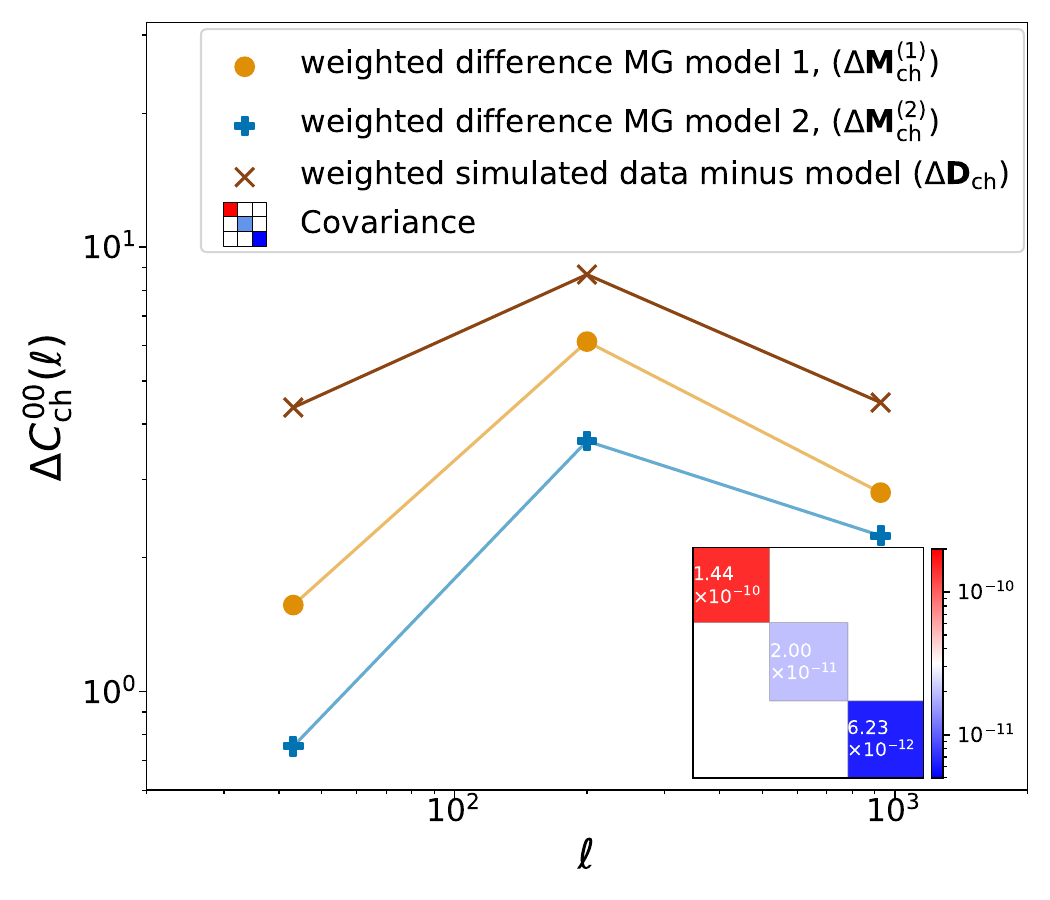}
\caption{Simplified example analysis: Cholesky-weighted difference vectors $\Delta C^{00}_{\kappa\kappa}(\ell)$ for the simulated data vector (brown, $(\times)$) and for the two data reduction models (orange $(\text{o})$ and blue $(+)$). We also display the Gaussian covariance matrix $\textbf{C}$ used for the Cholesky decomposition.}
    \label{fig:Cell_weighted_example}
\end{figure}

\begin{figure*}
    \centering
    \includegraphics[width=1.1\linewidth]{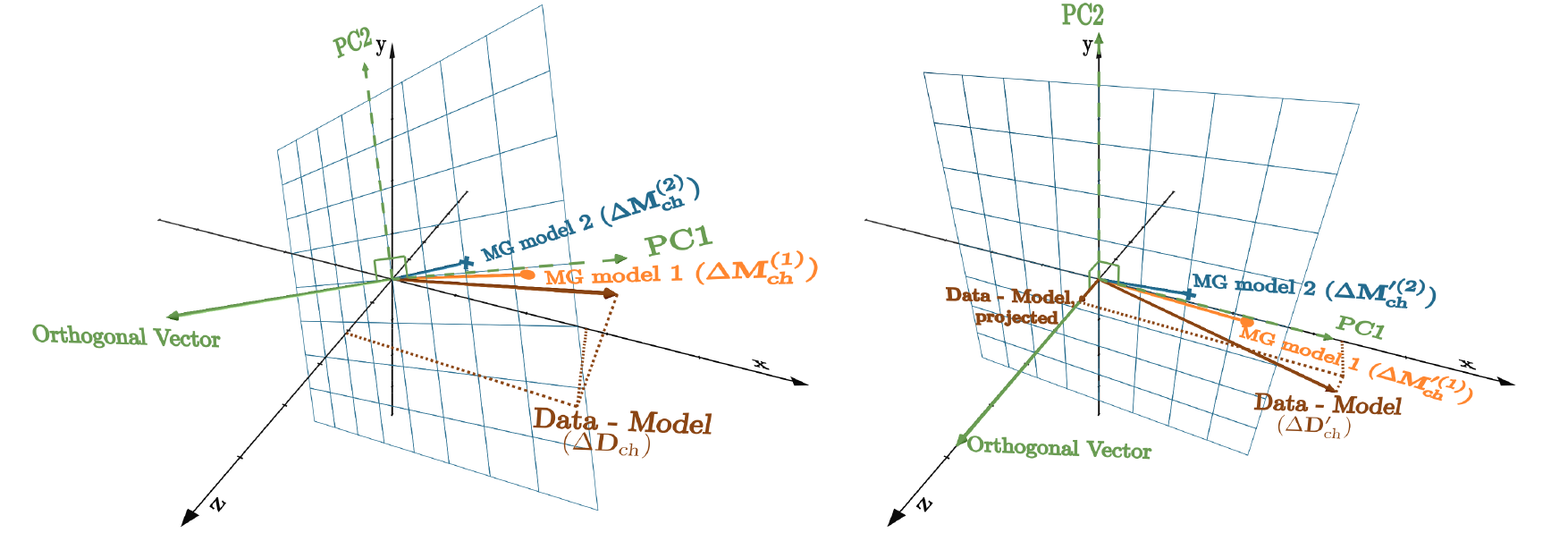}
    \caption{Simplified example analysis: The left panel displays a visualisation of the weighted data vector difference $\Delta\mathbf{D}_{\text{ch}}$ (brown) and the weighted model vector differences for the two data reduction models ($\Delta \mathbf{M}^{\text{(1)}}_{\text{ch}}$ in orange, $(\text{o})$, and $\Delta \mathbf{M}^{(2)}_{\text{ch}}$ in blue, $(+)$), in a 3D space in which the axes each represent the value of the data vector difference in one of three $\ell$ bins. These can be compared with the alternate visualisation of the identical quantities in Figure \ref{fig:Cell_weighted_example}. The Cartesian basis vectors $[x,y,z]$ are in black, the principal components (PCs) are the green dashed lines, and their orthogonal vector is in green. The equivalent quantities rotated by $\mathbf{U}_{\text{ch}}$ are displayed in the right panel. In a general problem, the space would be $n$ dimensional with $n$ the number of $\ell$ bins.}
    \label{fig:PCs_dataspace_all}
\end{figure*}

In Figure \ref{fig:Cell_weighted_example}, we display $\Delta\mathbf{M}^{(i)}_{\text{ch}}$ for each data reduction model within our simplified example. We also display the (similarly weighted) difference between the data vector and the model vector, this time computed within the linear $\mu \ \Sigma$ parameterisation (details of which are provided below in Section \ref{sec:parameterisations}):
\begin{equation}
\Delta\mathbf{D}_{\text{ch}} = \mathbf{D}_{\text{ch}}(\mathbf{p}^{\text{fid}}) - \mathbf{M}_{\text{ch}}(\mathbf{p}^{\text{mcmc}}).
\label{eq:Ddch_weighted}
\end{equation}
In our simple example, with a single angular power spectrum and only 3 $\ell$ bins, we can visualize our data vector difference $\Delta \mathbf{D}_{\text{ch}}([\ell_1, \ell_2, \ell_3])$ as a vector in $\mathbb{R}^3$, where its components correspond to the $x$-, $y$-, and $z$-coordinates in three-dimensional space (in general, the data vector can be thought of as a vector in a space with dimension equal to the number of datapoints). We can additionally populate this 3D space with two other points, each corresponding to the model data vector difference in one of our two data reduction models: $\Delta \mathbf{M}^{(1)}_{\text{ch}}([\ell_1, \ell_2, \ell_3])$ and $\Delta \mathbf{M}^{(2)}_{\text{ch}}([\ell_1, \ell_2, \ell_3])$. These are displayed in the left panel of Figure \ref{fig:PCs_dataspace_all}.

In order to find our data reduction matrix, we find the singular value decomposition (SVD) of the $3\times2$ difference matrix $\mathbf{\Delta} = \left[\Delta \mathbf{M}^{(1)}_{\text{ch}}([\ell_1, \ell_2, \ell_3]), \Delta \mathbf{M}^{(2)}_{\text{ch}}([\ell_1, \ell_2, \ell_3])\right]^T$:

$$\mathbf{\Delta} = \mathbf{U}_{\text{ch}} \mathbf{\Sigma} \mathbf{V}^T$$
where $\mathbf{\Sigma} = \left[\begin{matrix} \sigma_1 & 0 & 0 \\0&\sigma_2 & 0\end{matrix}\right]^T$ is a $3\times2$ matrix, where $\sigma_1$ and $\sigma_2$ are the square root of the eigenvalues of $\mathbf{\Delta}\mathbf{\Delta}^T$, and  $\mathbf{V}$ is a $2\times2$ square unitary matrix. Crucially, we find a $3\times3$ rotation matrix $\mathbf{U}_{\text{ch}}$:

$$\mathbf{U}_{\text{ch}} = \left[\begin{matrix}
    | & | \\
        PC_1 & PC_2\text{ }\\
        | & |
    \end{matrix} \begin{matrix}
    \text{silent} \\
        \text{ orthogonal} \\
        \text{vector}
    \end{matrix} \right].$$
The first two columns represent the first and second orthonormal directions where the Cholesky-weighted differences for our data reduction models are largest. These two vectors are displayed as $PC1$ and $PC2$ in the left panel of Figure \ref{fig:PCs_dataspace_all}. The third column is a vector orthonormal to the PCs. $\Delta \mathbf{M}^{(i)}_{\text{ch}}$ lie in the same plane as $PC1$ and $PC2$; this means that the differences for our data reduction models along the third orthogonal direction will be zero. Because of this, in the case where our data reduction models are representative of the space of alternative theories in question, linear modelling can be safely applied for a point in data space along this orthogonal vector. 

If we rotate each vector in our space to a new basis aligned with the the PC basis:
$$\Delta \mathbf{M}'^{(i)}_{\text{ch}} = \mathbf{U}_{\text{ch}}^T\Delta \mathbf{M}^{(i)}_{\text{ch}} $$
\vspace{-1em}
$$\Delta\mathbf{D}'_{\text{ch}} = \mathbf{U}_{\text{ch}}^T\Delta\mathbf{D}_{\text{ch}} $$
\vspace{-1em}
$$\begin{bmatrix}x'\\y'\\z'\end{bmatrix} = \mathbf{U}_{\text{ch}}^T\begin{bmatrix}PC1\\PC2\\\text{orthogonal vector}\end{bmatrix},$$
we rotate to the frame where $\Delta \mathbf{M}'^{(1)}_{\text{ch}}$ and $\Delta \mathbf{M}'^{(2)}_{\text{ch}}$ are perpendicular to the $z'$ axis (the direction of the rotated silent orthogonal vector, as displayed in the right panel of Figure \ref{fig:PCs_dataspace_all}). In the equations in this section, a prime is used to describe quantities in this rotated space.

We can see in the right  panel of Figure \ref{fig:PCs_dataspace_all} that, if we project the MG model 1 ($\Delta \mathbf{M}'^{(1)}_{\text{ch}}$, orange) and the MG model 2 ($\Delta \mathbf{M}'^{(2)}_{\text{ch}}$, blue) vectors onto the orthogonal vector, we have $\Delta \mathbf{M}'^{(i)}_{\text{ch}} \times [0,0,1]^T = 0$. This means that data projected along the orthogonal vector direction ($\mathbf{D}'_{\text{ch}}\times [0,0,1]^T$) should be well-modelled by linear theory (again, assuming our data reduction models are representative). If we project $\Delta\mathbf{D}'_{\text{ch}}$ onto the $z'$ axis (labelled \textit{Data - Model, projected} in the right panel of Figure \ref{fig:PCs_dataspace_all}) we recover a non-zero vector which should be minimally sensitive to nonlinear effects.

We define a data reduction matrix:
$$\mathbf{U}_{\text{ch,cut}} = \left[\begin{matrix}
    \text{silent} \\
     \text{orthogonal} \\
    \text{vector}
    \end{matrix} \right]$$    
that applies a cut to our data within our likelihood function to remove inaccuracies due to linear modelling (see Equation \eqref{eq:D_ch_cut} in Section \ref{sec:PCA_MGtheories} for details). 

In a realistic 3$\times$2pt analysis, the dimension of our data vector is not 3, but in the hundreds or thousands. Take, for example, the proof-of-concept analysis we will conduct in Section \ref{sec:PCA_MGtheories} below: in that case, we are working with a $254$-dimensional data space and 3 distinct data reduction models. This means that by removing the three principal components, we retain all the constraining power of the remaining $251$ rotated dimensions. 

\begin{figure*}
    \centering
    \includegraphics[width=0.9\linewidth]{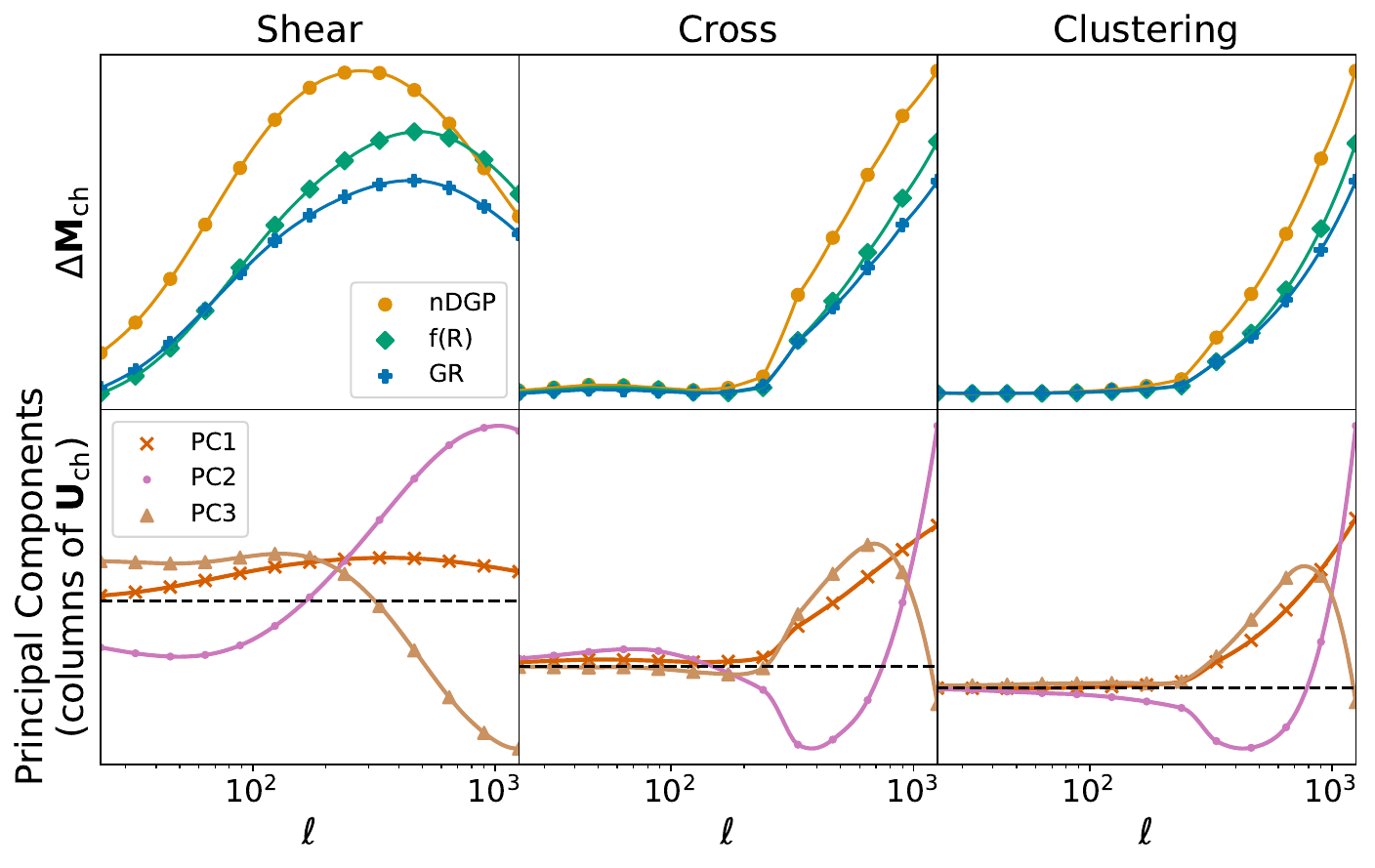}
    \caption{Weighted difference between linear and nonlinear model vectors for our three data reduction models ($\Delta\mathbf{M}_{\text{ch}}$, top; see Section \ref{sec:MG_theory} for details) and resulting principal components (bottom) for the $0^{\text{th}}$ source and lens redshift bins. The y-axis scale is not shown, as the scales of the shear, cross, and clustering subplots vary significantly.}
    \label{fig:PCs_ellspace}
\end{figure*}

While a $254$-dimensional space is impossible to visualise directly, we can still examine the PCs in $\ell$-space -- analogous to Figure \ref{fig:Cell_weighted_example} from our toy example -- to gain valuable insights into the effect of PCA data reduction in high dimensions. In Figure \ref{fig:PCs_ellspace}, we display principal components for the $00$ redshift bin combinations from our realistic analysis from Section \ref{sec:3x2pt_data} below. We can see that the principal components represent the most important common patterns in $\ell$-space of the difference model vectors $\Delta \mathbf{M}_{\text{ch}}$. Much like in the low-dimensional case which we have used as an illustrative example, once we identify the key data-space directions that contribute to the modelling discrepancies introduced by the linear approximation, we can remove them from both our data and theory vectors before computing the likelihood.

In Section \ref{sec:PCA_MGtheories}, we will explain in detail how we construct a realistic reduction matrix in high-dimensional space and situate this method in a likelihood estimation pipeline. However, before diving into the mechanics of this approach, we first introduce key aspects of modified gravity, both to clarify why this data reduction is necessary and to provide an overview of the parameterisations and theories that form the foundation of our analysis.

\section{Modified Gravity: theories and parameterisations}\label{sec:MG}

The method proposed in this work integrates realistic, theory-specific modelling at small scales while ultimately constraining a simple large-scale parameterisation of modified gravity. In this section, we introduce both the general modified gravity parameterisations we are constraining, and the specific modified gravity theories and screening mechanisms we use to inform (and validate) our data reduction method. 

Any viable alternative gravity theory must be theoretically consistent and meet stringent constraints, ensuring the background evolution closely resembles that of $\Lambda$CDM (e.g., \cite{2016}) and that the speed of gravitational waves remains approximately luminal (e.g., \cite{Abbott_2017b,Abbott_2017}, at least at high frequencies). Additionally, solar system constraints mean that any modification to GR must be suppressed at very small scales. Many viable models achieve this through so-called \textit{screening mechanisms} that ensure MG theories recover the GR limit at small scales. Despite the vast number of MG theories, only a few viable screening mechanisms have been identified.

We conjecture that selecting our data reduction models to correspond to theories with different screening mechanisms provides a sensisble starting point for capturing a wide variety of nonlinear gravity effects. We justify this choice in part because, for a general scalar-tensor theory with a potential and a conformal coupling to matter, the complete dynamics of the model are uniquely defined by specifying two time-dependent functions which are determined entirely by the cosmological background \cite{Lombriser_2016, Hassani_2020}. Furthermore, for luminal Horndeski theories, once we model the background expansion and modifications to the growth of structure ($\mu$), the power spectrum on nonlinear scales will differ because of the choice of screening mechanism \cite{Brando_2021}. Given this strategy, we select Hu-Sawicki $f(R)$ (Chameleon screening) and nDGP gravity  (Vainshtein screening) as our data reduction models (together with GR). Section \ref{sec:results_ESS} below will go some way towards validating this choice: we use one Vainshtein-screened theory (nDGP gravity) to inform our data reduction matrix and then another (ESS gravity) to test the resulting PCA-based data reduction matrix. 

We now introduce the linear modified gravity parameterisations used in this paper as well as the specific alternative gravity models of interest and the emulators available for the fast production of their nonlinear matter power spectra.

\subsection{Linear growth parameterisations of modified gravity}\label{sec:parameterisations}

A common practical way of conducting tests of gravity using late-time WL and LSS survey data involves introducing phenomenological functions $\mu$, $\gamma$ and $\Sigma$. Introducing these functions allows one to test deviations from GR across a wide range of models, avoiding the inefficient testing of each theory individually. This method also provides a broad phenomenological view of modified gravity signals in the data, reducing the chance of missing key signatures and mistaking deviations from GR as tensions in our data.

For the scalar-perturbed Friedmann-Lemaitre-Robertson-Walker metric,
$$
ds^2 = -(1 + 2\Psi)dt^2 + a^2(1 - 2 \Phi ) dx^2 , 
$$
and in the linear and quasi-static limits of the perturbation equations, these functions are defined as
\begin{ceqn}
\begin{align}\label{eq:mu_Sigma_parameterisations}
    \begin{aligned}
    k^2 \Psi &= -4\pi G \mu(a,k) a^2 \rho \delta \\
    \Psi &= \gamma(a,k) \Phi \\
    k^2 \frac{(\Phi + \Psi)}{2} &= -4\pi G \Sigma(a,k) a^2 \rho \delta \\
    \end{aligned}
\end{align}
\end{ceqn}
where $k$ is the Fourier wavenumber and $\delta$ is the comoving matter density contrast ($\delta \equiv \Delta \rho/\rho$). The  quasi-static approximation (QSA) is defined by $k \ll H/a$ and all temporal derivatives being much smaller than their spatial derivatives. $\mu$, $\gamma$ and $\Sigma$ are phenomenological functions equal to unity in $\Lambda$CDM, and any two of them are sufficient to fully characterise lensing and structure growth in a given theory. In this work, we choose to employ $\mu$ and $\Sigma$. This formalism is valid for most gravity theories of interest under the QSA and in the linear limit \cite{Silvestri_2013}. The functions  $\Sigma = G_{\text{light}}/G$ and $\mu = G_{\text{matter}}/G$ are directly linked to weak lensing and galaxy clustering respectively and have been widely used in cosmological tests of GR.

With the modified potentials in Equation \ref{eq:mu_Sigma_parameterisations}, the growth of linear density perturbations is given by \citep{Simpson_2012}:
\begin{ceqn}
\begin{align}
\Ddot{\delta}(a) + 2H\Dot{\delta}(a) - 4\pi G \rho \mu(a)\delta(a) = 0, 
\label{eq:growth_eq}
\end{align}
\end{ceqn}
where $G$ denotes Newton’s constant, $H$ is the Hubble parameter, $\rho$ the background density of matter and the dot denotes a derivative with respect to coordinate time $t$.

When constraining these parameterisations with realistic cosmological survey data, some choices have to be made about the functional forms of $\mu$ and $\Sigma$ with respect to $a$ and $k$. As a first simplification, it has been shown that the scale dependence of these functions is sub-dominant at scales probed by linear theory in large-scale structure surveys for most MG theories of interest (see \cite{Silvestri_2013, Baker_2014_musigmaprobes}). Choosing the dependence with respect to scale factor (or equivalently redshift) is less straightforward. 
In this first proof-of-concept investigation of the PCA-based data reduction method we implement a simple parameterisation used very commonly throughout the literature (see for example \cite{Abbott_2023,2016, Abbott_2019,Simpson_2012}): 
\begin{ceqn}
\begin{align}\label{eq:mu_sigma_ourparam}
    \begin{split}
        \mu(z) &= 1 + \mu_0 \frac{\Omega_{\Lambda}(z)}{\Omega_{\Lambda 0}} \\
    \Sigma(z) &= 1 + \Sigma_0 \frac{\Omega_{\Lambda}(z)}{\Omega_{\Lambda 0}}.
    \end{split}
\end{align}
\end{ceqn}
We will call this the ``Dark Energy (DE)-like'' or ``$\Lambda$-'' parameterisation throughout this work.

These linear phenomenological parameterisations can serve as indicators of deviations from GR, but any such deviations must ultimately be sourced by a specific modified gravity theory. Our simulated analysis reflects this, by using modified gravity theories to inform the non-linear data reduction process. In the following subsection, we discuss the specific modified gravity theories relevant to this work.

\subsection{Modified gravity theories}\label{sec:MG_theory}

We now introduce the alternative theories of gravity that we use in this work, either as data reduction models or as validation when assessing our method's efficacy.

For our data reduction models, we consider Hu-Sawicki $f(R)$ and nDGP gravity, the standard workhorses for modified gravity in the nonlinear regime. Even though they are not competitive alternatives to GR -- their parameters have been largely constrained to be unlikely to be cosmologically relevant (e.g., see \cite{Xu_2014, Cataneo_2015}) -- they remain valuable due to the extensive availability of N-body simulations and emulators, notably \texttt{nDGPemu} \citep{fiorini2023fast}, \texttt{MGLenS} \citep{harnoisdéraps2022mglensmodifiedgravityweak}, \texttt{MGemu} \citep{Ramachandra_2021}, \texttt{FREmu} \citep{Bai_2024}, \texttt{Sesame} \citep{Mauland_2024}, and \texttt{e-MANTIS} \citep{sáezcasares2023emantis}. Hu-Sawicki $f(R)$ and nDGP gravity have the additional advantage of displaying two different screening mechanisms, respectively Chameleon and Vainshtein screening.

\begin{figure}
    \centering
    \includegraphics[width=1.0\linewidth]{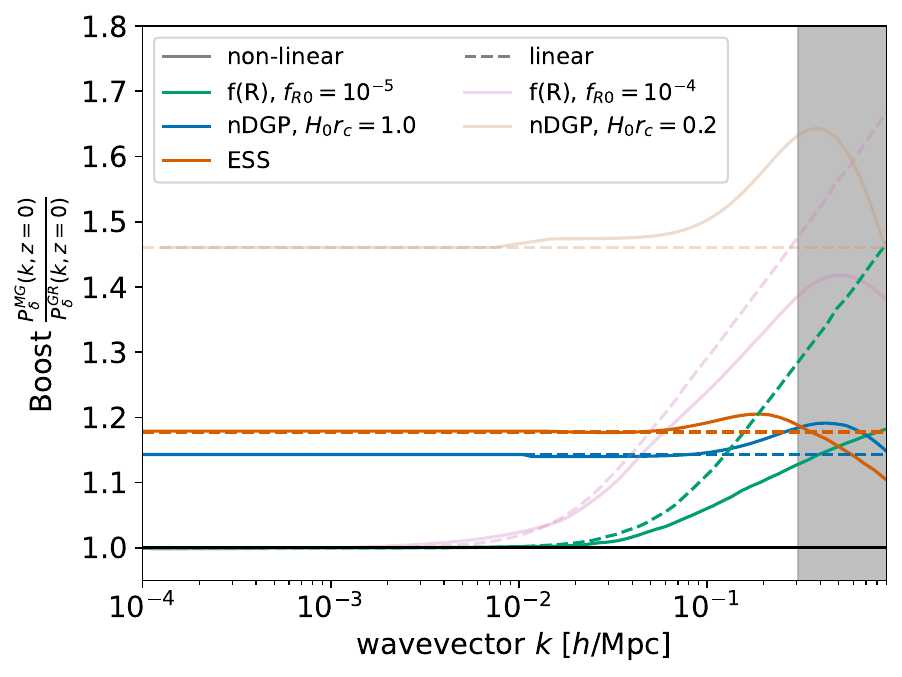}
    \caption{Linear and nonlinear power spectrum boosts for Hu-Sawicki $f(R)$, nDGP and ESS gravity for the simulated data used in Section \ref{sec:results} and Appendix \ref{Appendix:f(R)_nDGP_tests} (left-hand side of the legend) and for the data reduction models (right-hand side of the legend). The Boost is defined as the ratio of the modified gravity power spectrum to the GR power spectrum.}
    \label{fig:Boosts}
\end{figure}

Figure \ref{fig:Boosts} shows the ratio of the modified gravity power spectrum to the GR power spectrum for the gravity models considered in this paper both during the nonlinear data reduction process and in simulated scenarios. This diverse phenomenology underscores the importance of properly accounting for modified gravity nonlinear effects at the scales relevant to WL and LSS analyses.
 
\subsubsection{nDGP gravity and the Vainshtein Screening Mechanism}\label{sec:nDGP_theory}

Dvali-Gabadadze-Porrati (DGP) gravity is the simplest model of the braneworld class of modified gravity theories, which assumes we are living on a 4D brane in a higher dimensional spacetime. The DGP action is \citep{Dvali_2000}:

\begin{ceqn}
\begin{align}
    \mathcal{S} = \frac{1}{16\pi G_{(5)}} &\int \text{d}^5x \sqrt{-g_{(5)}} R_{(5)} \nonumber \\
    &\quad + \int \text{d}^4x \sqrt{-g} \left(\frac{R}{16 \pi G} + \mathcal{L}_m \right)
\end{align}
\end{ceqn}
where the subscript $(5)$ indicates the 5D generalization of the 4D quantity. The cross-over radius, defined as $r_c = \frac{G_{(5)}}{2G}$, is a transition scale between domination of the 5D ($r_c \geq 1$) and 4D ($r_c \leq 1$) part of the action.

The DGP Hubble parameter is given by \citep{Schmidt_2009}:
\begin{ceqn}
\begin{align}
    H(a) = \sqrt{\Omega_{m0} a^{-3} + \Omega_{rc}} \pm \sqrt{\Omega_{rc}}
\end{align}
\end{ceqn}
where $\Omega_{rc} \equiv \frac{1}{4 H_0^2 r_c^2}$. The $(-)$ solution of the expression above is called the normal branch of DGP (nDGP). It is typical to add a dark energy component to explain the late-time acceleration and to achieve a background consistent with $\Lambda$CDM:

\begin{ceqn}
\begin{align}
\Omega_{DE}(a) = \Omega_{\Lambda 0} - 2\sqrt{\Omega_{rc} (\Omega_{m 0} a^{-3} + \Omega_{\Lambda 0})}.
\end{align}
\end{ceqn}
This is the version of the theory of interest for our work, and the version to which we refer whenever we talk about nDGP gravity from this point onwards. The modified perturbation equations at the linear level and in the quasistatic regime can be described by \citep{Bose_2016}:
\begin{ceqn}
\begin{align}\label{eq:mu_nDGP}
    \mu(a) = 1 + \frac{1}{3\left[1 + \frac{H}{\Omega_{rc}} \left(1 + \frac{a}{3H} \left(\frac{\text{d}H}{\text{d}a}\right)\right)\right]}
\end{align}
\end{ceqn}
and $\Sigma = 1$ \citep{Koyama_2006,Koyama_2007}. Under the continued assumption that we are in the linear and quasistatic regime, we can find the 3D linear matter power spectrum in nDGP gravity via:
\begin{ceqn}
\begin{align}
P^{\text{nDGP}}_{\delta, {\rm lin}} (a) = \left(\frac{D_{\text{nDGP}}(a)}{D_{\text{nDGP}}(\bar{a})}  \times \frac{D_{\text{GR}}(\bar{a})}{D_{\text{GR}}(a)} \right)^2 P^{\text{GR}}_{\delta, {\rm lin}}(a)
\label{eq:pk_ndgp_lin}
\end{align}
\end{ceqn}
where $\bar{a}$ is a value of the scale factor during matter domination and we can find $D(a) \propto \delta (a)$ by solving the second order ODE from Equation \ref{eq:growth_eq} (in both GR and nDGP) with initial conditions $D(a) = a$ and $D'(a) = 1$ during matter domination.

nDGP gravity is part of a class of models where the nonlinear derivative interactions screen the fifth force. Specifically, Vainshtein screening operates when the second derivative of the potential is higher than some critical value and suppresses the fifth force (see \cite{Koyama_2016} for a review on the subject). In this work, we use the \texttt{nDGPEmu}\footnote{https://github.com/LSSTDESC/mgemu} emulator to compute the nonlinear 3D matter power spectrum for nDGP gravity with a $\Lambda$CDM background. This is accurate at 3\% for $k = 5h/\text{Mpc}$ and $0\leq z\leq 2$ compared to \texttt{MG-AREPO} full simulations \citep{fiorini2023fast}. 

\subsubsection{Hu-Sawicki f(R) gravity and the Chameleon Screening Mechanism}\label{sec:f(R)_theory}

In the $f(R)$ class of gravity theories, the Einstein-Hilbert action is augmented with a general function of the scalar curvature:
\begin{ceqn}
\begin{align}\label{eq:fR_action}
    S_G = \int \text{d}^4x\sqrt{-g}\left[\frac{R+f(R)}{16\pi G}\right]
\end{align}
\end{ceqn}

This is equivalent to a scalar-tensor theory, with additional degree of freedom $f_R = \text{d}f/\text{d}R$. 

We consider more specifically the Hu-Sawicki model \citep{Hu_2007}, with characteristic function:
\begin{ceqn}
\begin{align}
    f(R) = -2\Lambda \frac{R^n}{R^n+ \tilde{\mu}^{2n}}
\end{align}
\end{ceqn}
with three free parameters $n$, $\Lambda$ and $\tilde{\mu}$. This model does not contain a cosmological constant, but for $R >> \tilde{\mu}^2$, $f(R)$ can be approximated as
\begin{ceqn}
\begin{align}
    f(R) = -2\Lambda - \frac{f_{R0}}{n}\frac{\bar{R}_0^{n+1}}{R^n}
\end{align}
\end{ceqn}
with $f_{R0} = f_R(\bar{R}_0)= -2\Lambda \tilde{\mu}^2 / \bar{R}^2_0$ and $\bar{R}_0 = \bar{R}(z=0)$, where the overbar denotes the quantity for the background spacetime.

As for nDGP gravity, at the background level we artificially constrain $f(R)$ to look like $\Lambda$CDM by introducing an additional dark energy component. This version of the theory (with $n=1$) is the version to which we refer whenever we talk about $f(R)$ gravity from this point onwards in the paper.

At the level of the linear perturbation equations under the QSA, we can find \citep{Bose_2020}:
\begin{equation}\label{eq:mu_ka_fR}
    \mu(k,a) =  1 + \frac{(k/a)^{2}}{3\left((k/a)^2 + \frac{H_0^2(\Omega_{m0} a^{-3} +4(1-\Omega_{m0}))^3}{2f_{R0}(4-3\Omega_{m0})^2}\right)}
\end{equation}
and $\Sigma = 1$ \citep{Hojjati_2016}, which allows us to find the 3D linear matter power spectrum in $f(R)$ gravity valid in the quasistatic regime by using the same method outlined in Section \ref{sec:nDGP_theory} for nDGP.

$f(R)$ gravity can be thought of as a subclass of scalar-tensor theory with a single additional scalar field \citep{Sotiriou_2006}. It displays a Chameleon-type screening mechanism, where in dense environments the mass of the scalar field is large, causing the scalar field not to propagate and the additional force mediated by the scalar field to be suppressed. In this work, we use the \texttt{e-MANTIS}\footnote{https://pypi.org/project/emantis/} emulator to compute the nonlinear 3D matter power spectrum for $f(R)$ gravity with a $\Lambda$CDM background. This is accurate at 3\% for $k \approx 7h/\text{Mpc}$ and $0 \leq z \leq 2$ compared to full N-body simulations \citep{sáezcasares2023emantis}.

\subsubsection{Extended Shift-Symmetric (ESS) gravity}\label{sec:ESS_theory}

Extended Shift-Symmetric (ESS) gravity \cite{Traykova_2021} is a luminal subclass of Horndeski theories characterised by the action \citep{Wright_2023}:
$$
\mathcal{S}[g_{\mu\nu}, \phi] = \int \text{d}^4x \sqrt{-g} \left[\frac{M_p^2}{2}(R - 2\Lambda) + K(X) - G_3(X) \square\phi \right]
$$
with:
$$
K(X) \propto H_0^2k_1X + k_2 X^2, \hspace{1em} G_3(X) \propto H_0^2 g_{31}X + g_{32} X^2.
$$
Here, $\phi$ represents a new scalar field, $X \equiv \frac{1}{2}\nabla^\mu\phi\nabla_\mu\phi$, and $k_1, k_2, g_{31}$ and $ g_{32}$ are free parameters of the theory.

ESS belongs to shift-symmetric Horndeski theories, often termed ``weakly broken Galileons'' in the literature. In this work, we employ ESS not as a data reduction model for SVD decomposition (Section \ref{sec:PCA_introsec}), but rather as a test case to assess whether our method is flexible enough to constrain a true theory of gravity that is not within the data reduction set.

\begin{figure}
    \centering
    \includegraphics[width=1.0\linewidth]{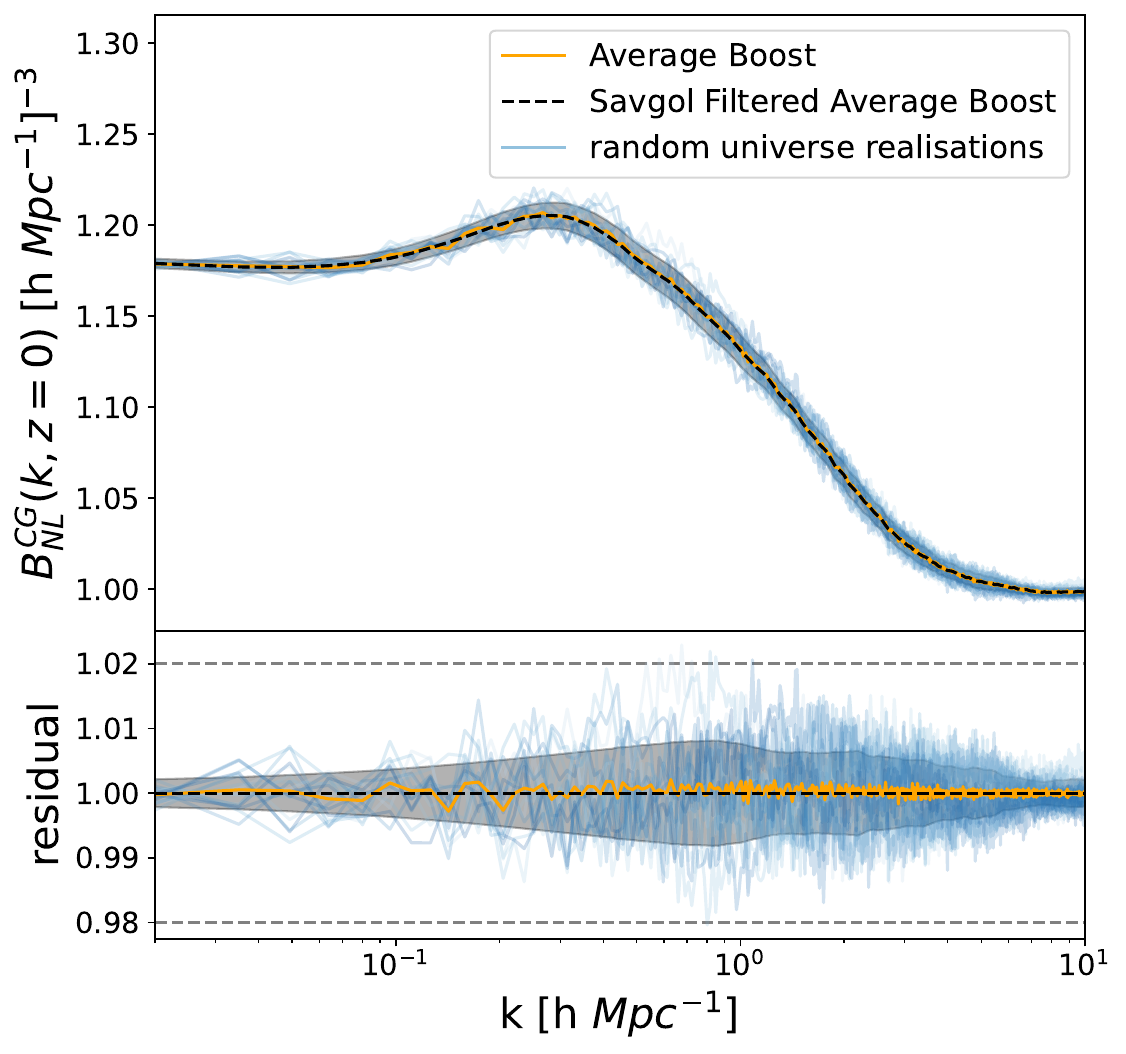}
    \caption{Ratios of real-space nonlinear matter power spectra at $z = 0$ for the ESS and $\Lambda$CDM models (the boost) from Table \ref{Table:ESS_params} described in Section \ref{sec:ESS_theory}. We define the `residual' as the ratio of a given quantity to the Savgol-filtered average boost.}
    \label{fig:ESS_Boost}
\end{figure}

Recent work in \cite{Wright_2023} has developed a codebase, \texttt{HiCOLA}\footnote{https://github.com/Hi-COLACode/Hi-COLA}, for calculating the nonlinear matter power spectrum using approximate N-body simulations for Horndeski theories, with ESS serving as an example. Although not a full N-body code, \texttt{HiCOLA} is one of the few available tools that provide a fast and versatile method to generate power spectra for a broad range of modified gravity theories with screening. The baryonic scale cuts necessarily imposed on our data (see Section \ref{sec:likelihood_scalecuts} below) mean that our analysis aims to capture Vainshtein-like modified gravity behaviour on mildly nonlinear scales, making this code suitable for our needs. 

To compute the theoretical ESS 3D matter power spectrum, we average over the individual spectra from a total of 20 \texttt{HiCOLA} simulations. These come from a set of 10 initial condition (IC) seeds (with GR initial conditions at $z=49$, see \cite{Wright_2023} for a characterization of the error due to GR initial conditions) and their phase pairs (i.e. pairs of simulations using phase-shifted initial conditions with matching amplitudes \cite{R_cz_2023}). Next, we compute the matter power spectra for 20 equivalent GR \texttt{HiCOLA} simulations. We then calculate the \textit{boost} ($B_i$) for each sampled IC ($i = 1,...,20$) using the MG and GR power spectra:

\begin{ceqn}
\begin{align}\label{eq:boost_definition}
    B_i(k,z,\textbf{p}) = \frac{P^{\text{MG}}_{\delta, i}(k,z,\textbf{p})}{P^{\text{GR}}_{\delta, i}(k,z,\textbf{p})}
\end{align}
\end{ceqn}

and then average over all boosts:
\begin{ceqn}
\begin{align}\label{eq:boost_definition_2}
    B(k,z,\textbf{p}) = \frac{1}{20} \sum^{20}_{i=1} B_i(k,z,\textbf{p}).
\end{align}
\end{ceqn}

Finally, we apply a Savitzky-Golay (Savgol) filter the boost, to obtain a smoothed power spectrum
ratio (for an example of Savgol-filtered boosts in the context of power spectra modelling for LSST, see \cite{Ramachandra_2021}). Once multiplied by a GR 3D matter power spectrum, this will give us a theoretical prediction of the ESS 3D matter power spectrum. The result is displayed in Figure \ref{fig:ESS_Boost} for the modified gravity parameters in Table \ref{Table:ESS_params} and the fiducial cosmological parameters in Table \ref{table:p_fid_and_priors}.

\begin{figure}
    \centering
    \includegraphics[width=1.0\linewidth]{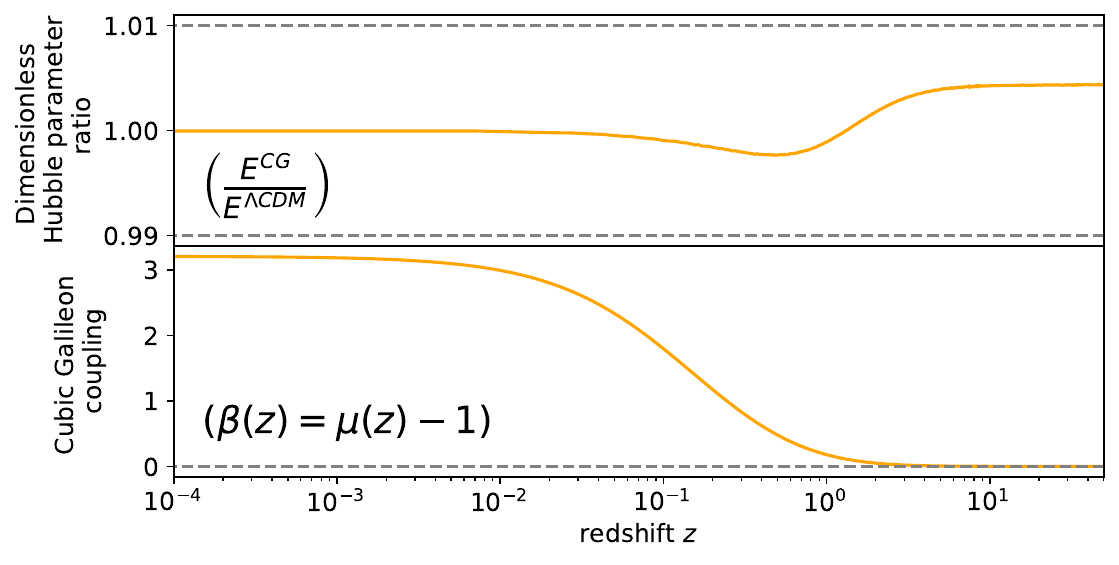}
    \caption{Evolution with redshift of the dimensionless Hubble parameter $E(a) = \frac{H(a)}{H_0}$ and modified growth parameter $\beta(a) = \mu(a) - 1$ for the ESS gravity model from Table \ref{Table:ESS_params} described in Section \ref{sec:ESS_theory}.}
    \label{fig:ESS_Background_and_lingrowth}
\end{figure}

We note that although \cite{Wright_2023} validates the \texttt{HiCOLA} Cubic Galileon power spectrum against N-body simulations, the ESS model has not undergone similar empirical validation. Cubic Galileon is a subset of ESS gravity (recovered for $k_2, g_{32} = 0$) which has been widely explored \citep{Barreira_2013,Nicolis_2009,Deffayet_2009a,Deffayet_2009b} and strongly constrained observationally \citep{Peirone_2018b,Frusciante_2020}, with full N-body simulations available in the literature \citep{Wyman_2013, Barreira_2013, Zhang_2020}. Nevertheless, we use a set of (relatively unexplored) ESS parameters that deviate from the Cubic Galileon case to test our novel data reduction method. This choice is motivated by the need to ensure minimal deviations from $\Lambda$CDM evolution in $H(a)$ (see Figure \ref{fig:ESS_Background_and_lingrowth}). In this paper we hereafter approximate $H_{\text{ESS}}(a) \approx H_{\Lambda \text{CDM}}(a)$. 

Finally, in this work, we artificially set the lensing modification in ESS gravity to $\Sigma(a) = 1$, rather than its theoretical value $\Sigma(a) = \mu(a)$ \citep{Peirone_2019,Frusciante_2020,Noller_2021}. This allows us to perform simulated constraints on a theory that has more similar properties to those used to create our data reduction matrix, given that $\Sigma=1$ in both $f(R)$ and nDGP. While this deviates from the true ESS prediction, this `modified ESS' model enables a controlled test of our proof-of-concept data reduction approach, given the limited availability of full N-body simulations for modified gravity theories of this nature.

\begin{table}
\centering
\begin{tabular}{|c|c|c|c|} 
 \hline
 $k_1$ & $g_{31}$ & $k_2$ & $g_{32}$ \\
 \hline
 $-0.45$ & $-30$ & $-9.085$ & $16.589$ \\
 \hline
\end{tabular}
\caption{$\mathbf{p}_{\text{ESS}}^{\text{fid}}$, ESS modified gravity parameters for the model we use to build our simulated data.}
\label{Table:ESS_params}
\end{table}

\section{Analysis set-up and methodology}\label{sec:Setup_Methodology}

Now that we have outlined our data reduction method (Section \ref{sec:PCA_introsec}) and introduced the modified gravity theories and parameterisations which we will make use of in this work (Section \ref{sec:MG}), we are in a position to deploy our method on simulated WL and LSS data. 

We will do so by constraining parameterised deviations from GR in an LSST Y1-like 3$\times$2pt simulated analysis, where 3$\times$2pt refers to a combined tomographic analysis of the cross-correlation of galaxy positions $\delta_g$ and galaxy lensing shear $\kappa$ (known as galaxy-galaxy lensing), as well as their respective autocorrelations (galaxy clustering and cosmic shear). In order to effectively constrain $\mu_0$ and $\Sigma_0$ with a 3$\times$2pt analysis, it is typical to also include auxiliary data to break parameter degeneracies. We therefore additionally include Cosmic Microwave Background (CMB) information through {\it Planck}-informed priors \citep{2020} and, where specified, employ simulated measurements of redshift-space distortions (RSD) from spectrosopic galaxy surveys.

In Section \ref{sec:priors}, we discuss the {\it Planck}-informed priors and fiducial cosmological parameters used in our analysis. Sections \ref{sec:MF_data}, \ref{sec:covariances} and \ref{sec:likelihood_formalism} describe the simulated data, the covariance, and the likelihood formalism for parameter inference, respectively. Finally, Section \ref{sec:PCA_MGtheories} details our data reduction method within the context of a realistic parameter inference analysis.

\subsection{Priors and Parameter Space}\label{sec:priors}

\begin{table*}
\centering
\scalebox{1.0}{\begin{tabular}{c c c c c} 
\hline\hline
 & Parameter & Fiducial Value & Baseline prior distributions & Additional prior distributions (from emulators)\\ 
\hline\hline
 & $\Omega_m$ & 0.31966 & $\mathcal{U}(0.1, 0.9)$ & $\mathcal{U}(0.28, 0.36)$ \\ 
 & $h$ & 0.6688 & $\mathcal{U}(0.55, 0.91)$ & $\mathcal{U}(0.61, 0.73)$ \\
\multirow{3}{*}{$\mathbf{p}_{\text{co}}$} 
 & $A_s$ & $2.092\times 10^{-9}$ & $\mathcal{U}(0.5\times 10^{-9}, 5\times 10^{-9})$ & $\mathcal{U}(1.7\times 10^{-9}, 2.5\times 10^{-9})$ \\
 & $\omega_b$ & 0.022383 & \multicolumn{1}{c}{$\mathcal{N}\left(\begin{bmatrix} 0.022383 \\ 0.9626 \end{bmatrix}, \begin{bmatrix} 2.0381\times10^{-5} & 3.6116\times10^{-7}  \\ 3.6116\times10^{-7} & 2.5265\times10^{-8} \end{bmatrix} \right)$} & $\mathcal{U}(0.04 h^2, 0.06h^2)$ \\
 & $n_s$ & 0.9626 & & $\mathcal{U}(0.92, 1.0)$ \\
\hline 
\multirow{5}{*}{$\mathbf{p}_{\text{nu}}$} 
 & $b_1$ & 1.562362 & $\mathcal{U}(0.8, 3.0)$ & \\
 & $b_2$ & 1.732963 & $\mathcal{U}(0.8, 3.0)$ & \\
 & $b_3$ & 1.913252 & $\mathcal{U}(0.8, 3.0)$ & \\
 & $b_4$ & 2.100644 & $\mathcal{U}(0.8, 3.0)$ & \\
 & $b_5$ & 2.293210 & $\mathcal{U}(0.8, 3.0)$ & \\
\hline 
 & $\mu_0$ & - & $\mathcal{U}(-1.5, 1.5)$ & \\
 & $\Sigma_0$ & - & $\mathcal{U}(-1.5, 1.5)$ & \\
\hline
\end{tabular}}

\caption{Priors and fiducial parameters ($\mathbf{p}^{\text{fid}}$) for the cosmological, modified gravity, and nuisance parameters in our simulated analysis. We quote the lower and upper limits of flat priors and the mean and standard deviation of Gaussian priors. The priors for the standard scale cuts (``Baseline Prior distributions'') match those in \cite{DES_Y3_results}.}
\label{table:p_fid_and_priors}
\end{table*}

Table \ref{table:p_fid_and_priors} provides a summary of the priors and fiducial parameters for the cosmology ($\mathbf{p}_{\mathbf{co}} = \{\Omega_{c}, A_s,h, n_s,\Omega_{b}\}$), modified gravity ($\mu_0$ and $\Sigma_0$) and nuisance ($\mathbf{p}_{\mathbf{nu}} = \{b_1,b_2,b_3,b_4,b_5\}$) parameters to be constrained in our simulated analysis. 
The fiducial cosmological parameters $\mathbf{p}^{\text{fid}}_{\text{co}}$ are selected to be the best-fit values from \cite{2020} (TT+lowE case), and will be used to generate simulated data in our analyses unless specified otherwise. 

Fiducial values for $b^i$, the constant galaxy bias for each lens redshift bin, are taken from the Y1 forecasts from the LSST Dark Energy Science Collaboration Science Requirements Document \citep[SRD,][]{DESCSRD2021lsst}. We refer to these as our set of nuisance parameters $\mathbf{p}^{\text{fid}}_{\text{nu}}$. This is not comprehensive of that required for a realistic LSST Y1 analysis (which would also be expected to include photo-z, intrinsic alignment, and shear calibration parameters at minimum), however, we do expect galaxy bias to be the most important nuisance parameter to include given its known degeneracy with $\mu_0$ (see e.g. \cite{abbott2019dark}). A more complex investigation of the power of our method in the context of a full systematics model is left for future work.

We use flat prior distributions corresponding to those used in \cite{DES_Y3_results} for our cosmological, nuisance and modified gravity parameters. We call these ``baseline'' priors -- they represent the prior ranges we would ideally hope to sample over if we had no limitations from emulator tools. Additional flat priors have to be imposed on some cosmological parameters during the PCA-based inference process because of unavoidable hard cutoffs from the emulator codes. 

The only informative prior in our analysis is the joint Gaussian prior on $\omega_b$ and $n_s$. This comes from the marginalised and shifted posterior distributions for the Planck TT,TE,EE+lowE+lensing data from an analysis using the late-time DE-like model parameterisation outlined in Equation \ref{eq:mu_sigma_ourparam} (see Fig. 31 and Section 7.4.2 in \cite{2020}).

\subsection{Simulated Data}\label{sec:MF_data}

\subsubsection{3$\times$2pt data vector}\label{sec:3x2pt_data}

\begin{figure}
    \centering
    \includegraphics[width=1.0\linewidth]{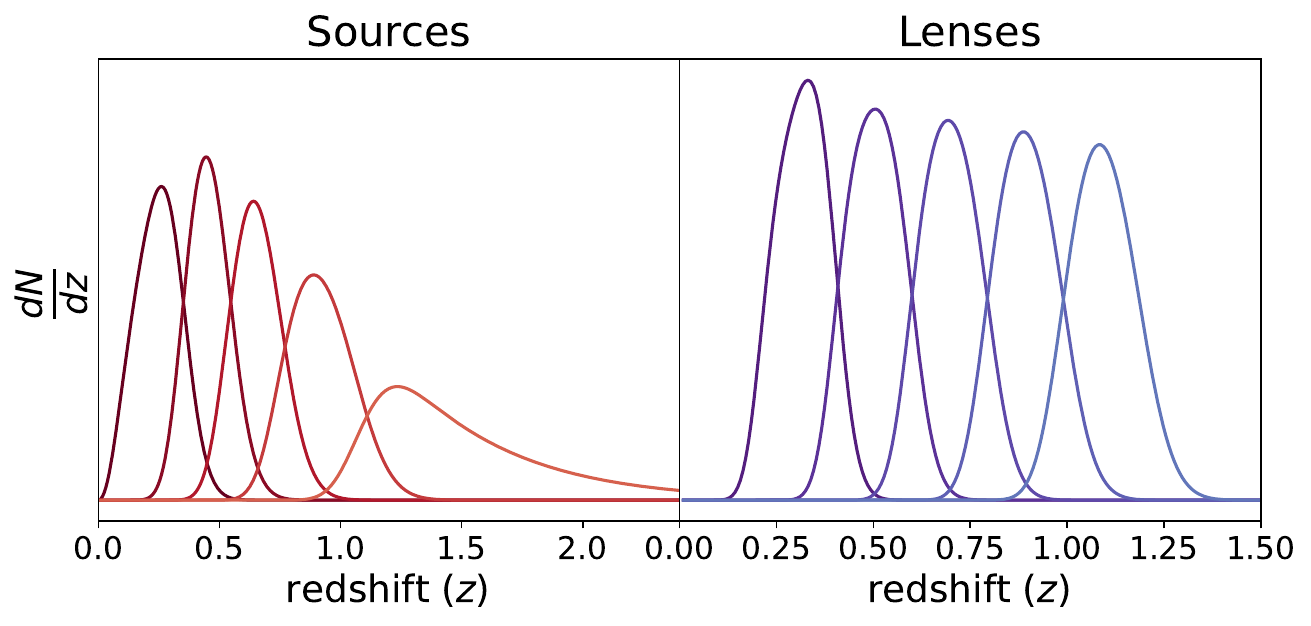}
    \caption{The normalised galaxy number density for the source (left) and lens (right) distributions for the 5 tomographic redshift bins from low-z (darkest) to high-z (lightest).}
    \label{fig:z_dist}
\end{figure}

In our proof-of-concept analysis we use 3$\times$2pt angular power spectra, with data vector $\textbf{D} = \{C^{ii}_{\delta_g\delta_g}(\ell),C^{ij}_{\kappa\delta_g}(\ell),C^{ij}_{\kappa\kappa}(\ell)\}$, where: 

\begin{itemize}
    \item $C^{ii}_{\delta_g\delta_g}(\ell)$ is the angular power spectrum of density contrast from galaxies in $n_{g\text{ bins}}$ lens redshift bins, with $i \in \{1, ..., n_{g\text{ bins}}\}$. We only include the autocorrelation in redshift bins.
    
    \item $C^{ij}_{\kappa\kappa}(\ell)$ is the angular power spectrum for galaxy shear in $n_{\kappa\text{ bins}}$ source redshift bins, with $i,j \in \{1,...,n_{\kappa \text{ bins}}\}$.
    
    \item $C^{ij}_{\kappa\delta_g}(\ell)$ is the angular power spectrum from the cross-correlation of galaxy shear in the $i^{th}$ source galaxy bin with the galaxy density contrast of the the $j^{th}$ foreground lens galaxy bin, with $i \in \{1,...,n_{g\text{ bins}}\}$ and $j \in \{1,...,n_{\kappa\text{ bins}}\}$.
\end{itemize}
To get the angular power spectrum from a 3D matter power spectrum we use the following equation:
\begin{ceqn}
\begin{align}\label{eq:C_ell_general}
    C^{ij}_{AB}(\ell) = \int d\chi \frac{q^i_A\left(\chi \right) q^j_B\left(\chi \right)}{\chi^2} P\left(\frac{\ell+\frac{1}{2}}{\chi}, z(\chi)\right)
\end{align}
\end{ceqn}
where the tracers $\text{AB}$ are either for cosmic shear ($\kappa \kappa$), galaxy-galaxy lensing ($\delta_g \kappa$) or galaxy-galaxy clustering ($\delta_g \delta_g$), and we have assumed the Limber approximation \citep{Limber_approx}. Here $\chi$ is the comoving radial distance, $i$ and $j$ denote redshift bin indices, and $q^i_\kappa$ and $q^i_\delta$ are line-of-sight kernels given by
\begin{eqnarray}
 q^i_\kappa (\chi) &=& \frac{3H_0^2 \Omega_m \Sigma(z(\chi))}{2a(\chi)/\chi} \int_{\chi}^{\chi_h} d\chi^\prime \left(\frac{\chi^\prime - \chi}{\chi}\right) n^i_s(z(\chi^\prime)) \frac{d z}{d\chi^\prime} , \nonumber \\
 q^i_\delta (\chi) &=& b^i\  n^i_l(z(\chi)) \frac{d z}{d\chi}\ ,
\end{eqnarray}
where $a$ is the scale factor and relevant quantities are defined in Section \ref{sec:priors}. We note here that the constant galaxy bias $b^i$ for each lens redshift bin can in theory be scale- as well as redshift- dependent. The function $\Sigma(z)$ is defined in Equation \eqref{eq:mu_Sigma_parameterisations} and is included here to allow for the possibility of deviations from GR in the lensing potential. Additionally, $n^i_{s,l}(z)$ denote the redshift distributions of the different source/lens redshift bins, normalised to unity (see Figure \ref{fig:z_dist}). 

Following \cite{DESCSRD2021lsst}, we assume a parametric redshift distribution for our overall underlying `true' source and lens distributions:
\begin{ceqn}
\begin{align}
    \frac{\text{d}N}{\text{d}z}\propto z^2\text{exp}[-(z/z_0)^\alpha].
\end{align}
\end{ceqn}
The source sample is separated in 5 bins with equal number counts, and with $\{z_0,\alpha\} = \{0.13, 0.78\}$ and photo-$z$ scatter $\sigma_z = 0.05(1 + z)$. The lens sample is separated in 5 bins separated by $0.2$ in photo-$z$ in the redshift interval $0.2 \leq z \leq 1.2$, with $\{z_0,\alpha\} = \{0.26,0.94\}$ and with photo-$z$ scatter $\sigma_z = 0.03(1 + z)$. 

Given this model for our total true source and lens distributions, we require a model for the distribution of source and lens galaxies which in practice end up in each tomographic source and lens bin, being aware that this assignment will be done on the basis of photo-z information. Our model for this follows \cite{DESCSRD2021lsst} in approximating the effect of photo-z uncertainty without directly emulating the practical process of photo-z distribution estimation. Starting with the reference redshift distribution for the entire sample of lenses ($l$) or sources ($s$) as appropriate $\left(\frac{dN_{l,s}}{dz}\right)$, we apply a top-hat selection function $\Theta^i(z)$ for each redshift bin $i$. We then convolve this with a Gaussian to produce the convolved redshift distribution for each bin, $n^i_{s,l}(z)$:
\begin{ceqn}
\begin{align}
n^i_{s,l}(\tilde{z}) &= \frac{\int\text{d}z \frac{\text{d}N_{s,l}}{\text{d}z}\Theta^i_{s,l}(z) p^i(z, \tilde{z})}{\int d\tilde{z} \int \text{d}z \frac{\text{d}N_{s,l}}{\text{d}z}\Theta^i_{s,l}(z) p^i(z, \tilde{z})},
\label{eq:pz_convolved}
\end{align}
\end{ceqn}
where the denominator serves as a normalization factor. The Gaussian $p^i(z, \tilde{z})$ is defined by: 
\begin{ceqn}
\begin{align}
p^i(z, \tilde{z}) \propto \exp\left(-\frac{1}{2}\left(\frac{z - \tilde{z}}{\sigma_z(1+z)}\right)^2\right). 
\label{eq:pz_gaussian}
\end{align}
\end{ceqn}
This construction is not a realistic model of the sample redshift inference, for which more detailed treatments are available (e.g., \cite{Myles_2021, Bilicki_2018, Rau_2023, Schmidt_2020}). Nevertheless, for our purpose, it enables us to construct a sufficient model which accounts appropriately for photo-z uncertainty at the level required for this proof-of-concept analysis.

We will use the Core Cosmology Library\footnote{https://github.com/LSSTDESC/CCL} (\texttt{CCL}) to compute all the above quantities \citep{Chisari_2019}. For galaxy clustering, we only consider the auto-correlations $C^{ii}_{\delta_g\delta_g}(\ell)$ for each tomographic bin since these carry a significant portion of the tracer's information. In the case of galaxy-galaxy lensing, we only consider combinations of lens and source bins where the source bins are at high enough redshift relative to the lens bins for a significant detection of galaxy-galaxy lensing: $( n^i_l, n^j_s) = \{(2,0), (3,0), (4,0), (3,1), (4,1), (4,2), (4,3)\}$ \citep{DESCSRD2021lsst}.

Our 3$\times$2pt data vector thus has dimension:
$$
    n_{\text{uncut}} = n_{\ell\text{ bins}} \times \left[n_{g \text{ bins}} + n_{\text{overlap}} + \frac{n_{\kappa \text{ bins}}(n_{\kappa \text{ bins}} + 1)}{2}\right]
$$
before scale cuts are applied, for some number of bins $n_{\ell\text{ bins}}$ logarithmically distributed between $\ell_{\text{min}}$ and $\ell_{\text{max}}$, where $\ell_{\text{max}}$ corresponds to a scale substantially smaller than any scale we are expecting to model. Our analysis choices here are the same as in \cite{DESCSRD2021lsst}, namely $20$ logarithmically spaced $\ell$ bins, covering $20 \leq \ell \leq 15000$. Once we apply cuts to our data, such as those explained in Section \ref{sec:likelihood_scalecuts} below, our data will have dimension $n \leq n_{\text{uncut}}$. For our standard linear scale cuts analysis, we choose to apply our $\Delta \chi^2 < 1$ criterion to the full datavector rather than to the datavector after baryonic cuts -- we expect this choice will have a negligible effect on the posteriors.

\subsubsection{Baryonic scale cuts}\label{sec:likelihood_scalecuts}

Even before enforcing data reduction because of the limits of our dark-matter-only theoretical modelling, we will need to implement some scale cuts in $\ell$ because of the effect of baryonic physics. While modelling baryonic physics is a major limiting factor in $\Lambda$CDM analyses, with extensive efforts dedicated to mitigating its effects, MG analyses face an even greater limitation in the modelling of nonlinear DMO behavior. As a result, we focus on addressing these challenges incrementally, prioritizing nonlinear DMO modelling as a foundational step before tackling the additional complexities of baryonic effects. 

There will be a maximum wavevector beyond which DMO modelling -- that is, ignoring baryonic effects -- is not a good model, which we call $k_{\text{max}}^{\text{DMO}}$ (see e.g. \cite{grandis_2024,Sunseri_2023,Chisari_2018}). We follow \cite{DESCSRD2021lsst} in fixing the scale for these baryonic cuts to be at $k^{\text{DMO}}_{\text{max}} = 0.3 h/\text{Mpc}$ and find $\ell^{\text{DMO}}_{\text{max}}$ using:
\begin{ceqn}
\begin{align}\label{eq:clustering_cut}
    \ell^{\text{DMO}}_{\text{max}} = k_{\text{max}}^{\text{DMO}} \chi(\langle z \rangle) - 0.5
\end{align}
\end{ceqn}
where $\langle z \rangle$ is the mean redshift of the relevant lens bin.

We use this method to implement scale cuts for $C^{ij}_{\delta_g\delta_g}(\ell)$ and $C^{ij}_{\delta_g\kappa}(\ell)$, but not for $C^{ij}_{\kappa\kappa}(\ell)$. This is because Equation \eqref{eq:clustering_cut} is only valid for narrow kernels -- given that in Equation \eqref{eq:C_ell_general}, when the kernel spans a small range of redshifts we can approximate the range of $\ell$ values affected by a real scale $k$ with a single value centred that the kernel mean redshift -- and the lensing kernel is broad. For the latter, we will instead use a method based on \cite{Doux_2022}. 
We compare two synthetic, noiseless data vectors
computed at the fiducial cosmology, one for a DMO $\Lambda$CDM universe and one where the power spectrum includes some baryonic effects by the rescaling of the matter power spectrum:
$$
    P_{\text{baryonic}}(\mathbf{p}^{\text{fid}},k,z)= \frac{P^{\text{sim}}_{\text{OWLS}}(k,z)}{P^{\text{sim}}_{\text{DMO}}(k,z)} \times P_{\text{DMO}}(\mathbf{p}^{\text{fid}},k,z). 
$$
Here, $P^{\text{sim}}_{\text{DMO}}(k,z)$ is the power spectrum measured in the DMO OWLS simulations \citep{van_Daalen_2011}, and $P^{\text{sim}}_{\text{OWLS}}(k,z)$ is the power spectrum measured in OWLS simulations with AGN feedback only\footnote{The matter power spectrum data can be found under the \texttt{OWLS\_AGN} entry and its DMO counterpart from \href{https://powerlib.strw.leidenuniv.nl/\#data}{https://powerlib.strw.leidenuniv.nl/\#data}.}. 
Using the data covariance matrix $\mathbf{C}$ we then compute:
$$
    \Delta \chi^2 = \left(\mathbf{D}_{\text{baryonic}}^{\Lambda\text{CDM}} - \mathbf{D}_{\text{DMO}}^{\Lambda\text{CDM}}\right)^T \mathbf{C}^{-1} \left(\mathbf{D}_{\text{baryonic}}^{\Lambda\text{CDM}} - \mathbf{D}_{\text{DMO}}^{\Lambda\text{CDM}}\right)
$$
at some fiducial cosmology for the 3$\times$2pt data vectors using Equation \eqref{eq:C_ell_general}. We then identify the point that contributes the most to $\Delta \chi^2$, remove it and repeat the process until $\Delta \chi^2 < 1$. Applying this procedure to an LSST Y1-like 3$\times$2pt data vector results in retaining the blue data points in Figure \ref{fig:Cuts_3x2pt}. We notice that the data vector with baryonic cuts applied retains many more data points than the typical linear-only modelling cuts upon which our novel PCA method seeks to improve. Still, we emphasize that this method for baryonic mitigation is conservative and leads to a significant degradation in constraining power. We adopt it here to enable a consistent comparison with the stringent linear-only scale cuts. Alternatively, less conservative data reduction methods -- such as the BNT transform \cite{Bernardeau_2014} -- could be applied to both sets of scale cuts. Importantly, we note here that a major advantage of our method is its compatibility with other PCA-based data reduction techniques targeting modelling uncertainties at different scales (specifically, the method outlined in \cite{Huang_2019}). One could create a larger set of data reduction difference vectors ($\mathbf{B}_i - \mathbf{M}_i$ in Section \ref{sec:PCA_MGtheories} below) for both baryonic mitigation and nonlinear mitigation, in order to retain as much constraining power as possible. This is possible because most analyses treat baryonic and MG effects as separable, based on the assumption that screening mechanisms suppress MG effects on the small scales at which drivers of baryonic physics play a role. 

\subsubsection{$f\sigma_8$}\label{sec:fsigma8_data}

As mentioned previously, 3$\times$2pt data must typically be combined with auxiliary data to break degeneracies between MG parameters, cosmological parameters, and galaxy bias, with a common choice for this being $f\sigma_8$ measurements. In order to include simulated $f\sigma_8$ measurements in our proof-of-concept analysis, we use noiseless simulated RSD data constructed to align with existing and upcoming surveys. We include simulated $f\sigma_8$ measurements corresponding to 5 different datasets (4 of which match the analysis in \cite{Ruiz_Zapatero_2022}) -- we have 3 $f\sigma_8$ datapoints corresponding to data from the BOSS Data Release (DR) 12 \citep{Alam_2017}, one from the BOSS DR16 quasar sample \citep{Hou_2020}, three from the WiggleZ Dark Energy Survey \citep{Blake_2012}, one from peculiar velocities of the Democratic Samples of Supernovae \citep{Stahl_2021}, and one which is a forecast for the DESI Luminous Red Galaxy (LRG) sample \citep{Zhou_2023}. 

Our simulated data consists of values for $f\sigma_8(z)$ at a small set of redshifts $z^{\text{dat}}$, where $\sigma_8$ is a measure of matter density fluctuations on a scale of $8 \text{ Mpc}/h$ and $f$ is the linear growth rate of structure. Recall that we will need to produce these simulated $f\sigma_8$ measurements under both GR and alternative theories of gravity. We can find the value of $\sigma_8$ at a given cosmology for a given modified gravity theory using the expression \citep{COORAY_2002}:
\begin{ceqn}
\begin{align}
    \sigma^{\text{MG}}_8(z) = \sqrt{\int \frac{k^2 P^{\text{MG}}(k,z)}{2\pi^2} \left[\frac{3 j_1\left(k\times 8\text{Mpc}/h\right)}{k\times 8\text{Mpc}/h}\right]^2 \text{d}k}
\end{align}
\end{ceqn}
where $j_1$ is the spherical Bessel function of the first kind. We can find the linear growth rate $f \equiv \text{dln}D/\text{dln}a$ (where $D$ is proportional to the growing mode of the linear matter density perturbation) by solving Equation \eqref{eq:growth_eq} with $D \propto \delta$ and with initial condition $D(a) = a$ during matter domination.

We stress here that we introduce $f\sigma_8$ as an auxiliary probe to 3$\times$2pt, and that the PCA-based data reduction process \textit{is not} applied to this dataset. When we mention a data reduction method, be it linear scale cuts or PCA-based reduction (e.g. Subsection \ref{sec:PCA_MGtheories}), this is in reference to 3$\times$2pt only.

\subsection{Covariances}\label{sec:covariances}

We make use of the full non-Gaussian Y1 3$\times$2pt covariance\footnote{\href{https://github.com/CosmoLike/DESC\_SRD}{https://github.com/CosmoLike/DESC\_SRD}} described in \cite{DESCSRD2021lsst} (created following the numerical prescription from \cite{Krause_2017}) as the 3$\times$2pt covariance in our simulated analysis. While this covariance is pre-computed at slightly different fiducial cosmological parameter values $\mathbf{p}^{\text{fid, SRD}} = \{\Omega_m = 0.3156, \sigma_8 = 0.831,n_s =0.9645,\Omega_b =0.0492,h =0.6727\}$ to our analysis, this should not have a significant effect on our final results (see \cite{Kodwani_2019}).

In computing the covariance of the simulated $f\sigma_8$ measurements, we assume that all surveys are mutually independent but allow for correlations between multiple $f\sigma_8$ measurements at different redshifts from the same survey. Covariances of the $f\sigma_8$ datapoints which correspond to existing measurements (i.e. all but the DESI LRG datapoint) are given by their existing estimated covariances as relayed in \cite{Ruiz_Zapatero_2022}. 
The error for the simulated $f\sigma_8$ data from the DESI LRG dataset is computed following the formalism in \cite{White_2009}, evaluated at our fiducial cosmological parameters. 

\begin{table*}
\centering
\scalebox{1.0}{
 \begin{tabular}{|c|c|c|c|c|c|c|c|c|c|} 
 \hline
  & \multicolumn{3}{|c|}{BOSS DR12} & BOSS DR16 & \multicolumn{3}{|c}{WiggleZ} & \multicolumn{1}{|c|}{DSS} & DESI LRGs \\
 \hline
 $z^\text{dat}$ & 0.38 & 0.51 & 0.61 & 1.48 & 0.44 & 0.6 & 0.73 & 0.0 & 0.72 \\ 
 \hline
 $f\sigma_8(z)$ (real) & 0.497 & 0.458 & 0.436 & 0.462 & 0.413 & 0.39 & 0.437 & 0.39 & - \\
 \hline
 $f\sigma_8(z, \mathbf{p}_{\text{co}}^{\text{fid}})$ (theory) & 0.484 & 0.482 & 0.476 & 0.381 & 0.484 & 0.477 & 0.466 & 0.437 & 0.467 \\
 \hline
\multirow{9}{*}{covariance $\mathbf{C}^{ij}_{f\sigma_8}$} 
 & 2.03    & 0.816  & 0.261 & 0     & 0     & 0      & 0      & 0     & 0     \\
 & 0.816   & 1.44   & 0.659 & 0     & 0     & 0      & 0      & 0     & 0     \\
 & 0.261   & 0.659  & 1.16  & 0     & 0     & 0      & 0      & 0     & 0     \\
 & 0       & 0      & 0     & 2.03  & 0     & 0      & 0      & 0     & 0     \\
 & 0       & 0      & 0     & 0     & 6.40  & 2.57   & 0      & 0     & 0     \\
 & 0       & 0      & 0     & 0     & 2.57  & 3.97   & 2.54   & 0     & 0     \\
 & 0       & 0      & 0     & 0     & 0     & 2.54   & 5.18   & 0     & 0     \\
 & 0       & 0      & 0     & 0     & 0     & 0      & 0      & 0.484 & 0     \\
 & 0       & 0      & 0     & 0     & 0     & 0      & 0      & 0     & 0.0577   \\
 \hline
 \end{tabular}}
 \caption{An example of simulated GR $f\sigma_8$ data used in our analysis for $\mathbf{p_{\text{co}}} = \mathbf{p}^{\text{fid}}_{\text{co}}$ (see Table \ref{table:p_fid_and_priors}) as well as the real $f\sigma_8$ data and its covariance. }
\label{table:fsigma8_data}
\end{table*}

\begin{figure}
    \centering
    \includegraphics[width=1.0\linewidth]{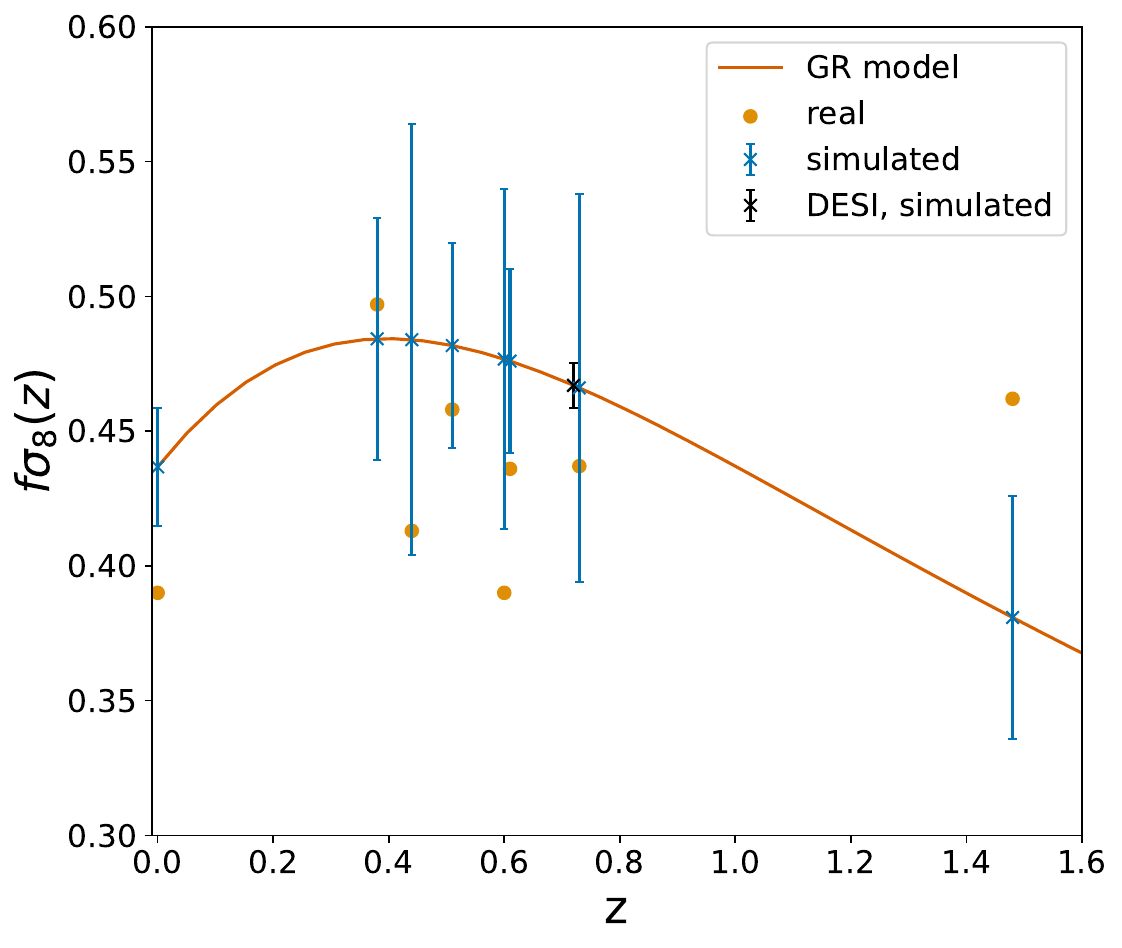}
    \caption{Example of the simulated noiseless $f\sigma_8$ data for a GR universe with $\mathbf{p} = \mathbf{p}_{\text{fid}}$. The blue $\times$ markers are for the simulated data, the orange circle markers for the original real data and the black point highlights the simulated DESI datapoint. The errors for the real and simulated data are the same.}
    \label{fig:fsigma8_data}
\end{figure}

All of the quantities related to our $f\sigma_8$ simulated data that are necessary for the likelihood analysis, including the covariance values, are displayed in Table \ref{table:fsigma8_data}.
An example of what the simulated data looks like for for our fiducial cosmology in a GR universe is displayed in Figure \ref{fig:fsigma8_data}.

We assume that there is no covariance between the $f\sigma_8$ data and the $3\times2$pt data because the former probes modes along the line-of-sight while our angular power spectra probe modes perpendicular to this. We can therefore write the full covariance of our data either as $\mathbf{C} = \mathbf{C}_{3\times2\text{pt}}$ when using 3$\times2$pt data only, or as:
$$\mathbf{C} = \left[\begin{matrix}
    \mathbf{C}_{3\times2\text{pt}} & 0 \\
    0 & \mathbf{C}_{f\sigma_8}
\end{matrix}\right]$$
when additionally using $f\sigma_8$ simulated data.

\subsection{Likelihood formalism}\label{sec:likelihood_formalism}

With all the necessary components in place, we now outline the general method for computing the likelihood in our simulated analysis. We first compute the relevant statistics as a function of model parameters $\mathbf{p}$. When completing analysis $1,2,5 $ and $6$ in Table \ref{table:Parameter_inference_models} (as described in Section \ref{sec:results} below), we create a mock data vector from Equation \eqref{eq:C_ell_general} and, using a 3D matter power spectrum computed for some cosmological parameters (labeled `$\text{fid}$', see the Table \ref{table:p_fid_and_priors}) and some modified gravity theory-specific parameters $\mathbf{p}_{\text{MG}}$:
\begin{ceqn}
\begin{align}
    \mathbf{D} &= \{C^{ii,\text{NL}}_{\delta_g\delta_g}(\ell\mid \mathbf{p}_{\text{co}}^{\text{fid}},\mathbf{p}_{\text{nu}}^{\text{fid}},\mathbf{p}_{\text{MG}}),\nonumber\\
    &C^{ij,\text{NL}}_{\kappa\delta_g}(\ell\mid \mathbf{p}_{\text{co}}^{\text{fid}},\mathbf{p}_{\text{nu}}^{\text{fid}},\mathbf{p}_{\text{MG}}),C^{ij,\text{NL}}_{\kappa\kappa}(\ell\mid \mathbf{p}_{\text{co}}^{\text{fid}},\mathbf{p}_{\text{nu}}^{\text{fid}},\mathbf{p}_{\text{MG}})\}.\nonumber
\end{align}
\end{ceqn}

We leave $\mathbf{p}_{\text{MG}}$ general, but for our validation of this proof-of-concept analysis we are either referring to $\{k_1, g_{31}, k_2, g_{32}\}$ for ESS, or modelling a GR universe. When instead completing analysis $3,4,7 $ and $8$ in Table \ref{table:Parameter_inference_models}, we also similarly compute the simulated $f\sigma_8$ data vector, $f\sigma_8(z^\text{dat}, \mathbf{p}^{\text{fid}})$ using the method described in Section \ref{sec:fsigma8_data}:
\begin{ceqn}
\begin{align}
    \mathbf{D} = \{C^{ii,\text{NL}}_{\delta_g\delta_g}(\ell\mid \mathbf{p}_{\text{co}}^{\text{fid}},\mathbf{p}_{\text{nu}}^{\text{fid}},\mathbf{p}_{\text{MG}}), C^{ij,\text{NL}}_{\kappa\delta_g}(\ell\mid \mathbf{p}_{\text{co}}^{\text{fid}},\mathbf{p}_{\text{nu}}^{\text{fid}},\mathbf{p}_{\text{MG}}),&\nonumber\\
    C^{ij,\text{NL}}_{\kappa\kappa}(\ell\mid \mathbf{p}_{\text{co}}^{\text{fid}},\mathbf{p}_{\text{nu}}^{\text{fid}},\mathbf{p}_{\text{MG}}), f\sigma_8(z^\text{dat}, \mathbf{p}_{\text{co}}^{\text{fid}},\mathbf{p}_{\text{MG}})\}.&\nonumber
\end{align}
\end{ceqn}

Likewise, our model vector (under $\Lambda$CDM+$\mu_0$+$\Sigma_0$) is either:

$$\mathbf{M}(\mathbf{p}) = \{C^{ii,\text{lin}}_{\delta_g\delta_g}(\ell,\mathbf{p}),C^{ij,\text{lin}}_{\kappa\delta_g}(\ell,\mathbf{p}),C^{ij,\text{lin}}_{\kappa\kappa}(\ell,\mathbf{p})\},$$ 
a 3$\times$2pt data vector only, when completing analysis $1,2,5 $ and $6$, or:

\begin{ceqn}
\begin{align}
    \mathbf{M}(\mathbf{p}) = \{C^{ii,\text{lin}}_{\delta_g\delta_g}(\ell,\mathbf{p}),C^{ij,\text{lin}}_{\kappa\delta_g}(\ell,\mathbf{p}),C^{ij,\text{lin}}_{\kappa\kappa}(\ell,\mathbf{p}),&\nonumber\\
    f\sigma_8(z^\text{dat}, \mathbf{p})\}.&\nonumber
\end{align}
\end{ceqn}
a combination of a 3$\times$2pt data vector and a 9-element $f\sigma_8$ data vector, when completing analysis $3,4,7 $ and $8$.

We then model the likelihood as Gaussian in the data vector and as a function of the parameters $\mathbf{p}$:
\begin{ceqn}
\begin{align}\label{eq:likelihood}
    \begin{split}
        \text{ln} \mathcal{L} (\mathbf{D}|\mathbf{p})\propto &-\frac{1}{2} \left[ (\mathbf{D} - \mathbf{M}(\mathbf{p}))^T \mathbf{C}^{-1} (\mathbf{D} - \mathbf{M}(\mathbf{p}))\right].
    \end{split}
\end{align}
\end{ceqn}
It will be useful to reformulate the likelihood by applying a Cholesky decomposition as described in Section \ref{sec:PCA_introsec}:
\begin{ceqn}\label{eq:likelihood_Ch}
\begin{align}
    \text{ln} \mathcal{L} (\mathbf{D}|\mathbf{p})&\propto -\frac{1}{2} \left[ (\mathbf{D} - \mathbf{M}(\mathbf{p}))^T \mathbf{L}^{-1}(\mathbf{L}^{T})^{-1} (\mathbf{D} - \mathbf{M}(\mathbf{p}))\right] \nonumber\\
        \propto&-\frac{1}{2} \left[ (\mathbf{D}_{\text{ch}} - \mathbf{M}_{\text{ch}}(\mathbf{p}))^T (\mathbf{D}_{\text{ch}} - \mathbf{M}_{\text{ch}}(\mathbf{p}))\right]
\end{align}
\end{ceqn}
where $\mathbf{C} = \mathbf{L} \mathbf{L}^{T}$ (see Equation \ref{eq:Choleski_decomposition}). This is the form of the likelihood we will use when considering the PCA-based data reduction method.

We do parameter inference using Markov Chain Monte Carlo (MCMC) sampling through the package \texttt{emcee}\footnote{https://github.com/dfm/emcee} \citep{emcee} using likelihoods and priors described in Section \ref{sec:Setup_Methodology}.  To ensure convergence, we run the sampler so that the number of iterations for each chain is at least 100 times its autocorrelation time (as defined in \citep{Sokal1996MonteCM}). 

\subsection{Finding the data reduction matrix during parameter inference}\label{sec:PCA_MGtheories}

In this subsection, we describe the steps for computing the 3$\times$2pt likelihood using the PCA-based data reduction method. We stress here that this procedure applies only to 3$\times$2pt data and does not affect the $f\sigma_8$ component, which is computed using the standard approach. The $f\sigma_8$ data serves solely as an auxiliary probe.

The following steps occur at each point when sampling the parameter space $\mathbf{p}$ described in Section \ref{sec:priors}, which does not include any theory-specific MG parameters (only $\mu_0$ and $\Sigma_0$). First, we find $m$ linear and $m$ nonlinear MG 3D matter power spectra corresponding to the $m$ different data reduction models:
$$
    P_{\delta, \text{NL}}^{\text{MG}^{m}}(k,z\mid \mathbf{p}_{\text{co}}, \mathbf{p}_{\text{MG}^m}), P_{\delta, \text{lin}}^{\text{MG}^{m}}(k,z\mid \mathbf{p}_{\text{co}},\mathbf{p}_{\text{MG}^m})
$$
for all $z$ and $k$. We compute these using emulators and growth factor ODEs (see Equation \ref{eq:growth_eq}) described in Section \ref{sec:MG}. We note here that the modified gravity theory-specific parameters $\mathbf{p}_{\text{MG}^m}$, which will be different in each data reduction theory, don't vary during our sampling process. 

In practice, in this proof-of-concept analysis, we have three nonlinear and three linear power spectra: one each for GR, $f(R)$ and nDGP. We select the theory-specific parameter values of $f_{R0} = 10^{-4}$ (for $f(R)$) and $H_0r_c = 0.2$ (for nDGP) for the construction of the reduction matrix. We choose these parameters because they are the largest deviations from GR for which we can find accurate power spectra with the emulator tools at our disposal. We stress here that all steps of the parameter inference process will have the same theory-specific modified gravity parameters -- this means that throughout our analysis, $f_{R0} = 10^{-4}$ and $H_0r_c = 0.2$ always for building the reduction matrix. A more complete analysis would vary the modified gravity parameters along with the cosmological parameters, but in this proof-of-concept work there is no self-consistent way to do this, given there is no direct relationship between the MG parametrised parameters $\mu_0$ and $\Sigma_0$ and the theory-specific parameters $f_{R0}$ and $H_0r_c$. We refer the reader to Section \ref{sec:conclusion} for discussion of future approaches to this setback.

Next, we find the 3D matter power spectrum for the MG linear $\mu-\Sigma$ parameterisation (see Equations \ref{eq:growth_eq} and \ref{eq:mu_sigma_ourparam}):
$$
    P_{\delta, \text{lin}}^{\text{MG}}(k,z\mid \mathbf{p}_{\text{co}}, \mu_0).
$$

We then compute the model angular power spectrum as well as the individual linear and nonlinear angular power spectra for the $m$ different data reduction theories using their respective 3D matter power spectra in Equation \ref{eq:C_ell_general} (following here and throughout this susbsection the notational style of \cite{Huang_2019}):

\begin{eqnarray*}
    &\mathbf{M}(\mathbf{p}) = C^{ij}_{\text{lin}}(\ell \mid \mathbf{p}_{\text{co}},\mathbf{p}_{\text{nu}}, \mu_0, \Sigma_0) \nonumber \\
    &\mathbf{M_m} = C^{ij}_{\text{lin}, m}(\ell \mid \mathbf{p}_{\text{co}},\mathbf{p}_{\text{nu}},\mathbf{p}_{\text{MG}^m})\nonumber \\
    &\mathbf{B_m} = C^{ij}_{\text{NL}, m}(\ell\mid \mathbf{p}_{\text{co}},\mathbf{p}_{\text{nu}},\mathbf{p}_{\text{MG}^m})
\end{eqnarray*}
each with dimension $n$.

We use $\mathbf{M_m}$ and $\mathbf{B_m}$ to build our PCA reduction matrix. We first find our difference matrix:
\begin{eqnarray*}
    \mathbf{\Delta}(\mathbf{p}_{\text{co}},\mathbf{p}_{\text{nu}},\mathbf{p}_{\text{MG}^m}) = \left[ \begin{matrix}
    | &  & | \\
        \mathbf{B_1} - 
        \mathbf{M_1} 
        & ... & \mathbf{B_m} -
        \mathbf{M_m} \nonumber\\
        | &  & |
    \end{matrix} \right] \\
    \hfill(m \times n)
\end{eqnarray*}
and then our Cholesky-weighted difference matrix:
\begin{equation*}
    \mathbf{\Delta}_{\text{ch}} = \mathbf{L}^{-1}\mathbf{\Delta},
\end{equation*}
where $\mathbf{L}$ is the Cholesky decomposition of the covariance $\mathbf{C}$ as defined in Equation \ref{eq:Choleski_decomposition}.

As described in Section \ref{sec:PCA_introsec} above, using SVD, we build a unitary rotation matrix $\mathbf{U}_{\text{ch}}(\mathbf{p}_{\text{co}},\mathbf{p}_{\text{nu}},\mathbf{p}_{\text{MG}^m})$ with shape $(n \times n)$:

$$
     \mathbf{U}_{\text{ch}} = \left[ 
     \begin{matrix}
    | &  & | \\
        PC_1 & ... & PC_m\\
        | &  & |
    \end{matrix} \begin{matrix}
    &\text{silent} \\
     &   \text{orthogonal} \\
     &   \text{vectors}
    \end{matrix} \right]
$$
The first $m$ columns of this matrix represent the principal components (PCs) of our data space (of dimension $n$). The remaining columns are silent orthogonal vectors that make the matrix unitary. We can then find the reduction matrix $\mathbf{U}_{\text{ch,cut}}$ with shape $(n\times n -m)$:
$$
     \mathbf{U}_{\text{ch,cut}} = \left[ \begin{matrix}
    \text{silent} \\
       \text{orthogonal} \\
       \text{vectors}
    \end{matrix} \right].
$$

We then use this matrix to compute the reduced data vectors:

\begin{ceqn}
\begin{align}
    &\mathbf{D}_{\text{ch}} = \mathbf{L}^{-1}\mathbf{D} \nonumber\\
    &\mathbf{M}_{\text{ch}} = \mathbf{L}^{-1}\mathbf{M} \nonumber\\
    &\mathbf{D}_{\text{ch,cut}} = \mathbf{U}_{\text{ch,cut}}^T\mathbf{D}_{\text{ch}} \label{eq:D_ch_cut} \\
    &\mathbf{M}_{\text{ch,cut}} = \mathbf{U}_{\text{ch,cut}}^T\mathbf{M}_{\text{ch}}
\end{align}
\end{ceqn}
from which we evaluate our likelihood:
\begin{equation*}
\begin{split}
    \text{ln} \mathcal{L}&(\mathbf{D}|\mathbf{p}) \propto \\ & -\frac{1}{2} \left[ \mathbf{D}_{\text{ch,cut}} - \mathbf{M}_{\text{ch,cut}}(\mathbf{p})\right]^T \left[\mathbf{D}_{\text{ch,cut}} - \mathbf{M}_{\text{ch,cut}}(\mathbf{p}))\right]. \\
\end{split}
\end{equation*}
by using the fact that $\mathbf{C}_{\text{ch}} = \mathds{1} \implies \mathbf{C}_{\text{ch,cut}} = \mathbf{U}_{\text{ch,cut}}^T\mathbf{C}_{\text{ch}}\mathbf{U}_{\text{ch,cut}} = \mathds{1}$. Combining this with the $f\sigma_8$ likelihoods in Eq.  \eqref{eq:likelihood_Ch}, we can compute the total likelihood at each point in the parameter inference process. Therefore, by applying the reduction matrix we are cutting our data in a way that minimises the impact of the small-scale nonlinear effects from our training MG theories.
We update our PCA basis at each point in parameter space.\footnote{This is equivalent to introducing a parameter-dependent covariance matrix $\mathbf{C}^{-1}(\mathbf{p}) = (\mathbf{U}_{\text{ch,cut}}(\mathbf{p}) \mathbf{L}^{-1})^T (\mathbf{U}_{\text{ch,cut}}(\mathbf{p}) \mathbf{L}^{-1})$. In theory, the covariance matrix should be fixed in parameter space when using Gaussian likelihoods, to guarantee posterior estimates are unbiased and observables are assigned a sufficiently conservative information content \cite{Carron_2013}. In practice, this effect is small in the large-$\ell$ limit, which is what we expect from the majority of the information content of Stage-IV surveys \cite{Kodwani_2019}.}

We expect the PCA method will operate effectively if certain underlying assumptions are true. The most important assumption is that the reduction matrix built assuming theory-specific, fixed modifications to gravity (measured at fixed parameters $\mathbf{p}_{\text{MG}^m}$) adequately captures a broad class of nonlinear behaviour. The results in the next section indicate that, for a 3$\times$2pt LSST Y1-like analysis, the PCA-based data reduction method appears to be robust to that assumption for the cases considered. We refer the reader to Appendix \ref{Appendix:convergence_trainingtheories} for an investigation into the completeness of our minimal PCA basis for theories displaying the Vainshtein screening mechanism.

We are also assuming that the modified gravity DE-like parameterisation ($\mu_0$ and $\Sigma_0$) is good at capturing the linear-regime modified gravity signatures of the theories under consideration, and will therefore not introduce model mis-specification biases in our cosmological parameter estimation. As we will see, this assumption is not strictly valid for all tested theories, as discussed in Section \ref{sec:MG_param_bias}. The latter is a standard assumption and is independent of our PCA method, also existing within the standard scale-cut case, but we state it here for completeness.

\section{Results and performance}\label{sec:results}

We now present results of our simulated analyses based on the methodology described above. We consider a total of 8 simulated data vectors, with analysis choices in Table \ref{table:Parameter_inference_models}. 

\begin{table}
\centering
 \scalebox{0.8}{\begin{tabular}{|c|c|c|c|c|} 
 \hline\hline
 Index & True Theory & Data Reduction & Data & MG parameters \\ [0.5ex] 
 \hline\hline
 1 & $\Lambda$CDM & Linear Scale Cuts & $3\times2$pt & - \\ 
 2 & $\Lambda$CDM & PCA method & $3\times2$pt & - \\
 3 & $\Lambda$CDM & Linear Scale Cuts & $3\times2$pt$+f\sigma_8$ & -  \\ 
 4 & $\Lambda$CDM & PCA method & $3\times2$pt$+f\sigma_8$ & - \\ 
 5 & modified ESS & Linear Scale Cuts & $3\times2$pt & \multirow{4}{*}{$\left\{\begin{array}{c}
    k_1 = -0.45 \\ g_{31} = -30 \\ k_2 = -12.456 \\ g_{32} = 17.151
 \end{array}\right\}$} \\ 
 6 & modified ESS & PCA method & $3\times2$pt &  \\ 
 7 & modified ESS & Linear Scale Cuts & $3\times2$pt$+f\sigma_8$ &  \\ 
 8 & modified ESS & PCA method & $3\times2$pt$+f\sigma_8$ &  \\ 
 \hline
 \end{tabular}}
 \caption{Gravity model choices for the simulated analyses discussed in Section \ref{sec:results}. This table lists the modified gravity parameters, while the cosmological and nuisance parameters follow the fiducial values given in Table \ref{table:p_fid_and_priors}. We display the posteriors for our key results in Figures \ref{fig:corner_GR_nofsigma8} and \ref{fig:corner_ESS_nofsigma8}; all remaining posteriors are used for further validation of the PCA method.}
\label{table:Parameter_inference_models}
\end{table}

To compare how good a fit our model is to the data under the different data reduction methods, we will be quoting the change in the $\chi^2$, $\Delta \chi^2$:

\begin{equation}\label{eq:chisquared}
    \Delta \chi^2 =  [\mathbf{D} - \mathbf{M}(\mathbf{p})]^T \mathbf{C}^{-1} [\mathbf{D} - \mathbf{M}(\mathbf{p})].
\end{equation}

A shift of $\Delta \chi^2 \leq 1$ in the case of a Gaussian likelihood implies that the resulting parameter bias, for a linear parameter, is at most $\sim1\sigma$, indicating that the model remains a statistically acceptable fit to the data \cite{DESCSRD2021lsst}. However, the mapping between parameters and observables is often nonlinear, and biases may predominantly affect nuisance parameters, and so have limited impact on the inferred cosmological parameters of interest. Therefore, large deviations in $\Delta \chi^2$ are not necessarily indicative of significant biases to cosmological parameters in the final analysis, as we will see in Section \ref{sec:results_ESS}.

In Sections \ref{sec:results_GR} and \ref{sec:results_ESS}, we show that the PCA-based data reduction method provides a significant improvement in the constraining power of 3$\times$2pt(+$f\sigma_8$) data when compared to linear scale cuts, and reduces the misestimation in cosmological parameters due to linear-only modelling, for the case of a true underlying model of both GR and ESS gravity. In Appendix \ref{Appendix:f(R)_nDGP_tests}, we provide additional validation for further examples of the true underlying theory of gravity (nDGP and $f(R)$) for a 3$\times$2pt$+f\sigma_8$ data vector. We display the full posterior distributions in Appendix \ref{Appendix:contours}.

\subsection{Improved constraints and degeneracy breaking for GR gravity with the PCA method}\label{sec:results_GR}

\begin{figure*}
    \centering
    \includegraphics[width=0.8\linewidth]{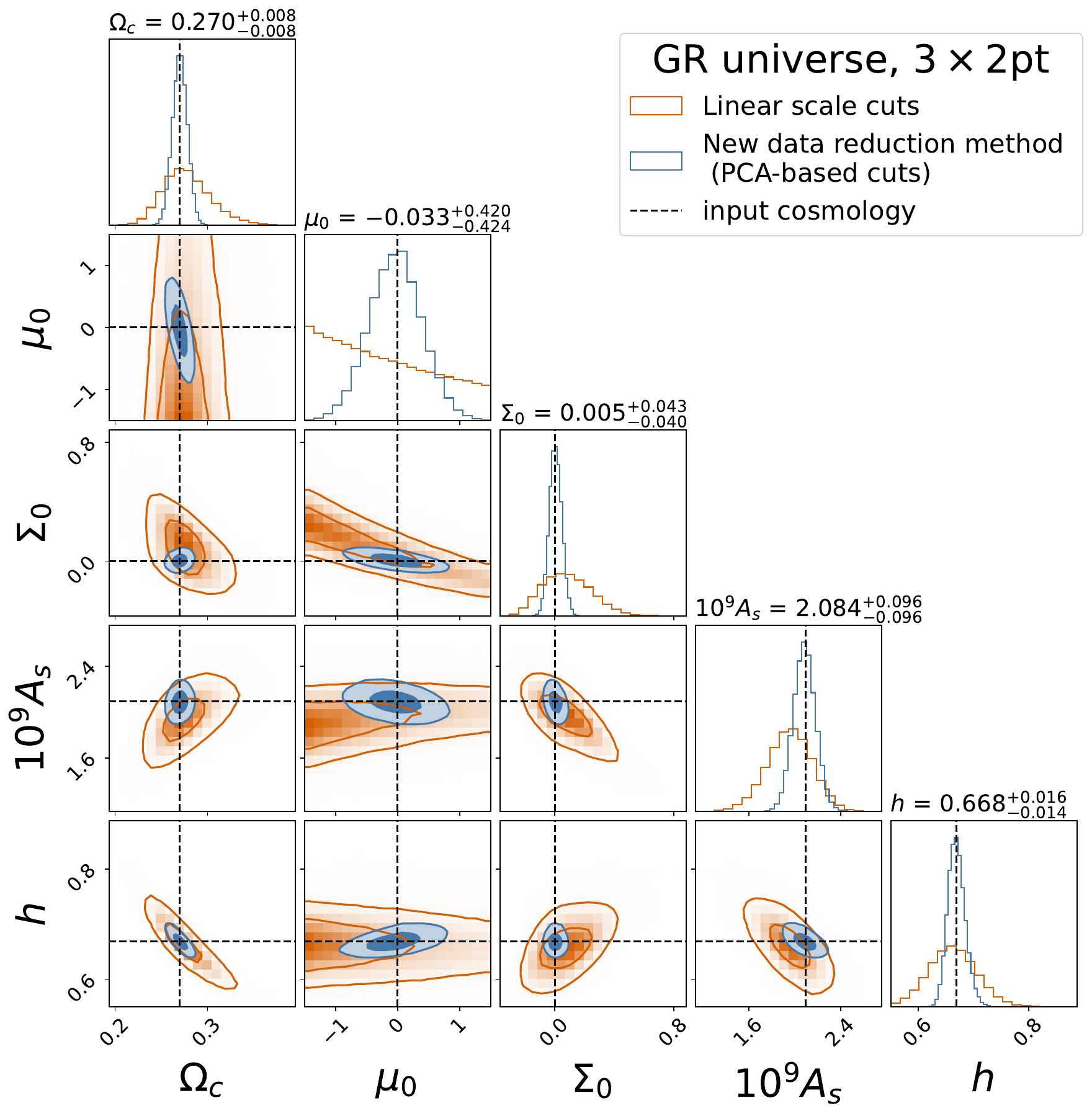}
    \caption{\textbf{Posteriors 1 and 2 - $\Lambda$CDM}  Contours of the posteriors for our cosmological parameter inference over $\mathbf{p} = \{\mathbf{p}_{\text{co}}, \mathbf{p}_{\text{nu}},\mu_0, \Sigma_0\}$, marginalised over $\omega_b$, $n_s$ and nuisance parameters $\mathbf{p}_{\text{nu}}$. We are using noiseless 3$\times$2pt GR simulated data. The orange contours show constraints when using scale cuts from Section \ref{sec:conservative_scalecuts}, and blue contours when using the data reduction method from Section \ref{sec:likelihood_PCA}. We see a substantial improvement in the constraining power for 3$\times$2pt data using our PCA method as opposed to standard linear-only methods.}
    \label{fig:corner_GR_nofsigma8}
\end{figure*}

We present the results of a simulated analysis constraining a $\Lambda$CDM$+\mu_0+\Sigma_0$ model using a General Relativistic 3$\times$2pt data vector, comparing data reduction techniques by applying either linear scale cuts or our PCA-based method. In Figure \ref{fig:corner_GR_nofsigma8}, we show marginalised posterior distributions for the parameter subset of interest. We see immediately that the PCA method maintains unbiased cosmological parameter constraints while notably breaking the degeneracy between modified gravity parameters $\mu_0$ and $\Sigma_0$ for 3$\times$2pt-only data. This degeneracy has been previously recorded in the literature as a serious limitation of this probe (see for example in \cite{ishak2019modified}); the PCA method solves this issue self-consistently in a 3$\times$2pt analysis without needing to introduce additional datasets, by retaining more information during data reduction. Additionally, the PCA-based data reduction decreases the $\Delta\chi^2$ (defined by \ref{eq:chisquared}) to 0.71 from 1.2 when using linear scale cuts, showing that the model under these new cuts provides a better fit to the data when compared to standard linear scale cuts.

\begin{figure}
    \centering
    \includegraphics[width=1.0\linewidth]{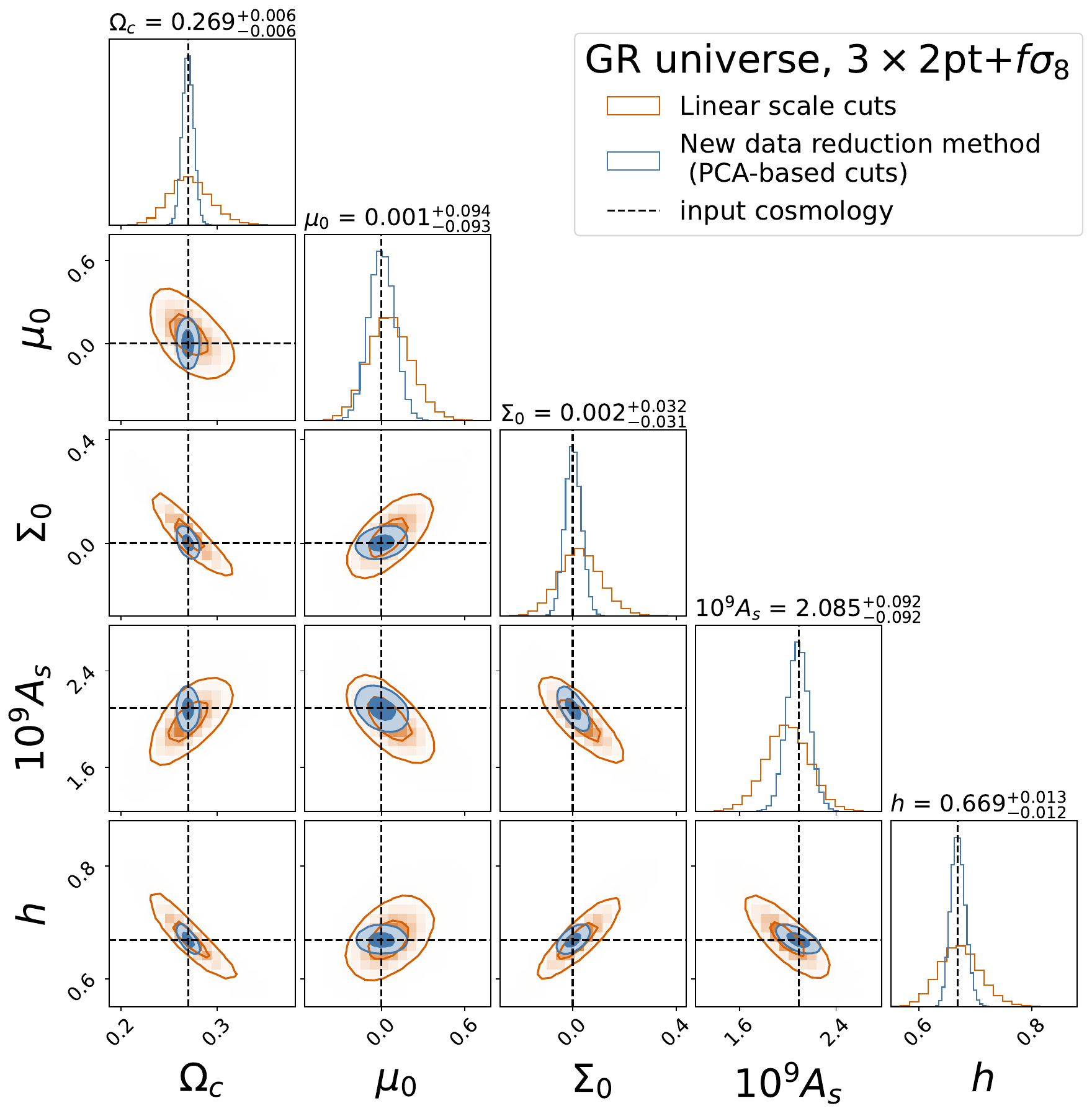}
    \caption{\textbf{Posteriors 3 and 4 - $\Lambda$CDM} Contours of the posteriors for our cosmological parameter inference over $\omega_b$, $n_s$ and $\mathbf{p} = \{\mathbf{p}_{\text{co}}, \mathbf{p}_{\text{nu}},\mu_0, \Sigma_0\}$, marginalised over nuisance parameters $\mathbf{p}_{\text{nu}}$. We are using noiseless 3$\times$2pt$+f\sigma_8$ GR simulated data. The orange contours show constraints when using scale cuts from Section \ref{sec:conservative_scalecuts}, and blue contours when using the data reduction method from Section \ref{sec:likelihood_PCA}.}
    \label{fig:corner_GR}
\end{figure}

While these results are already very promising, we note that a more realistic analysis would include additional data from auxiliary cosmological probes. This would increase the significance of a modified gravity detection, but also require a higher accuracy threshold from our data reduction and modelling techniques. We also want to measure quantitatively the relative constraining power of the two methods for data reduction, and in order to do so in a fair way, we should compare constraints from posteriors with less pronounced degeneracy directions and more Gaussian likelihoods. In Figure \ref{fig:corner_GR}  we therefore show marginalised posteriors for a GR 3$\times$2pt$+f\sigma_8$ dataset. This shows that the significance of the detection of the modified gravity parameter $\mu_0$, measured as a ratio of the width of one standard deviation in the marginalised posterior, is 1.65 times larger in the analyses which implemented PCA-based cuts than those implementing linear scale cuts for GR gravity. The $\Delta\chi^2$ (defined by \ref{eq:chisquared}) is 0.40, compared to 1.5 when using linear scale cuts, showing that the model under these new cuts provides a significantly better fit to the data. For the interested reader, Appendix \ref{Appendix:f(R)_nDGP_tests} shows the equivalent contours for nDGP and $f(R)$ simulated data, further validating the PCA method on theories used to inform data reduction. 

Overall, these results show that the PCA method is more effective at maintaining the data's constraining power when compared to linear scale cuts. We now want to test whether the PCA method is able to provide precise and unbiased constraints when applied to datasets from theories outside its training set.

\subsection{Unbiased constraints in ESS gravity}\label{sec:results_ESS}

While we have shown that the PCA method provides improved and unbiased constraints in a simulated Universe based on one of the data reduction models from which it was developed, it may still lack the flexibility to provide unbiased constraints when applied to a MG theory not included in the data reduction process. To be of general use, a novel data reduction method should be versatile enough to constrain a broader class of theories beyond those used to compute the Principal Components in the reduction matrix. We address this concern by demonstrating that the PCA method accurately identifies modified gravity behaviour without introducing unacceptable levels of cosmological parameter bias, even when testing a strong MG signature that is not explicitly modelled in the data reduction process.

\begin{figure*}
    \centering
    \includegraphics[width=0.8\linewidth]{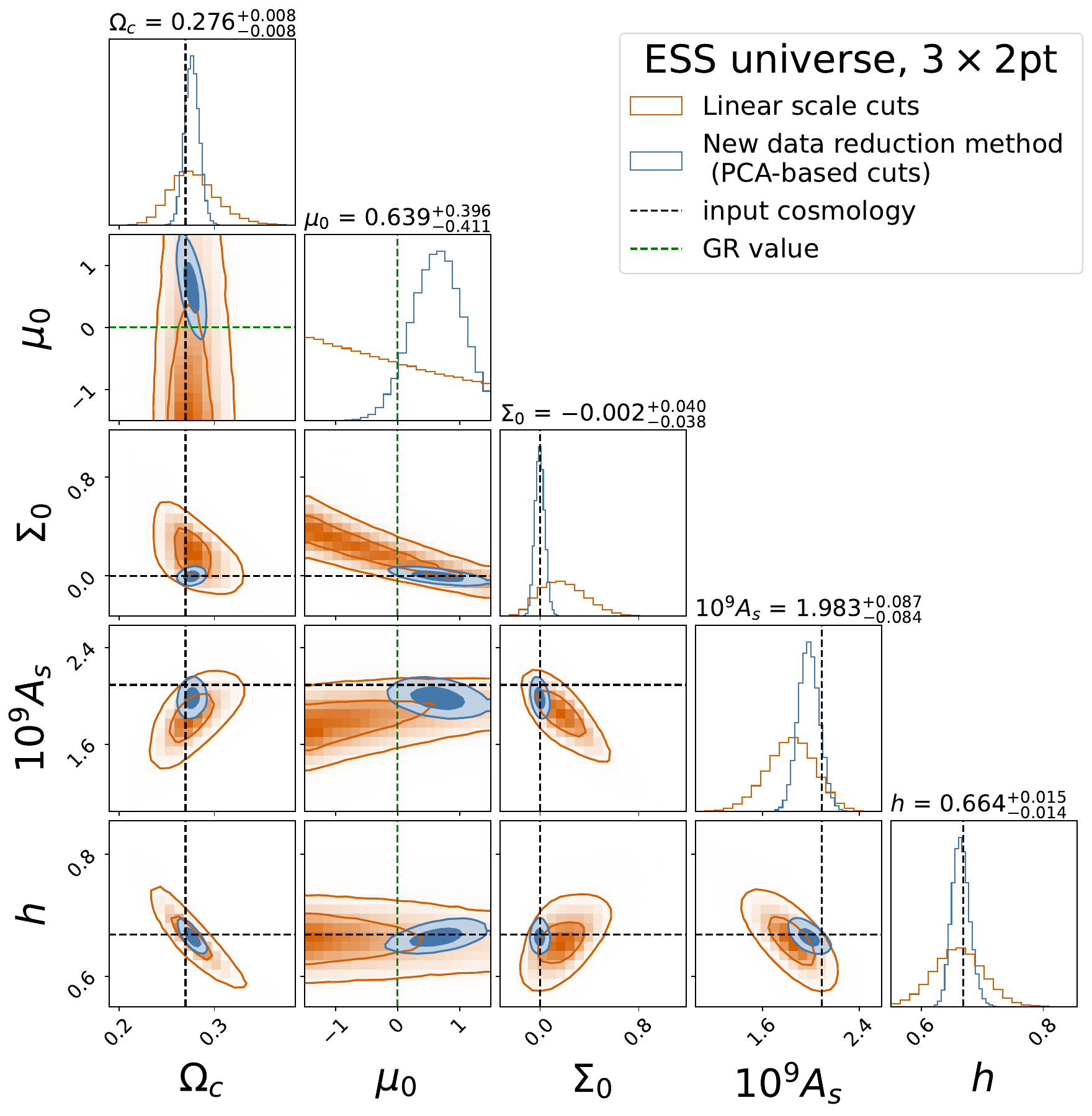}
    \caption{\textbf{Posteriors 5 and 6 - ESS} Demonstration of the fundamental improvement in MG signature detection for 3$\times$2pt data. Contours of the posteriors for our cosmological parameter inference over $\omega_b$, $n_s$ and $\mathbf{p} = \{\mathbf{p}_{\text{co}}, \mathbf{p}_{\text{nu}},\mu_0, \Sigma_0\}$, marginalised over nuisance parameters $\mathbf{p}_{\text{nu}}$. We are using noiseless 3$\times$2pt `modified ESS' simulated data. The orange contours show constraints when using scale cuts from Section \ref{sec:conservative_scalecuts}, and blue contours when using the data reduction method from Section \ref{sec:likelihood_PCA}.}
    \label{fig:corner_ESS_nofsigma8}
\end{figure*}

Specifically, we consider the case where the true Universe is described by the `modified ESS' gravity model introduced in Section \ref{sec:ESS_theory}, with a $\Lambda$CDM background, scale-independent structure growth modifications to GR, Vainshtein screening, and additionally modified to have a GR-like Weyl potential ($\Sigma = 1$). We show the posterior distributions of the parameter subset of interest in Figure \ref{fig:corner_ESS_nofsigma8} for a noiseless simulated data vector. Most notably, this shows a remarkable improvement in the power of the MG detection when using the PCA-based data reduction method, compared to using linear scale cuts. For a 3$\times$2pt LSST Y1-like analysis, the PCA method can effectively pick up the signature of `modified ESS' gravity without significant bias in the estimated cosmological parameters (when compared to linear scale cuts). We see that a non-GR value of $\mu_0$ is preferred at the $1.6\sigma$ level when using the PCA-based data reduction method, compared to a negligible preference with linear scale cuts. We note here that because there is no direct relation between $\mu_0$ and the ESS modified gravity parameters, we have no way of comparing whether this is the expected value from an unbiased analysis. Still, this is a powerful proof of the versatility and potential of this data reduction method. 

Taking a more detailed look at Figure \ref{fig:corner_ESS_nofsigma8}, we observe a $1.2\sigma$ deviation from the true value in the parameter $A_s$. We note there is a bias in $A_s$ for posteriors in both data reduction methods ($1.3\sigma$ for linear scale cuts). The $\Delta\chi^2$ is 1.8 for PCA-based data reduction, and 1.7 when using linear scale cuts, further suggesting that our best-fit model is not a particularly good fit to the data under either form of data reduction. Upon further analysis, we found that these biases are in part due to the specific choice of redshift-dependent ansatz in the $\mu-\Sigma$ parameterisation -- i.e., Equation \ref{eq:mu_sigma_ourparam} does not accurately model ESS gravity (for details, we refer the reader to Appendix \ref{Appendix:noparam_ESS_MM}). We have tested that these biases remain $\lesssim 1\sigma$ when the effects of the redshift parameterisation choice are accounted for. We emphasize that the bias introduced by the redshift parameterisation choice is \textit{not} the same as that introduced by our data reduction techniques. It cannot be solved by choosing between the different nonlinear mitigation schemes we are investigating in this paper. A more detailed discussion of this topic is provided in Section \ref{sec:MG_param_bias}.

\begin{figure}
    \centering
    \includegraphics[width=1.0\linewidth]{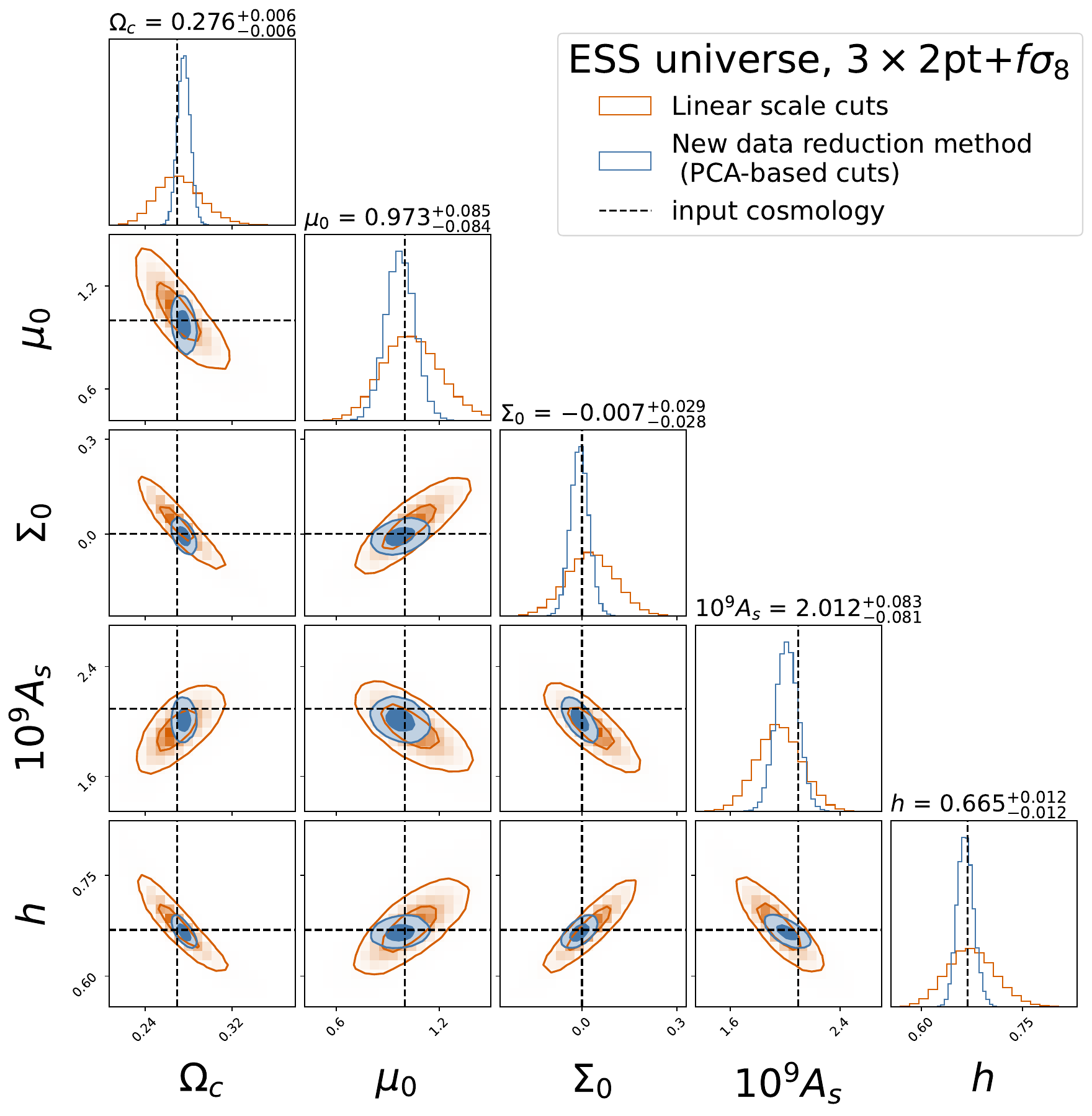}
    \caption{\textbf{Posteriors 7 and 8 - ESS} Contours of the posteriors for our cosmological parameter inference over $\mathbf{p} = \{\mathbf{p}_{\text{co}}, \mathbf{p}_{\text{nu}},\mu_0, \Sigma_0\}$, marginalised over $\omega_b$, $n_s$ and nuisance parameters $\mathbf{p}_{\text{nu}}$. We are using noiseless 3$\times$2pt$+f\sigma_8$ `modified ESS' simulated data. The orange contours show constraints when using scale cuts from Section \ref{sec:conservative_scalecuts}, and blue contours when using the data reduction method from Section \ref{sec:likelihood_PCA}.}
    \label{fig:corner_ESS-C}
\end{figure}

Following Section \ref{sec:results_GR}, we now consider the scenario of a (more constraining) 3$\times$2pt$+f\sigma_8$ data vector. In Figure \ref{fig:corner_ESS-C}, we show posterior distributions for this case. In order to remove any model misspecification biases, the linear MG parameterisation used to create the model vector will be a `theory-specific' parameterisation (defined in Equation \ref{eq:mu_Sigma_ESSparam} below) rather than the DE-like parameterisation from Equation \ref{eq:mu_sigma_ourparam} (see Section \ref{sec:MG_param_bias} for an explanation of this). This new parametrization has an additional advantage of having a direct correspondence between our ESS modified gravity parameters and $\mu_0$, meaning that in this case we get an input truth value for our modified gravity parameters ($\mu_0 = 1, \Sigma_0 = 0$) as well as our cosmological parameters. We stress here that this is not representative of a real analysis, but a fine-tuned parameterisation used purely for the purposes of validation in order to isolate the causes of misestimation of cosmological parameters.

The $\Delta\chi^2$ (defined by \ref{eq:chisquared}) for the PCA-based data reduction is 1.75, compared to 1.74 when using linear scale cuts. This is an improvement in the goodness of fit compared to using the DE-like parametrization, and now the PCA-based data reduction results in as good a fit to the data as to the standard scale cuts method. Still, the parametrization choice is not the only factor affecting the goodness-of-fit mismatch between the model and simulated data.  We are considering an extreme deviation from the standard model, so the biases introduced by the lack of non-linear modelling remain significant even after PCA reduction. We note here that our simulated datavector represents a worse-case scenario far beyond the modifications expected in a realistic analysis.

Additionally, as we have outlined in the previous Section, large deviations in $\Delta \chi^2$ are not necessarily indicative of significant systematic error in the final analysis.
The posteriors display a $\leq1\sigma$ deviation from the truth value for the parameters $\Omega_c$ and $A_s$, and a negligible deviation for all other parameters. 

Additionally, the modified gravity parameter $\mu_0$ is detected to be non-zero at the $11.5\sigma$ level in the posteriors which implemented PCA-based cuts when compared to a $6.2\sigma$ detection for posteriors implementing linear scale cuts, a $1.85 \times$ improvement in the significance of the detection. The results from Figure \ref{fig:corner_ESS-C} therefore indicate that, for an $f\sigma_8+$3$\times$2pt LSST Y1-like analysis, the PCA method can effectively constrain ESS gravity without significant bias in the estimated cosmological parameters of interest, while considerably improving parameter constraints when compared to linear scale cuts. 

\subsection{Model misspecification: a note on the choice of redshift parameterisation for $\mu$ and $\Sigma$}\label{sec:MG_param_bias}

We find that the DE-like redshift parameterisation of $\mu(z)$ and $\Sigma(z)$ as given in Equation \ref{eq:mu_sigma_ourparam} biases some of our results, especially when also introducing $f\sigma_8$ data. This reflects an existing body of literature which highlights the inadequacy of the DE-like parameterisation at detecting cosmological MG signatures, e.g. \cite{Linder_2017, Denissenya_2017, Denissenya_2022, Raveri_2023,Wen_2023}. While this effect is minimal for most realistic modifications from GR, which are small because of existing gravity constraints in the literature (see, e.g., \cite{Raveri_2023}), it becomes more pronounced for MG models with stronger modifications to structure growth, namely our `modified ESS' gravity model (with $\sim 18\%$ more structure than GR at large scales and low redshifts). This can already be ruled out with high significance using existing $f\sigma_8$ data alone, and we see sigificant bias when this data is analysed with the DE-like parameterisation.

When performing the equivalent analysis shown in Figure \ref{fig:corner_ESS-C} using the DE-like parameterization from Equation \ref{eq:mu_sigma_ourparam}, we observed substantial biases in most cosmological parameters relative to their simulated true values ($\geq 1\sigma$). Appendix \ref{Appendix:noparam_ESS_MM} quantifies the residual cosmological parameter biases explicitly. Additionally, Figure \ref{fig:corner_ESS_paramcompare} provides the PCA-reduced contours from \ref{fig:corner_ESS-C} alongside the equivalent contours under the assumption of a DE-like parametrization. Notably, this bias persisted regardless of the data reduction method used, but becomes $\lesssim 1\sigma$ when adopting a more suitable parameterisation. This suggest that part of the biases we see in cosmological parameters are indeed caused by an inadequate redshift parameterisation choice for $\mu(z)$.

To address the poor performance of the DE-like parameterization in describing $f\sigma_8$ signatures for ESS gravity, we introduce a parameterisation explicitly designed to model deviations from our input ESS model:\begin{eqnarray}\label{eq:mu_Sigma_ESSparam}
    \mu(a) = \mu_0 \left(\mu_{\text{ESS}}(a, \mathbf{p}^{\text{fid}}_{\text{co}},\mathbf{p}_{\text{ESS}}^{\text{fid}}) -1\right) + 1 \\
    \Sigma(a) = 1 + \Sigma_0 \frac{\Omega_{\Lambda}(a)}{\Omega_{\Lambda 0}}
\end{eqnarray}
where $\mu_{\rm ESS}$ represents the ESS-specific modification to structure growth, as defined in Equation \ref{eq:mu_Sigma_parameterisations}, for our specific choice of `true' input cosmological parameters and ESS modified gravity parameters. In principle, this function can be derived by solving for $\beta$ in equation (3.13) of \cite{Wright_2023}. This parameterisation is inherently unbiased around the input parameters, as $\mu_{\text{ESS}}(a)$ exactly matches the form of $\mu(a)$ in Equation \eqref{eq:mu_Sigma_parameterisations} for our specific test model (given our choice of modified gravity and cosmological parameters \textit{only}). We compute $\mu_{\text{ESS}}(a)$ (shown in Figure \ref{fig:ESS_Background_and_lingrowth}) by interpolating the output from the \texttt{HiCOLA} background computation module. We emphasize that this would never be a viable option in a real analysis; we use this `theory-specific' parameterisation only due to the lack of more suitable immediate alternatives in our proof-of-concept analysis. 

For completeness, we note that Figure \ref{fig:corner_nDGP} in the Appendices also displays this behaviour -- for nDGP gravity, the cosmological parameter $A_s$ is offset from its true value by $1-2\sigma$ when using PCA-based data reduction. Repeating the above analysis with nDGP gravity -- using Equation \ref{eq:mu_nDGP} as the parameterisation -- confirms that these small cosmological parameter biases vanish when adopting a more appropriate redshift parameterisation.

\section{Discussion and Conclusions}\label{sec:conclusion}

In this paper, we have introduced a new method for data reduction when constraining linear modified gravity parameterisations with $3\times 2$pt data, which decreases the information loss when compared to the standard data reduction method used in the literature, while still avoiding cosmological parameter bias. In this proof-of-concept analysis, we have demonstrated that this method provides strong, unbiased constraints for the case of a Universe described by GR, as well as for a beyond-$\Lambda$CDM model with fine-tuned yet significant deviations from GR, which was not included in our data reduction model set. By employing a Principal Component Analysis (PCA)-based approach, we identify and remove data components that are maximally biased from the effect of differences in linear and nonlinear modelling, offering an alternative to traditional linear scale cuts that often discard significant amounts of viable data. Crucially, this approach maintains a substantial amount of the constraining power from intermediate and small scales without increasing model complexity or introducing additional theoretical assumptions to those already present in the linear regime. In doing so, it removes degeneracies between modified gravity parameter posteriors which are otherwise common and problematic in $3\times2$pt-only data constraints (see e.g. \cite{DES_Y3_results}).

Through LSST Y1-like simulated analyses, we have demonstrated that this PCA-based method outperforms traditional linear scale cuts in a simplified simulated analysis across a variety of MG theories, both those used to construct the reduction matrix (GR, as well as nDGP and Hu-Sawicki $f(R)$ in Appendix \ref{Appendix:f(R)_nDGP_tests}) and an additional Horndeski theory that shares key characteristics with these data reduction models (ESS gravity).

Our analysis assumes all modified gravity theories of interest have a $\Lambda$CDM background, so any conclusions are only valid if the dark energy behaviour is accurately determined.
Additionally, our validation in this work has been limited to models with scale-independent modification to growth and no additional modification to lensing, which is a restricted but important subclass of modified gravity theories. 
While this study represents a minimal proof of concept, it highlights the potential of PCA-based data reduction to increase the constraining power in linear modified gravity parameterisation analyses without compromising accuracy. 

We note several limitations of our final analysis. The number of training theories (i.e. the data reduction models used in building our reduction matrix) is limited to three. These include the null case (GR) and two theories which cover the two most well-known screening mechanisms in the literature. While this selection is sufficient to demonstrate the potential improvement of our data reduction method over the standard linear scale cuts approach, a broader range of models on which to base our reduction matrix would offer more complete protection against biased results across a more comprehensive range of modified gravity behaviour. In a similar vein, an important step for future work will be to validate our method over a more extended sample of simulated data vectors from the MG theory space (not just the training theories and one additional theory, ESS gravity). Additionally, to get this method ready for application to real data in Stage-IV surveys, we need to account for nuisance cosmological parameters more comprehensively (crucially, photo-$z$ uncertainty and intrinsic alignment have been neglected here), and include data from auxiliary probes in a more realistic way (including redshift-space galaxy clustering and CMB probes, which in this analysis are either modelled under simplifying assumptions or included through the means of a prior only).

Despite these limitations, our findings highlight the potential of this data reduction method for improved parameter inference in WL and LSS analyses of modified gravity. While further validation and method refinement are necessary to address the outlined challenges, the demonstrated improvements in parameter constraints and the minimal parameter bias -- albeit for specific MG scenarios only -- are promising.

Beyond our approach, alternative methods exist that aim to improve upon traditional linear scale cuts. For instance, the BNT transform has been proposed for use in Euclid (see \cite{Taylor_2021, Taylor_2022}) by introducing a natural basis for cosmic shear kernels for mitigating the effects of baryonic physics and nonlinear structure growth. Incorporating such approaches could enhance both standard linear scale cuts and PCA-based data reduction, potentially increasing the constraining power of both. Future work should explore integrating these methods with our data reduction approach.

Alternative approaches for extracting information from the nonlinear regime when testing modified gravity parameterisations are actively being developed (e.g., \cite{Thomas_2020, srinivasan_2024b, Hassani_2020}). These primarily focus on building novel theoretical frameworks to model nonlinear MG effects. Our approach here offers a parallel solution: rather than constructing nonlinear MG parameterisations that model a wide range of MG theories, we develop a data reduction method capable of accommodating any MG theory, provided the necessary modelling tools are incorporated into the analysis. The key relative advantage of our method is that it only requires MG parametrizations to be modelled for the linear theory, enabling us to leverage well-validated assumptions, including scale-independence and specific functional forms. This can strike a balance between maintaining constraining power by having a small parameter space and ensuring we are still able to constrain a theoretically motivated and broad class of MG theories. 

Furthermore, our approach can be interpreted as an implicit marginalization over a subset of nonlinear parameters, which should asymptotically be model-independent as the sample of training MG theories grows. This means that the results coming from our method and from methods explicitly modelling these gravity-agnostic nonlinear MG effects should give the same results if a large enough theory and parameterisation space are respectively sampled. A direct comparison with nonlinear mitigation schemes of this kind in future work would be highly informative, shedding light on the interplay between theoretical modeling and simulation-driven data reduction in MG analyses. Bridging the gap between parameterisations and models is essential -- if we trust that our parameterisations serve as meaningful proxies for our models, we should be confident in integrating tools from both approaches to enhance our understanding and constraints of modified gravity.

Further extensions of this work could implement a more comprehensive treatment of the $\mu-\Sigma$ functions as a function of redshift in order to deal with the model misspecification issues brought up in Section \ref{sec:MG_param_bias}. The most rigorous testing of theoretically motivated deviations from GR would involve modelling these functions non-parametrically -- e.g., treating $\mu$, $\Sigma$ and possibly an additional dark energy component (such as $\Omega_X$ from \cite{pogosian2022imprints}) as Gaussian Processes or cubic splines with a Horndeski prior, while still keeping them functions of $a$ only \cite{Ruiz_Zapatero_2022}. 

Implementing such approaches presents several challenges, including the limited availability of emulators that simultaneously vary MG parameters and $\omega_0-\omega_a$ (as an exception, see the \href{https://github.com/renmau/Sesame\_pipeline}{\texttt{Sesame}} emulator). Tools like \texttt{ReACT}\footnote{https://github.com/nebblu/ReACT} or \texttt{HiCOLA} could be valuable for future work by providing fast, semi-accurate nonlinear MG power spectra to be used in the reduction matrix, either directly (if computationally feasible) or by aiding in the construction of emulators for specific theories. Applying this novel method to non-parametric constraints on $\mu(a)$, $\Sigma(a)$ and $\Omega_X(a)$ using linear-only probes and $3\times2$pt data would be a worthwhile endeavour. 

As a further generalisation, one could substitute our training theories for more general non-linear parametrised models with different screening mechanisms \cite{Bose_2023}. This would ensure there exists a direct relation between the parameters $\mu(a)$ and $\Sigma(a)$ and a subset of the modified gravity parameters in the non-linear training theories.

With further validation and development, the PCA-based data reduction method proposed here will provide a valuable addition to the standard toolkit for upcoming WL and LSS surveys, enhancing our ability to probe gravitational physics on small scales and to detect extensions to the $\Lambda$CDM model.

\appendix

\section{Additional Tests -- constraints on data reduction models \MakeLowercase{n}DGP and $\MakeLowercase{f}(R)$}\label{Appendix:f(R)_nDGP_tests}

\begin{table}
\centering
 \scalebox{0.8}{\begin{tabular}{|c|c|c|c|c|c|} 
 \hline\hline
 Index & True Theory & Data Reduction & Data & MG parameter & Value \\ [0.5ex] 
 \hline\hline
 9  & nDGP  & Linear Scale Cuts & $3\times2$pt$+f\sigma_8$ & $H_0r_c$ & $1.0$ \\ 
 10 & nDGP  & PCA method       & $3\times2$pt$+f\sigma_8$ & $H_0r_c$ & $1.0$ \\ 
 11 & $f(R)$ & Linear Scale Cuts & $3\times2$pt$+f\sigma_8$ & $f_{R0}$ & $10^{-5}$ \\ 
 12 & $f(R)$ & PCA method       & $3\times2$pt$+f\sigma_8$ & $f_{R0}$ & $10^{-5}$ \\ 
 \hline
 \end{tabular}}
 \caption{Gravity model choices for the simulated parameter inference models discussed in Appendix \ref{Appendix:f(R)_nDGP_tests}. This table lists the modified gravity parameters, while the cosmological and nuisance parameters follow the fiducial values given in Table \ref{table:p_fid_and_priors}.}
\label{table:Parameter_inference_models_fR_nDGP}
\end{table}

\begin{figure}
    \centering
    \includegraphics[width=0.8\linewidth]{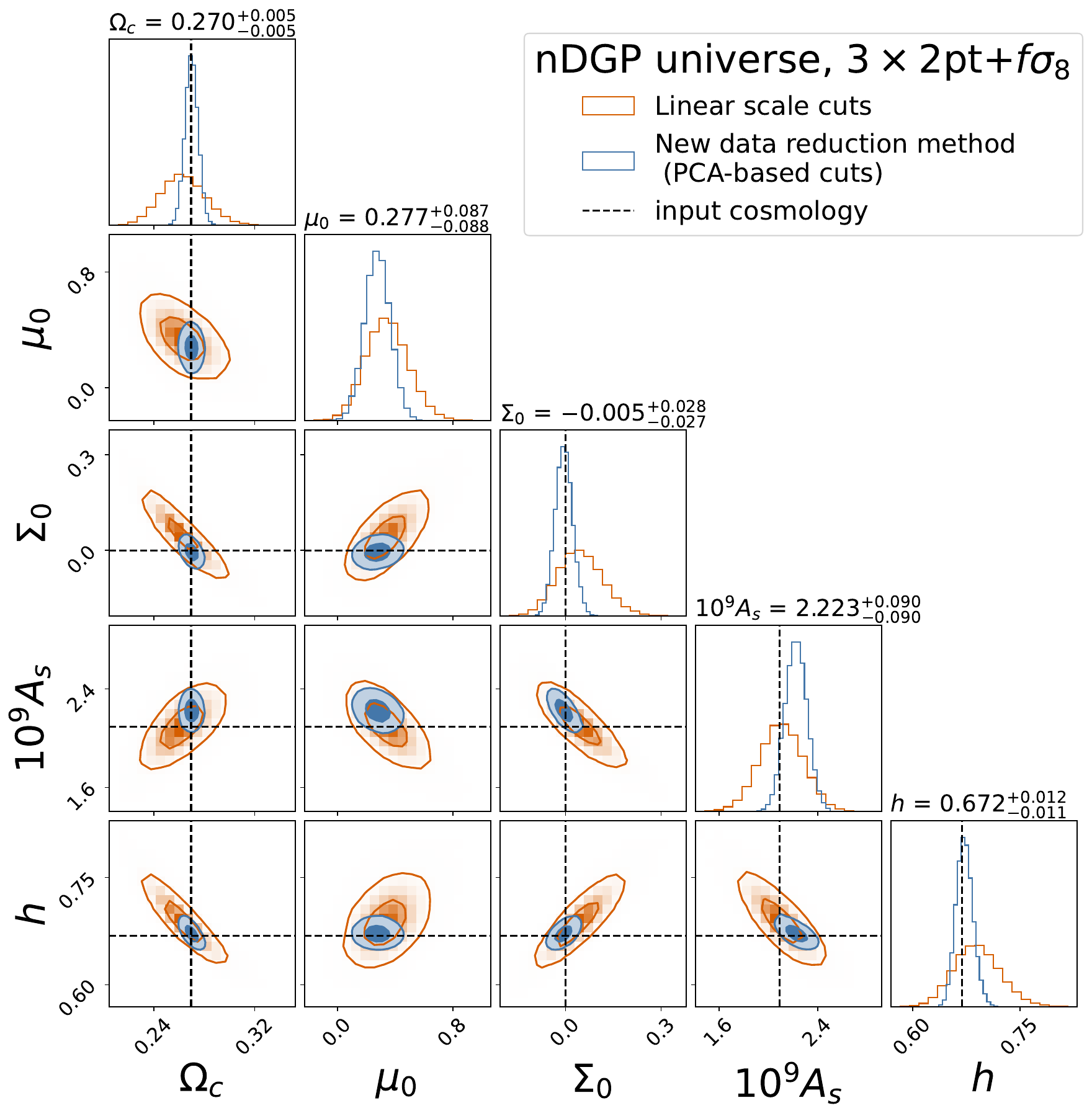}
    \caption{\textbf{Posteriors 9 and 10 - nDGP} Cosmological parameter inference over $\mathbf{p} = \{\mathbf{p}_{\text{co}}, \mathbf{p}_{\text{nu}},\mu_0, \Sigma_0\}$, marginalised over $\omega_b$, $n_s$ and $\mathbf{p}_{\text{nu}}$. We are using noiseless 3x2pt nDGP simulated data, with $H_0r_c = 1.0$. The orange contours show constraints when using scale cuts from Section \ref{sec:conservative_scalecuts}, and blue contours when using the data reduction method from Section \ref{sec:likelihood_PCA}.}
    \label{fig:corner_nDGP}
\end{figure}

\begin{figure}
    \centering
    \includegraphics[width=0.8\linewidth]{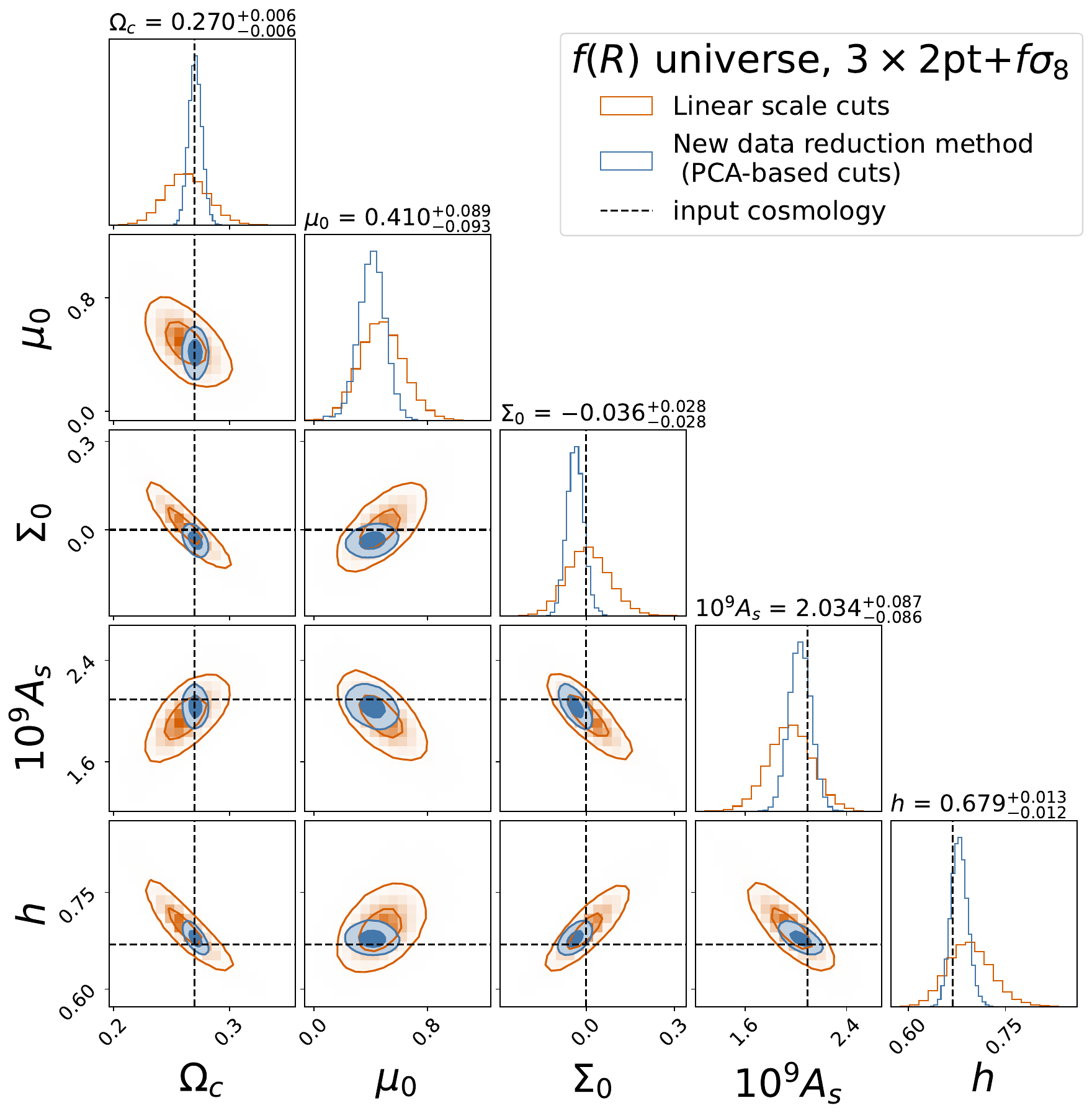}
    \caption{\textbf{Posteriors 11 and 12 - $\mathbf{f(R)}$} Contour plots for our cosmological parameter inference over $\mathbf{p} = \{\mathbf{p}_{\text{co}}, \mathbf{p}_{\text{nu}},\mu_0, \Sigma_0\}$, marginalised over $\omega_b$, $n_s$ and $\mathbf{p}_{\text{nu}}$. We are using noiseless 3x2pt $f(R)$ simulated data, with $f_{R0} = 10^{-5}$. The orange contours show constraints when using scale cuts from Section \ref{sec:conservative_scalecuts}, and blue contours when using the data reduction method from Section \ref{sec:likelihood_PCA}.}
    \label{fig:corner_fR}
\end{figure}

As further validation, we provide in Figures \ref{fig:corner_nDGP} and \ref{fig:corner_fR} the equivalent constraints as those described in Figure \ref{fig:corner_GR}, but instead modelling a universe described by the two remaining data reduction models, respectively nDGP and Hu-Sawicki $f(R)$. For this, we choose to simulate MG universes with large MG signatures, namely $H_0r_c=1.0$ ($\sim14\%$ deviation from GR on linear scales) and $f_{R0} = 10^{-5}$ (see Table \ref{table:Parameter_inference_models_fR_nDGP}). We do this because looking at more extreme (even if unrealistic) cases ensures a more rigorous validation. We display the full posterior distributions in Appendix \ref{Appendix:contours}. 

The posteriors from the PCA analysis for the $f(R)$ simulated data vector display a $\sim1\sigma$ deviation from the truth value in the $\Sigma_0$ parameter, and $<1\sigma$ in all cosmological parameters. The equivalent posteriors for nDGP have a $1-2\sigma$ deviation from the `true' input in the parameter $A_s$. Similarly to in the ESS case, we find that these small biases arise from the choice of redshift parameterisation of $\mu$ and $
\Sigma$ from Equation \ref{eq:mu_sigma_ourparam} rather than the application of the PCA data reduction (see Section \ref{sec:MG_param_bias} for details). Taking this into account, the PCA data reduction method provides unbiased constraints on nDGP and $f(R)$ gravity.

Additionally, the significance of the detection of the modified gravity parameter $\mu_0$, measured as a ratio of the width of one standard deviation in the marginalised posterior, is $\sim1.7$ ($\sim3.7$) times larger in the analyses which implemented PCA-based cuts than those implementing linear scale cuts for $f(R)$ (nDGP) gravity. 

We note that $f\sigma_8$ exhibits scale dependence in Hu-Sawicki $f(R)$ gravity. We assume this scale dependence is negligible in our analysis at the values of $f_{R0}$ under consideration and at the relevant scales. Equation \eqref{eq:mu_ka_fR} indicates that this will be the case only at sufficiently large scales (where modified gravity effects are anyways negligible). While this needs to be validated and might not be a realistic approximation for a general $f(R)$ universe, in our simulated analysis, the data reduction method itself does not extend to the $f\sigma_8$ data, which is rather included for the purpose of increasing the overall constraining power of our dataset. Given that we are primarily concerned with the relative improvement in constraining power between data reduction methods, we defer proper modelling of scale-dependent $f\sigma_8$ to future work.

We also note briefly that, while the DE-like parameterisation assumes that the linear growth factor and growth rate are scale-independent, this is not the case for the $f(R)$ model. Fortunately, the scale-dependent features in $f(R)$ become more prominent on small scales, and effectively start to matter only for $k > 0.1 \text{ Mpc}^{-1}$ at $f_{R0} \leq 10^{-6}$ -- this justifies our assumption of approximate scale-independence for the large-scale application of this parameterisation in this analysis (see \cite{Mirzatuny_2019}). We additionally show that this assumption doesn't bias our results for our analysis setup for $f_{R0}= 10^{-5}$ (see Figure \ref{fig:corner_fR}).

These results are further evidence that the PCA method is effective at increasing the constraining power without introducing any significant bias for the theories it was trained on.

\section{Modified Gravity Linear parameterisations -- Model Misspecification}\label{Appendix:noparam_ESS_MM}

\begin{figure}
    \centering
    \includegraphics[width=0.8\linewidth]{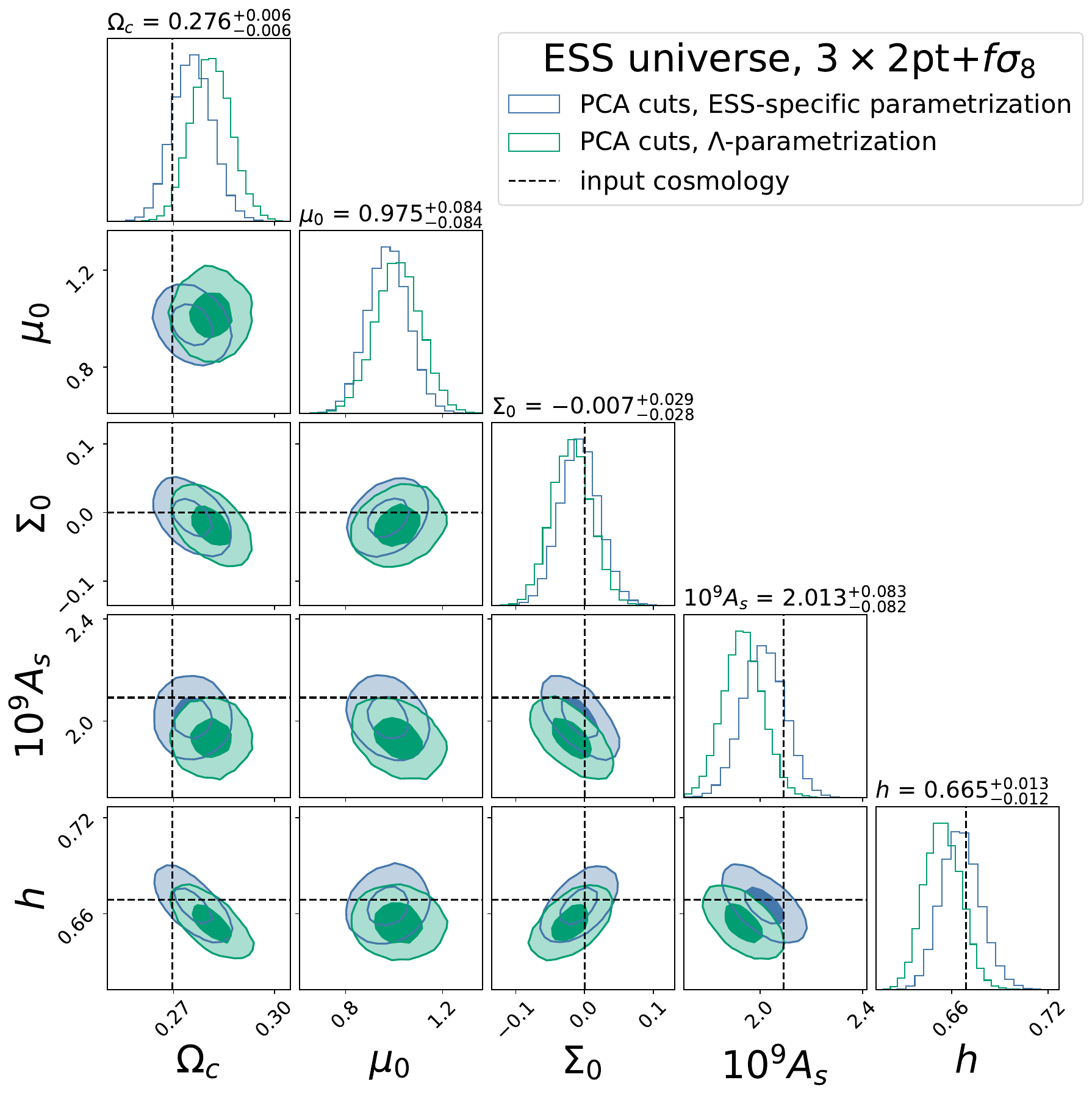}
    \caption{\textbf{Model misspecification -- ESS gravity} Marginalised posteriors for our cosmological parameter inference from a noiseless $3\times2$pt `modified ESS' simulated data vector. The blue contours represent the PCA-reduced posterior from Figure \ref{fig:corner_ESS-C}, while the teal contours correspond to the same analysis but use equation \ref{eq:mu_sigma_ourparam} for our linear modified gravity parameterisation instead of equation \ref{eq:mu_Sigma_ESSparam}.}
    \label{fig:corner_ESS_paramcompare}
\end{figure}

We show here an example of the biases introduced by the DE-like parameterisation when testing a modified gravity theory with a large deviation from GR. We present in Figure \ref{fig:corner_ESS_paramcompare} the marginalised posteriors when testing ESS gravity 3$\times$2pt$+f\sigma_8$ data, while constraining the DE-like parameterisation from Equation \ref{eq:mu_sigma_ourparam} compared to the ESS-specific parameterisation from Equation \ref{eq:mu_Sigma_ESSparam} with the PCA-based data reduction method. All cosmological parameters are biased at $>1\sigma$ level when using the DE-like parameterisation ($\Omega_c$ by $1.95\sigma$, $A_s$ by $2.05\sigma$ and $h$ by $1.25\sigma$) and at $\lesssim 1\sigma$ when using the ESS-specific parameterisation ($\Omega_c$ by $1.04\sigma$, $A_s$ by $0.97\sigma$ and $h$ by $0.25\sigma$). Additionally, the $\Delta\chi^2$ as defined in \ref{eq:chisquared} improves from $4.5$ to $1.75$. 

\begin{figure}
    \centering
    \includegraphics[width=1.0\linewidth]{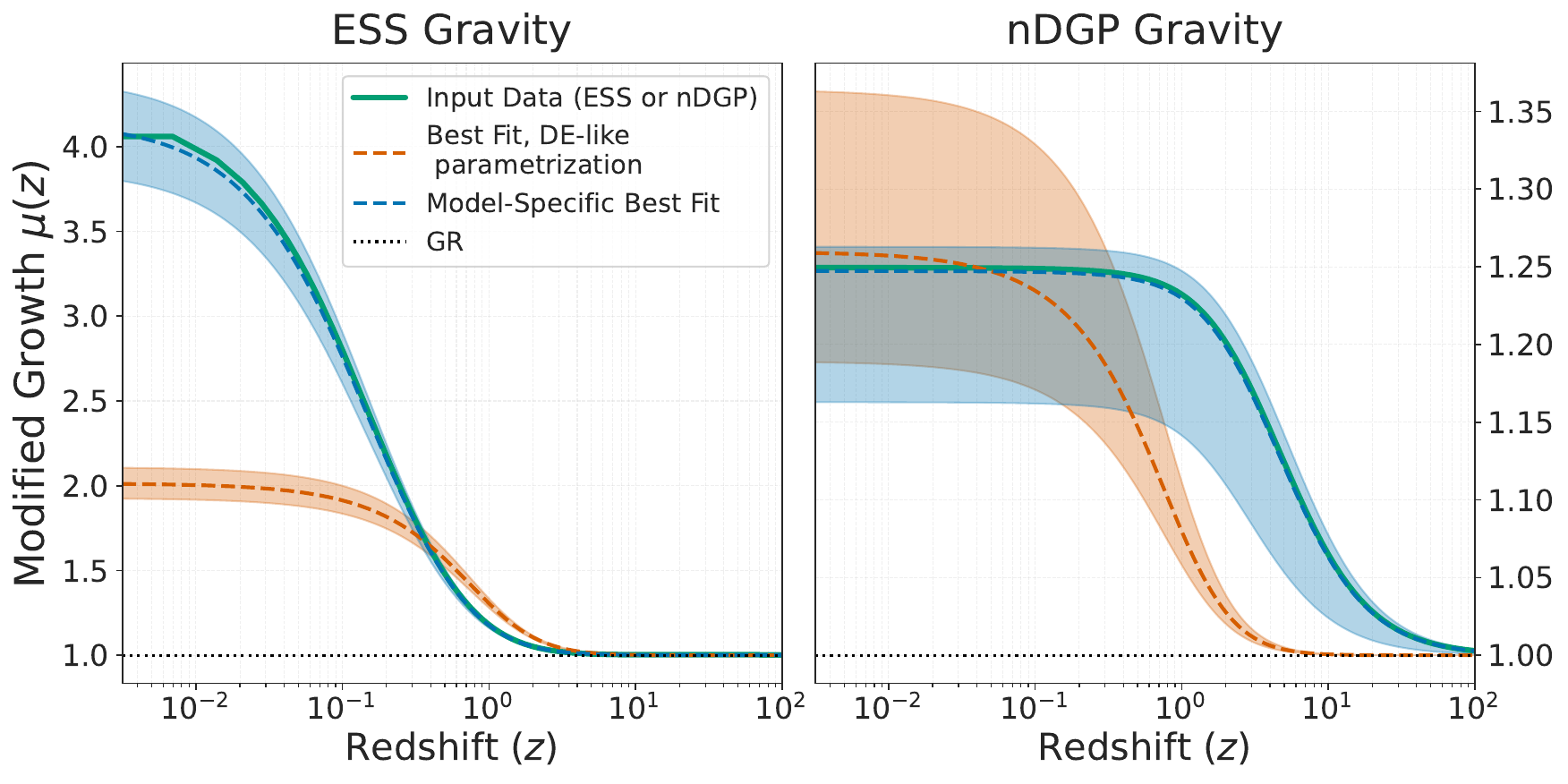}
    \caption{Comparison of the function $\mu(z)$ between that used to create the input data vector and those estimated through parameter inference for various parameterisation choices, in ESS and nDGP gravity.}
    \label{fig:muz_ESS_paramcompare}
\end{figure}

We also compare in Figure \ref{fig:muz_ESS_paramcompare} how well the output of these parameter inference chains fares at recovering the input functional form of $\mu(z)$. As well as the `modified ESS' example from Fig, \ref{fig:corner_ESS_paramcompare}, we repeat the equivalent analysis for nDGP gravity. We compare the best fit functional forms from chains that use the DE-like parameterisation to those from chains using the theory-specific parameterisation for both gravity theories. We can see that the DE-like parameterisation is very bad at recovering the correct functional form because it is quite limited in range. These biases in estimated $\mu(z)$ are what introduce biases in our estimated cosmological parameters.

\section{Sensitivity to Number of Data Reduction Models in PCA Basis}\label{Appendix:convergence_trainingtheories}

\begin{figure}
    \centering
    \includegraphics[width=0.8\linewidth]{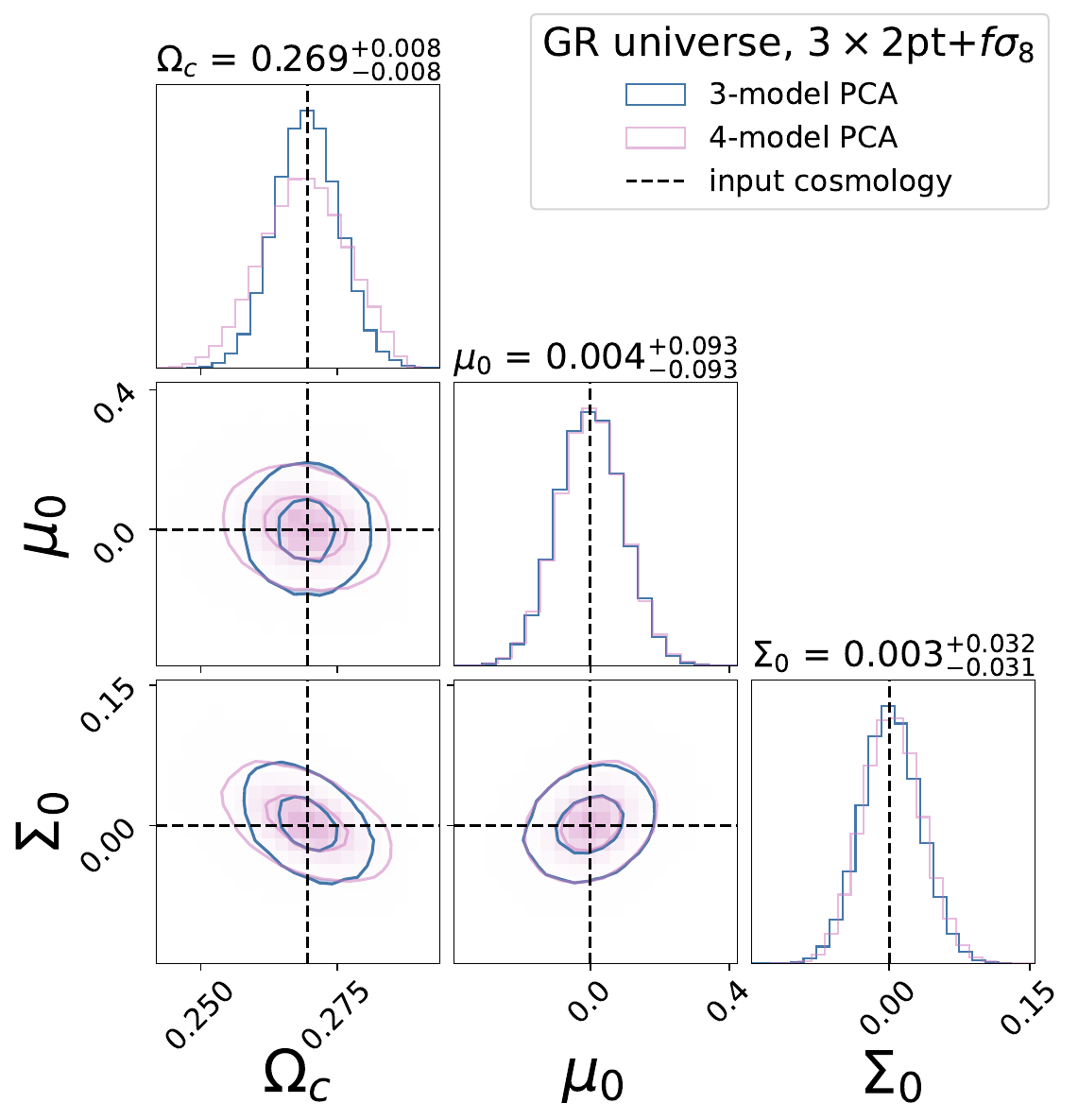}
    \caption{Posterior contours for our cosmological parameter inference over $\mathbf{p} = \{\mathbf{p}_{\text{co}}, \mathbf{p}_{\text{nu}},\mu_0, \Sigma_0\}$ using PCA marginalisation with three (blue) and four (pink) data reduction models. We are using noiseless $3\times2$pt GR simulated data -- the blue contour is the same as the blue contour in Figure \ref{fig:corner_GR}. The additional model corresponds to a nonlinear parametrisation with Vainshtein-type screening as described in Appendix \ref{Appendix:convergence_trainingtheories}.}
    \label{fig:Contours_4theories}
\end{figure}

The goal of this appendix is to quantify the completeness of our minimal PCA basis, which spans two screening mechanisms: Chameleon and Vainshtein. We extend the analysis from Section \ref{sec:results_GR} -- which results in Figure \ref{fig:corner_GR} -- by including an additional data reduction model in our PCA-based framework, as shown in Figure \ref{fig:Contours_4theories}. We test whether the posterior results are significantly affected when this additional data reduction model is included. 

We note that in this work, we consider simulated data produced using only theories which exhibit at most one of Vainshtein or Chameleon screening. We do not consider broader screening mechanisms (e.g., k-mouflage), as we expect that incorporating data reduction models with qualitatively new screening behavior could significantly enlarge the posteriors -- since the PCA basis would then be manifestly incomplete. We acknowledge that this is a significant limitation of the current work and one that will need to be addressed before this method can be reliably deployed on real data.

Given the current limitations of available emulators for MG theories, incorporating an entirely new, physically motivated MG model into the PCA basis is technically challenging. As a compromise, we introduce a non-linear parametrization model based on the Time-independent Growth Index Parametrisation with Screening using \texttt{ReACT-emus}\footnote{\href{https://github.com/nebblu/MGEmus/tree/main}{github.com/nebblu/MGEmus},\href{https://github.com/nebblu/ReACT-emus/tree/main}{github.com/nebblu/ReACT-emus}}, following the methodology outlined in \cite{Tsedrik_2024}. While this model is not tied to a complete underlying MG theory, it displays a Vainshtein-type screening mechanism in the nonlinear regime via the parameter $q_1$, and thus acts as an additional simple Vainshtein-like model. The linear power spectrum can then be found under QSA using the same procedure outlined in Section \ref{sec:nDGP_theory} -- by solving for $D(a)$ with Equation \ref{eq:growth_eq} -- using Equation (2.13) in \cite{Tsedrik_2024}. 

We adopt the fiducial choices of \cite{Tsedrik_2024}, fixing $\gamma^\text{fid}=0.4$ and $q_1^\text{fid}=0.76$, and apply the same extrapolation strategy for modes outside the emulator's $k$ range, with $k_\text{min} =0.016 \; h/\text{Mpc}$.

We then reconstruct a new PCA basis from four models (adding this parametrized model to the original three in our difference matrix $\mathbf{\Delta}$) and repeat the likelihood analysis of Section \ref{sec:results_GR}. The resulting posterior contours show negligible deviation from those obtained using three models. A representative comparison is shown in Figure \ref{fig:Contours_4theories}.

This result indicates that the parameter constraints are already close to convergence with three models: adding an additional mode to the PCA basis does not appreciably alter the posterior distribution, provided the true underlying screening behaviour is represented within the data-reduction model set. All marginalized posterior distributions remain broadly unchanged, with the exception of the $\Omega_{c}$ posterior, which has a mildly broader standard deviation (a factor of $\times 1.35$). We interpret this as an indication that the variation in nonlinear modelling due to the Vainshtein screening mechanism is already well-captured by the original basis. We do not claim full generality from this result -- however, it supports the working assumption that the set of three models used in the main analysis spans the relevant space of deviations in a way sufficient for the goals of this study.

\section{A note on $\mathit{A_s}$ and $\mathit{\sigma_8}$} \label{sec:As_sigma8_bias}
Usually, sampling the cosmological parameter space in $A_s$ rather than $\sigma_8$ makes no difference, given there is a direct relation between the two (all other cosmological parameters being equal). In our case, it makes a difference because it results in different power spectra contributing to our reduction matrix. For example, if we sample with respect to $A_s$, because our theory-specific MG parameters $\mathbf{p}_{\text{MG}^{m}}$ don't vary at each step of the chain, the value of $\sigma_8^{\text{MG}^m}(\mathbf{p}_{\text{co}},\mathbf{p}_{\text{MG}^{m}})$:

 $$
\sigma^{\text{MG}^{m}}_8 = \sqrt{\int \frac{k^2 P^{\text{MG}^m}(k,\mathbf{p}_{\text{MG}^{m}})}{2\pi^2} \left[\frac{3 j_1\left(k\times 8\text{Mpc}/h\right)}{k\times 8\text{Mpc}/h}\right]^2\, \text{d}k}
 $$

will be different from the value of $\sigma_8^{\mu- \Sigma}(\mathbf{p}_{\text{co}},\mu_0)$:
$$
 \sigma^{\mu- \Sigma}_8 = \sqrt{\int \frac{k^2 P(k,\mu_0)}{2\pi^2} \left[\frac{3 j_1\left(k\times 8\text{Mpc}/h\right)}{k\times 8\text{Mpc}/h}\right]^2\, \text{d}k}
$$
for most functional forms of $\mu(a)$ and values of $\mu_0$ in our analysis. In general, the matter power spectrum of our modified gravity linear parameterisation will be different to the linear power spectrum for the specific modified gravity theory (this becomes more obvious if one compares e.g. Eq.s \ref{eq:mu_sigma_ourparam} and \ref{eq:mu_nDGP}).
We have decided to vary over $A_s$ directly in our parameter exploration. We verify that this does not introduce any biases in our parameter inference, as shown in Section \ref{sec:results} and Appendix \ref{Appendix:f(R)_nDGP_tests}.

\section{Unmarginalised contour plots}\label{Appendix:contours}

We display the full (unmarginalised) contour plots for the GR, nDGP, $f(R)$ and ESS gravity models analysed in Section \ref{sec:results} and Appendix \ref{Appendix:f(R)_nDGP_tests}.

\begin{figure*}
    \centering
    \includegraphics[width=0.55\linewidth]{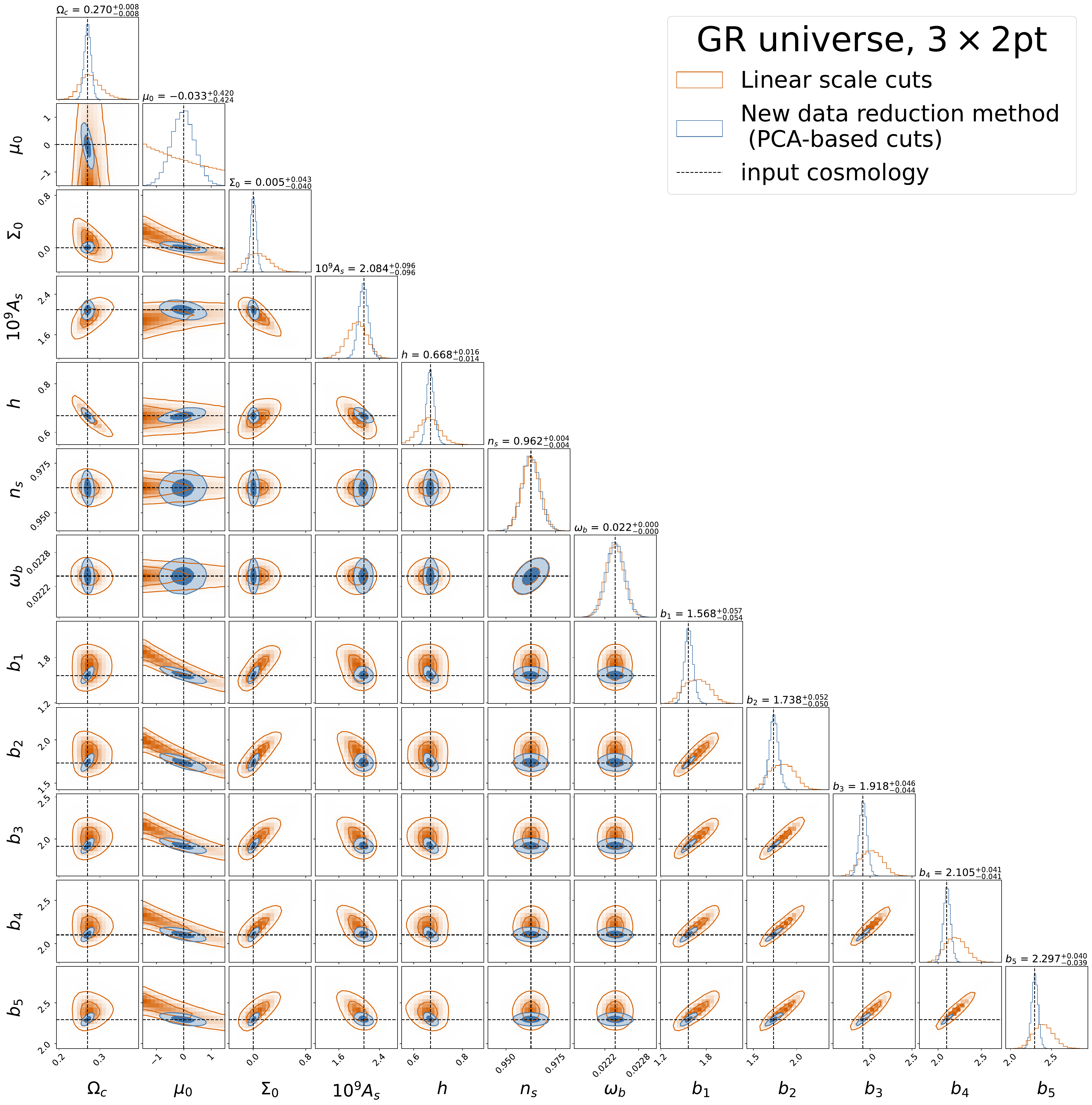}
    \caption{\textbf{Models  1 and 2 - $\Lambda$CDM} Full contour plots for our cosmological parameter inference over $\mathbf{p} = \{\mathbf{p}_{\text{co}}, \mathbf{p}_{\text{nu}},\mu_0, \Sigma_0\}$. The orange contours show constraints when using scale cuts described in Section \ref{sec:conservative_scalecuts}. The light blue contours show constraints using the data reduction method described in Section \ref{sec:likelihood_PCA}. The marginalised plots can be found in Figure \ref{fig:corner_GR_nofsigma8}.}\label{fig:corner_GR_nofsigma8_unmarg}
\end{figure*}

\begin{figure*}
    \centering
    \includegraphics[width=0.55\linewidth]{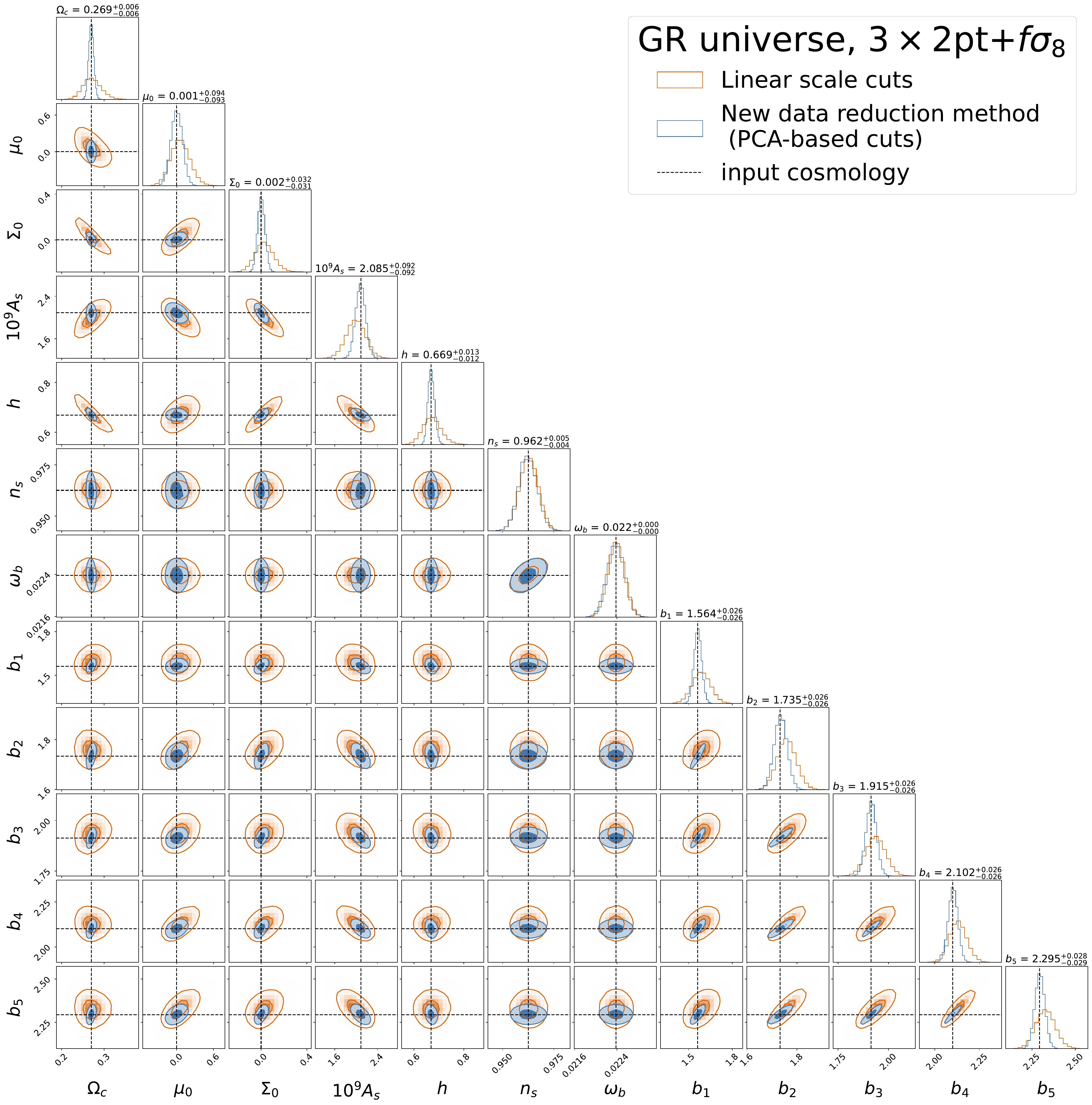}
    \caption{\textbf{Models 3 and 4 - $\Lambda$CDM} Full contour plots for our cosmological parameter inference over $\mathbf{p} = \{\mathbf{p}_{\text{co}}, \mathbf{p}_{\text{nu}},\mu_0, \Sigma_0\}$. The orange contours show constraints when using scale cuts described in Section \ref{sec:conservative_scalecuts}. The light blue contours show constraints using the data reduction method described in Section \ref{sec:likelihood_PCA}. The marginalised plots can be found in Figure \ref{fig:corner_GR}.}
    \label{fig:corner_GR_unmarg}
\end{figure*}

\begin{figure*}
    \centering
    \includegraphics[width=0.55\linewidth]{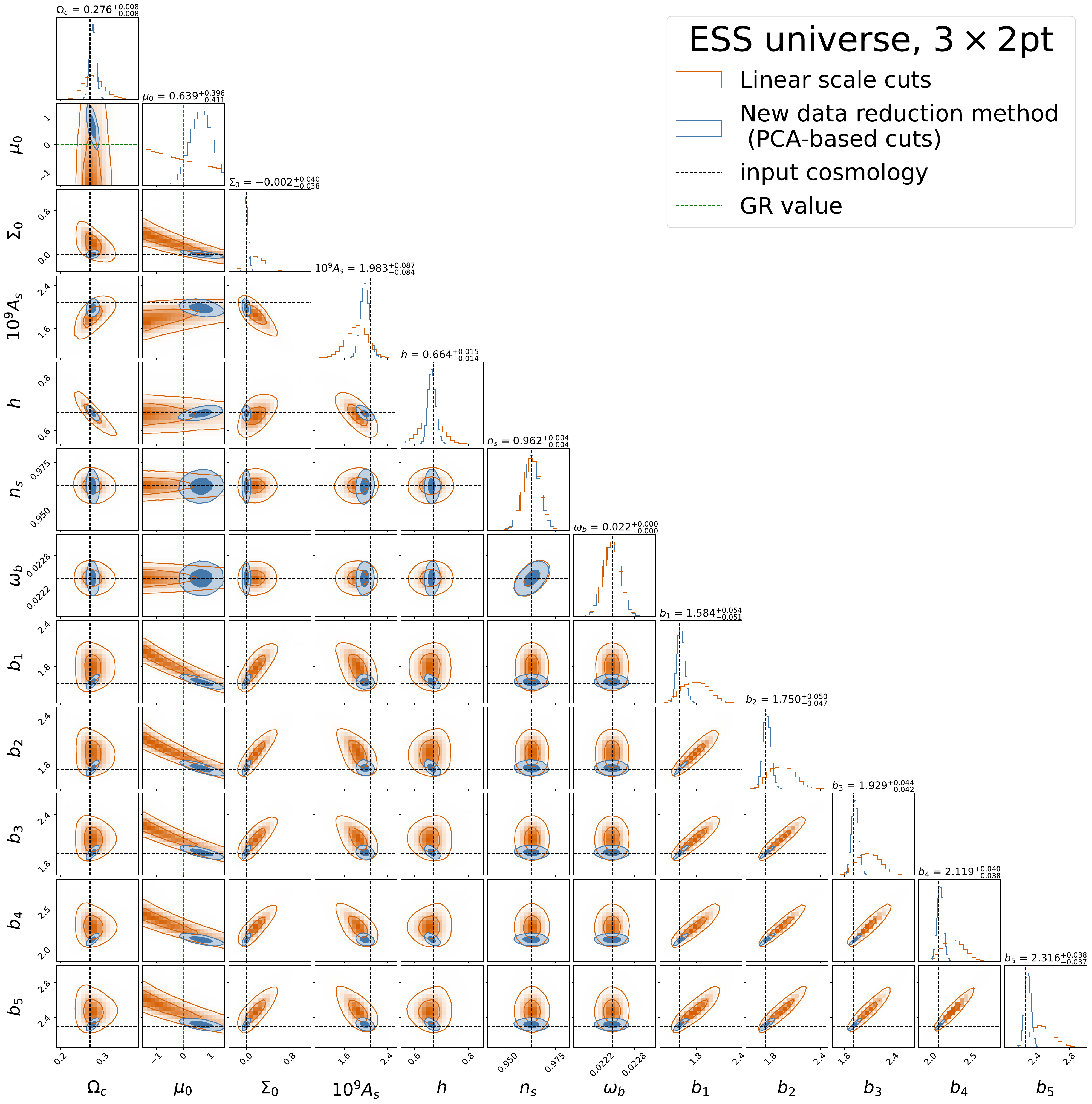}
    \caption{\textbf{Models  5 and 6 - ESS} Full contour plots for our cosmological parameter inference over $\mathbf{p} = \{\mathbf{p}_{\text{co}}, \mathbf{p}_{\text{nu}},\mu_0, \Sigma_0\}$. The orange contours show constraints when using scale cuts described in Section \ref{sec:conservative_scalecuts}. The light blue contours show constraints using the data reduction method described in Section \ref{sec:likelihood_PCA}. The marginalised plots can be found in Figure \ref{fig:corner_ESS_nofsigma8}.}
    \label{fig:corner_ESS_nofsigma8_unmarg}
\end{figure*}

\begin{figure*}
    \centering
    \includegraphics[width=0.55\linewidth]{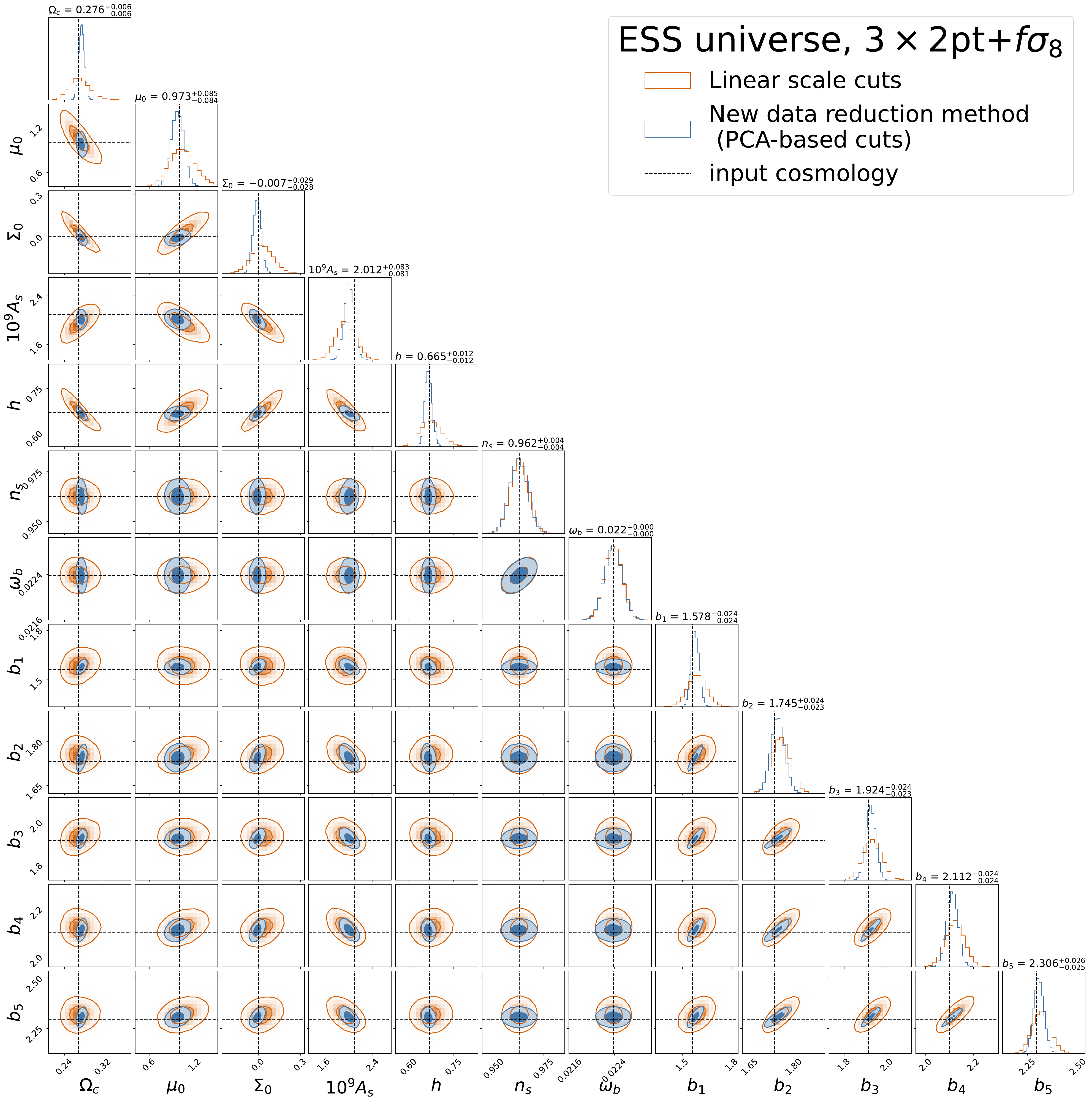}
    \caption{\textbf{Models 7 and 8 - ESS} Full contour plots for our cosmological parameter inference over $\mathbf{p} = \{\mathbf{p}_{\text{co}}, \mathbf{p}_{\text{nu}},\mu_0, \Sigma_0\}$. The orange contours show constraints when using scale cuts described in Section \ref{sec:conservative_scalecuts}. The light blue contours show constraints using the data reduction method described in Section \ref{sec:likelihood_PCA}. The marginalised plots can be found in Figure \ref{fig:corner_ESS-C}.}
    \label{fig:corner_ESS_unmarg}
\end{figure*}

\begin{figure*}
    \centering
    \includegraphics[width=0.55\linewidth]{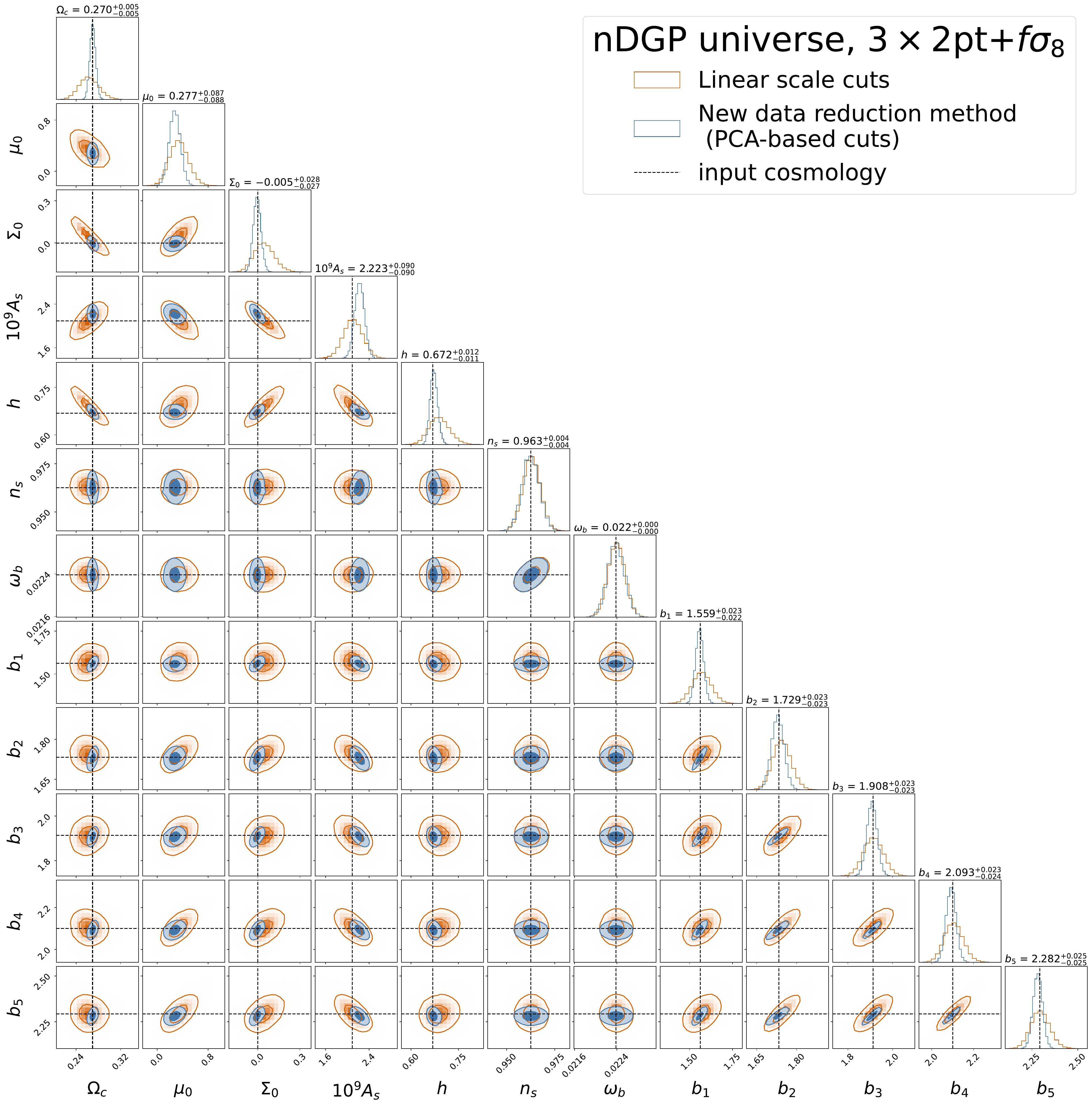}
    \caption{\textbf{Models 9 and 10 - nDGP} Full contour plots for our cosmological parameter inference over $\mathbf{p} = \{\mathbf{p}_{\text{co}}, \mathbf{p}_{\text{nu}},\mu_0, \Sigma_0\}$. The orange contours show constraints when using scale cuts described in Section \ref{sec:conservative_scalecuts}. The light blue contours show constraints using the data reduction method described in Section \ref{sec:likelihood_PCA}. The marginalised plots can be found in Figure \ref{fig:corner_nDGP}.}
    \label{fig:corner_nDGP_unmarg}
\end{figure*}

\begin{figure*}
    \centering
    \includegraphics[width=0.55\linewidth]{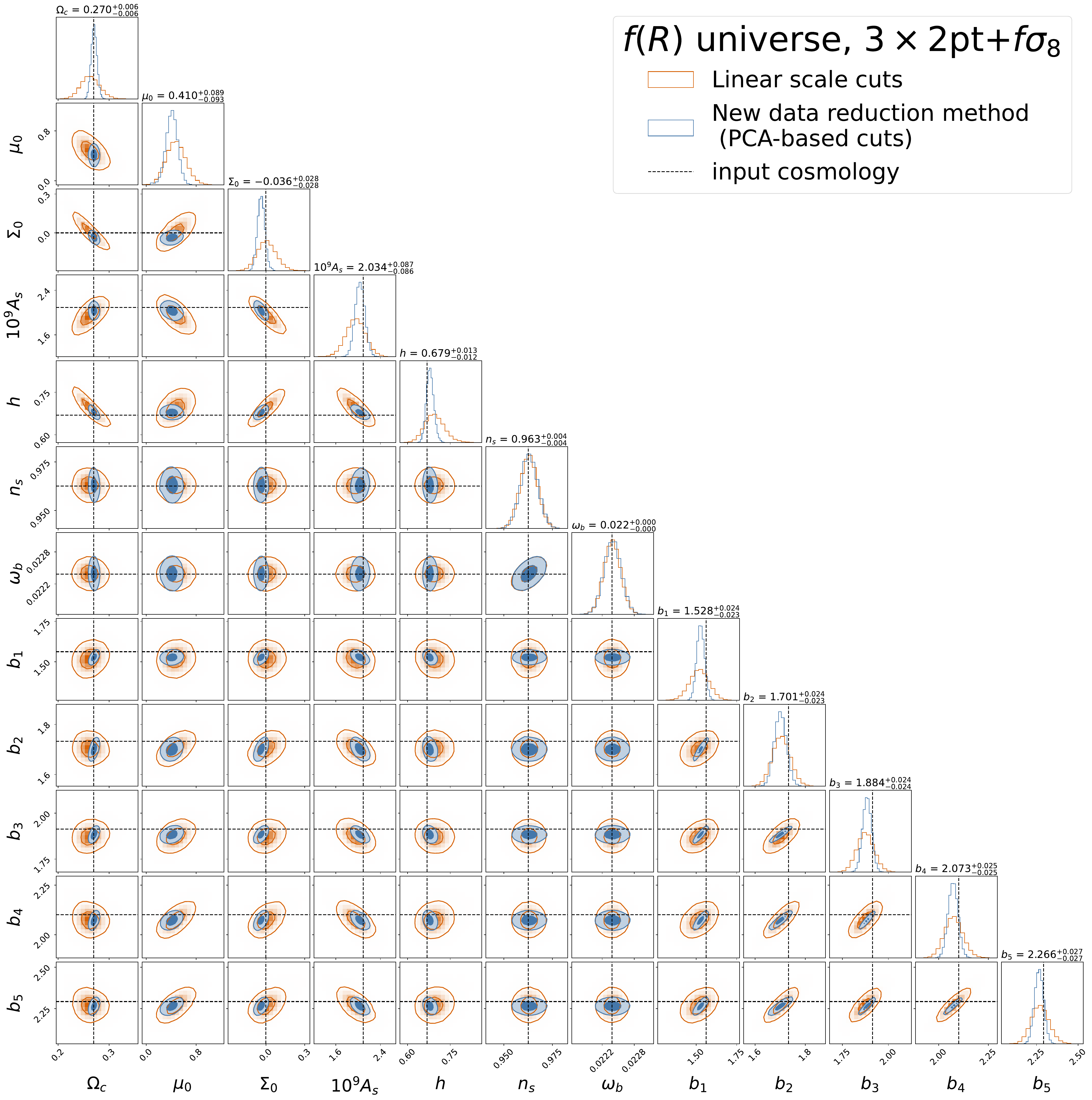}
    \caption{\textbf{Models 11 and 12 - $f(R)$} Full contour plots for our cosmological parameter inference over $\mathbf{p} = \{\mathbf{p}_{\text{co}}, \mathbf{p}_{\text{nu}},\mu_0, \Sigma_0\}$. The orange contours show constraints when using scale cuts described in Section \ref{sec:conservative_scalecuts}. The light blue contours show constraints using the data reduction method described in Section \ref{sec:likelihood_PCA}. The marginalised plots can be found in Figure \ref{fig:corner_fR}.}
    \label{fig:corner_fR_unmarg}
\end{figure*}

\begin{acknowledgments}
This work was supported by a Philip Robinson Cosmology PhD studentship. We thank Markus Rau, Ashim Sen Gupta, Bartolomeo Fiorini, Guillerme Brando, Charlie MacMahon-Gellér and Nikolina \v{S}ar\v{c}evi\'{c} for helpful insights and discussions. We thank Valeria Pettorino and Matteo Martinelli for providing helpful assistance with accessing Planck data products.

The Python libraries SciPy \cite{2020SciPy-NMeth}, NumPy \cite{harris2020array},
PyCCL\footnote{https://github.com/LSSTDESC/CCL} \cite{Chisari_2019}, corner.py \cite{corner} \texttt{TJPCov}\footnote{\hyperlink{https://github.com/LSSTDESC/tjpcov}{https://github.com/LSSTDESC/tjpcov}.} (LSST DESC, in preparation), \texttt{e-MANTIS}\footnote{\hyperlink{https://gitlab.obspm.fr/e-mantis/e-mantis}{gitlab.obspm.fr/e-mantis/e-mantis}} \cite{sáezcasares2023emantis} and \texttt{nDGPEmu} \citep{fiorini2023fast} were significant in enabling this work to be done.

For the purpose of open access, the author has applied a Creative Commons Attribution (CC BY) licence to any Author Accepted Manuscript version arising from this submission.
\end{acknowledgments}

\section*{Data Availability}
The data that support the findings of this article are openly available \cite{zanoletti2025mgpca}. 

\bibliographystyle{apsrev4-1}
\bibliography{bibliography}

\end{document}